\documentclass[preprint,12pt]{elsarticle}

\usepackage{amsmath}
\usepackage{graphicx}
\usepackage{enumerate}
\usepackage{booktabs} 
\usepackage{bm}
\usepackage[table,xcdraw]{xcolor}
\usepackage{changepage}
\usepackage{bbm}
\usepackage{amssymb}
\usepackage{upgreek}
\usepackage{geometry}
\usepackage{siunitx}
\usepackage{multirow}
\usepackage{url}

\usepackage{enumitem}
\setlist[itemize]{label=$\triangleright$}

\usepackage[normalem]{ulem}

\journal{}

\begin{document}

\begin{frontmatter}



\title{CAESar:\\Conditional Autoregressive Expected Shortfall}


\author[label1]{Federico Gatta}
\author[label1,label2]{Fabrizio Lillo}
\author[label3]{Piero Mazzarisi\corref{cor1}}
\ead{piero.mazzarisi@unisi.it}

\address[label1]{Scuola Normale Superiore, Pisa, Italy}
\address[label2]{Dipartimento di Matematica, Universit\`{a} di Bologna, Italy}
\address[label3]{Dipartimento di Economia Politica e Statistica, Universit\`{a} di Siena, Italy}
\cortext[cor1]{Corresponding author.}

\begin{abstract}
In financial risk management, Value at Risk (VaR) estimates potential portfolio losses but fails to account for losses beyond a certain threshold.  Expected Shortfall (ES) addresses this limitation by providing the conditional expectation of such exceedances, providing a better measure of tail risk. However, ES is not elicitable on its own, meaning that it cannot be estimated by minimizing some scoring function, although its joint elicitability with VaR allows for combined estimation. Building on this property, we propose the Conditional Autoregressive Expected Shortfall (CAESar) model, which flexibly handles dynamic patterns and heteroskedasticity, without making distributional assumptions on price returns. The optimization of CAESar coefficients involves three steps: fitting the VaR component via CAViaR regression, formulating ES as an autoregressive process, and jointly estimating VaR and ES coefficients while ensuring a monotonicity constraint to avoid crossing quantiles. Through extensive backtesting, CAESar outperforms existing methods, proving highly effective for risk forecasting.

\end{abstract}


\end{frontmatter}



\section{Introduction}
Risk management is a crucial aspect of the financial literature and industry, with Value at Risk (VaR) historically serving as the predominant measure. VaR, a tail quantile of the conditional distribution of portfolio returns, has been extensively studied and widely used by banks and other financial institutions for setting regulatory capital requirements over the past two decades. Despite its intuitive appeal, VaR lacks information regarding exceedances beyond the quantile. Furthermore, it is not a coherent risk measure as it is not sub-additive. To address this limitation, Expected Shortfall (ES) has emerged as a complementary measure, providing the conditional expectation of these exceedances. ES offers several theoretical advantages \cite{acerbi2002coherence} as it is a {\it coherent} risk measure \cite{artzner1999coherent}; in particular, in contrast to VaR, ES is {\it subadditive}, meaning that ES for a portfolio cannot be larger than the sum of ES's for the single assets in the portfolio. Coherence makes ES a more comprehensive measure of tail risk compared to VaR, and consequently, regulatory frameworks like Basel IV are increasingly emphasizing it \cite{feridun2020basel}. Basel IV is a set of regulations for banks established by the Basel Committee on Banking Supervision (BCBS), an international organism that includes bank supervisors from about 30 countries. Recently, with the introduction of the Basel IV framework, the BCBS has stated the shift from VaR to ES as the standard risk measure for the banking sector.

ES poses unique challenges, particularly because it is not {\it elicitable} on its own, meaning it cannot be directly estimated as the minimum of a scoring function \cite{weber2006distribution, gneiting2011making}. However, recent research has shown that VaR and ES are jointly elicitable, enabling their combined estimation using suitable scoring functions \cite{fissler2016higher}. This joint elicitability facilitates the evaluation and comparison of forecasting methods, a significant advancement in risk management. Building upon this research, some recent works have proposed a number of methods that can be grouped into static estimation, relying on Monte Carlo simulations \cite{rockafellar2000optimization} or high-frequency data \cite{dimitriadis2022realized}, and regressions, which can be seen as general extensions of the Conditional Autoregressive Value-at-risk by Regression quantiles (CAViaR) \cite{engle2004caviar}, in turn based on the general idea of quantile/expectile regression \cite{taylor2008estimating,de2009quantiles,bellini2017risk}. Everything stems from a simple intuition, which originates from the optimization of some properly defined scoring function permitting the inference of the pair (VaR, ES). In a regression context, price returns are observed as a time series on a daily time scale, and the two risk measures are regressed upon such observations (or nonlinear functions of these) and other covariates. Directly modeling VaR and ES avoids the need for a distributional assumption on price returns, achieving a more flexible description. This is key for financial time series, displaying fat tails and skewness, indicating that extreme returns are more common than predicted by a normal distribution and that the distribution may be asymmetrical. Additionally, while returns exhibit zero autocorrelation, volatility shows long-term memory and clustering; see \cite{cont2001empirical} for a review. Moreover, higher-order moments, like skewness and kurtosis, can be time-varying as well, see e.g. \cite{johnson2002volatility,wilhelmsson2009value}. These facts may have a crucial impact on the description of the tail of the distribution of returns, thus posing unique challenges for the estimation of VaR and ES \cite{bams2017volatility,pankratova2017method}. For example, when considering VaR, a Bayesian approach including nonlinear effects has been proposed for the general quantile regression problem \cite{gerlach2011bayesian}. Regression methods for conditional autoregressive Value at Risk (VaR) and Expected Shortfall (ES), similar to the approach of Engle and Manganelli \cite{engle2004caviar}, provide a well-suited and flexible framework for capturing time-varying patterns associated with tail risk, thanks to the inclusion of heteroskedastic effects.
In this context, it is worth mentioning the recent work of Patton et al. \cite{patton2019dynamic}, which proposes a GAS framework based on the model by Creal et al. \cite{creal2013generalized}. This framework imposes a parametric structure for the dynamics similar to the CAViaR model and defines a scoring rule that drives the evolution based on a proper loss function for jointly elicitable VaR and ES. Two model specifications are considered: one with a single factor and another with two factors, where the factors are score-driven parameters determining the joint dynamics of VaR and ES. The proposed Monte Carlo and empirical studies demonstrate the superior performance of the one-factor model, suggesting co-movement dynamics for the two tail risk measures. Within a similar context, Barrera et al. \cite{barrera2022learning} propose a general convergence analysis of a two-step approach to learn VaR and ES via neural networks, minimizing a proper loss function. Leveraging machine learning permits one to naturally account for general nonlinear patterns between regressors and tail risk. Finally, a natural approach for ES estimation is based on its definition as a tail mean, suggesting the adoption of a quantile regression approach for VaR estimation at different tail levels and then obtaining a consistent estimation of the ES as the mean of the VaR estimates, see \cite{emmer2013best,kratz2018multinomial}. See the section below for a detailed review of these works, which are benchmarks for our proposed method. 

In this paper, we contribute to the literature on regression methods for the joint estimation of the Value-at-Risk and the Expected Shortfall by proposing a \textcolor{black}{forecasting tool}, named Conditional Autoregressive Expected Shortfall (CAESar), inspired by the CAViaR model that can handle general dynamic patterns in a flexible way, without distributional assumptions on price returns. We include heteroskedastic effects for VaR and ES, whose dynamics are driven by regressors defined as nonlinear responses to the observed process. The methodology is in line with the aim of Patton et al. \cite{patton2019dynamic} \textcolor{black}{and Taylor \cite{taylor2019forecasting}}; in particular, we consider the same loss function \textcolor{black}{as in \cite{patton2019dynamic}} for jointly elicitable VaR and ES in the estimation process. However, our formulation is more flexible in the description of the tail risk by relaxing the constraint of the constant VaR-to-ES ratio of the one-factor specification \textcolor{black}{in \cite{patton2019dynamic} and the multiple specification for the ES in \cite{taylor2019forecasting}}. Furthermore, our VaR and ES forecasts are updated by always taking into account information about the price movement, in contrast to the \textcolor{black}{models in \cite{taylor2019forecasting} and \cite{patton2019dynamic}} that only use extreme events -i.e., those breaking the quantile. CAESar parameters estimation is a three-step approach. First, the conditional quantile VaR estimator for a given probability level is obtained using the well-known CAViaR regression \cite{engle2004caviar}. Second, we propose an autoregressive formulation for the ES based on quantile information. Third, VaR and ES are jointly estimated by exploiting joint elicitability.
We impose the inequality $ES_t\le VaR_t$ for all times $t$, for a fixed probability level (the ``monotonicity'' constraint), using a soft constraint on the loss function. In other words, a penalized estimation process for CAESar is adopted to ensure ``monotonicity'' constraint, thus avoiding the so-called 'crossing quantiles' problem, a typical issue in financial time series models \cite{han2016cross}. This results in more realistic dynamics for the VaR-ES pair.

Backtesting expected shortfall is a long-standing problem. In fact, non-elicitability has long been considered a limitation for backtesting \cite{carver2013mooted}, only recently surpassed, as pointed out in \cite{acerbi2014back}. A number of testing procedures have been proposed in recent years, see e.g. \cite{deng2021backtesting, du2023powerful}, even if no general agreement exists on the best approach. We corroborate our methodology on simulations and daily financial data by using a plethora of testing procedures (reviewed in detail below) recently proposed for backtesting ES, each one validating tail risk estimates from different perspectives. Interestingly, after an extensive evaluation and comparison with the state-of-the-art, we find a general agreement in terms of forecasting performance ranking, with CAESar outperforming any other regression method.

This paper is organized as follows. Section \ref{sec:problem} reviews the related literature and formalizes the problem. Section \ref{sec:model} introduces the CAESar methodology with a discussion of the theoretical advantages with respect to previous works. Section \ref{sec:tests} reviews the testing procedures for ES backtesting adopted in the experimental stage. Section \ref{sec:experiments} shows the Monte-Carlo and empirical results on the stock indices dataset. Section \ref{sec:conclusion} concludes. The Supplementary Material is organized as follows.  \ref{app:ms} discusses the specific functional form of our estimator. In  \ref{app:os}, the parameters optimization scheme is explained. \textcolor{black}{ \ref{app:srs} contains a simulation study within a regime-switching model that strengthens the novelty of CAESar and its outperformance with respect to competing methods.} \ref{app:eld} provides further insights into the results obtained by the statistical tests used to validate our proposal. Finally, in \ref{app:sd} we show the CAESar performance on the stock dataset.
\section{Problem Formulation and Related Literature}
\label{sec:problem}
The target problem involves the dynamic prediction of the ES for univariate time series. Consider a time series $\bm{y}=\{y_t\}_{t=1}^T$ and a probability level $\theta\in[0,1]$. If $\bm{y}$ represents the asset returns and $F_t$ the distribution function of $y_{t}$ conditional on the information set up to time $t-1$ ($\mathcal{F}_{t-1}$), then the VaR is defined as the conditional quantile of $\bm{y}$, that is, \textcolor{black}{$VaR_t(\theta) := \inf\left\{q\in\mathbb{R} \ s.t. \ F_t(q) \ge \theta \right\}$}. Instead, the ES (historically known as Conditional Value-at-Risk) is defined as the tail mean, that is, \textcolor{black}{$ES_t(\theta):=\frac{1}{\theta} \, \int_0^\theta VaR_t(\tau) \, d\tau$. Assuming the distribution function is strictly monotone and continuous, we can rewrite the above definitions as} $VaR_t(\theta) := F_t^{-1}(\theta)$ \textcolor{black}{and} $ES_t(\theta):=\mathbb{E}_{t-1}[y_t | y_t\le VaR_t(\theta)]$\textcolor{black}{\footnote{\textcolor{black}{In the case of a discontinuous distribution function at $VaR_t(\theta)$, the formulation of $ES_t(\theta)$ requires an adjustment term\textcolor{black}{; the formula reads $ES_t(\theta):=\mathbb{E}_{t-1}[y_t | y_t\le VaR_t(\theta)] + \frac{1}{\theta} VaR_t(\theta) \left(\theta - F_{t}(VaR_t(\theta))\right)$}.}}}, where $\mathbb{E}_{t-1}[\cdot]$ represents the mean conditional on $\mathcal{F}_{t-1}$. In the following, \textcolor{black}{we consider $\theta \in \{0.01, 0.025, 0.05\}$, as widely done in related works. Furthermore,} when no confusion can arise, we omit the explicit dependence of VaR and ES from $\theta$.\\
\linebreak
In the study of risk measures, a key role is played by the concept of elicitability \cite{brehmer2017elicitability, jaimungal2023risk}. In a nutshell, a functional is said to be elicitable if it can be obtained as the minimum of a loss function. To be more precise, we consider a set $A\subset\mathbb{R}^n$ with the Borel $\sigma$-algebra. Let $\mathcal{Y}$ be a family of $A$-valued random variables, and a functional $R:\mathcal{Y}\rightarrow \mathbb{R}^m$. The intuition is that $A$ represents the observable quantity, $Y\in\mathcal{Y}$ is a distribution for the elements in $A$, and $R$ is the risk measure we want to estimate. Next, consider a function $\mathcal{L}: A \times \mathbb{R}^m\rightarrow \mathbb{R}$. $\mathcal{L}$ is said to be a scoring function if $\forall z\in\mathbb{R}^m, \ \forall Y\in\mathcal{Y}$, we have that $\mathbb{E}\left[ |\mathcal{L}(Y, z)| \right]<\infty$.
Specifically, we are interested in a loss function that measures how well our guess $z$ fits the target quantity $R$. So, we call a scoring function $\mathcal{L}$ strictly consistent for the functional $R$ if $\forall z\in\mathbb{R}^m, \ \forall Y\in\mathcal{Y}$ with $z\neq R(Y)$, we have that $\mathbb{E}\left[ \mathcal{L}(Y, z) \right] > \mathbb{E}\left[ \mathcal{L}(Y, R(Y)) \right]$. Finally, the functional $R$ is said to be elicitable if it has a strictly consistent scoring function, that is if there exists a loss function that is minimized by $R$.\\
\linebreak
The estimation of VaR in a time series context has been widely discussed in the literature, starting from the seminal work on regression quantiles \cite{koenker1978regression}. In fact, VaR is elicitable thanks to the well-known Pinball (Quantile) loss function \cite{koenker1978regression}, defined as:
\begin{equation}
\label{eq:0127_1911}
\mathcal{L}^\theta_q(\bm{\hat{q}}, \bm{y}) := \frac{1}{T} \sum_{t=1}^T (y_t - \hat{q}_t)\left(\theta - \mathbbm{1}_{\{y_t < \hat{q}_t\}}\right)
\end{equation}
where $\bm{\hat{q}}=\{\hat{q}_t\}_{t=1}^T$ is the estimator of the Value-at-Risk time series $\bm{VaR}=\{VaR_t\}_{t=1}^T$. Among the different models for $\bm{\hat{q}}$, it is noteworthy to mention the CAViaR \cite{engle2004caviar}. It is an autoregressive model, where the target variable $\hat{q}_t$ is regressed against a (possibly multivariate) non-linear transformation of the returns $\bm{f}(y_{t-i})\in\mathbb{R}^d, \ i=1,\cdots,p$ and the estimates at previous time steps $\hat{q}_{t-j}, \ j=1,\cdots,u$ that is:
\begin{equation}
    \label{eq:0501_0922}
    \hat{q}_t = \beta_0 + \sum_{i=1}^p \bm{\beta}_{(i-1)d+1:id} \bm{\cdot} \bm{f}(y_{t-i}) + \sum_{j=1}^u \beta_{j+pd} \hat{q}_{t-j}
\end{equation}
where $\bm{\beta}_{(i-1)d+1:id}=\left(\beta_{(i-1)d+1}, \cdots, \beta_{id}\right)$, and $\bm{\cdot}$ represents the usual scalar product. Despite the great flexibility of CAViaR, researchers usually focus on the case $p=u=1$, as it is empirically shown to provide fast and reliable (that is, not prone to overfitting) dynamics estimation. Furthermore, $\bm{f}$ is usually chosen as the Asymmetric Slope (AS), which accounts for the asymmetry in asset returns by mapping $y_{t-1}$ into the pair (hence $d=2$) $\left( (y_{t-1})^+, (y_{t-1})^- \right)$, where $(y_{t-1})^+:=\max(0, y_{t-1})$ and $(y_{t-1})^-:=\max(0, -y_{t-1})$, respectively. Thus, the CAViaR specification (also used below as a baseline model) is:
\begin{equation}
\label{eq:0127_1913}
\hat{q}_t = \beta_0 + \beta_1 (y_{t-1})^+ + \beta_2 (y_{t-1})^- + \beta_3 \hat{q}_{t-1}
\end{equation}
\textcolor{black}{Since its proposal, there have been some attempts to extend the CAViaR formulation to make it more flexible. Specifically, one of the main points addressed is how to deal with jumps in the returns process. A widespread approach is to extend the autoregressive formulation by incorporating realized measures for the volatility and jumps. Noticeable examples are \cite{vzikevs2015semi} and \cite{gotz2020rq}. However, according to the results of these works, it is not clear the advantage of considering jumps in terms of forecasting performance.}\\
\linebreak
As for the ES, the problem is far more complex. Indeed, estimating ES in a dynamic context is an open problem in the literature, as ES is not elicitable \cite{gneiting2011making}. However, it has been proved the joint elicitability of VaR and ES pair \cite{fissler2016higher} (that is, the pair (VaR, ES) is an elicitable functional with values in $\mathbb{R}^2$), giving rise to plenty of works in this field. For example, \cite{emmer2013best} opens to two-step approaches. First, the VaR is estimated thanks to the Pinball loss. Then, it is fixed, and the ES is computed. Within this class of approaches, it is worth citing the work by Barrera et al. in \cite{barrera2022learning}, which is named as \textbf{BCGNS} in the following. Given a probability space, the conditioning event $X$ and the asset return $Y$ are seen as random variables with values in a proper Polish (separable and completely metrizable) space $\mathcal{S}$ and in $\mathbb{R}$. Then, the conditional VaR and ES are seen as Borel measurable functions $\mathcal{S}\rightarrow\mathbb{R}$, $VaR(S)$ and $ES(S)$ respectively. The authors in \cite{barrera2022learning} prove that the difference $r(S):=ES(S)-VaR(S)$ is in the argmin, among all the Borel measurable non-positive functions, of the following loss:
\begin{equation}
\label{eq:0227_1139}
\mathbb{E} \left[ \left(r(S) + \frac{1}{\theta}\left(VaR(S)-Y\right)^+ \right)^2 \right]
\end{equation}
Thus, the estimator $\bm{\hat{r}}$ of the time series difference $\bm{r} := \bm{ES}-\bm{VaR}$, is the argmin, within a fixed hypothesis class $\mathcal{R}$, of the following loss function:
\begin{equation}
\label{eq:0127_1912}
\mathcal{L}^\theta_B(\bm{\hat{r}}, \bm{y}; \bm{VaR}) := \frac{1}{T} \sum_{t=1}^T \left( \hat{r}_t + \frac{1}{\theta}\left(VaR_t-y_t\right)^+ \right)^2
\end{equation}
The loss can be interpreted as the mean square error between the $\bm{\hat{r}}$ prediction and the "excess loss" weighted by the confidence level $\theta$. The excess loss is the difference between the asset value and the quantile when there is a tail event, or 0 otherwise. As for the hypothesis class, the authors in \cite{barrera2022learning} use feedforward neural networks fed with previous $y_{t}$ values. It is important to observe that this loss requires separate training for VaR and ES estimators. The quantile needs to be estimated in the first step. This task is achieved using a neural network and minimizing the Pinball loss. Specifically, they use a network with 3 hidden layers, each one with the double of the input layer neurons. Then, $\bm{\hat{q}}$ is fixed to obtain the $\bm{ES}=\{ES_t\}_{t=1}^T$ estimates as $\bm{\hat{e}}=\{\hat{e}_t\}_{t=1}^T$. It is noteworthy that the computation cost in terms of training time is reduced by initializing the ES network weights as equal to the ones of the VaR network, (provided that the two networks share the same architecture). Despite the possibility of a joint approach for the estimation of VaR and ES (as discussed below), such a two-step approach represents a limitation, as claimed in \cite{barrera2022learning}, but the choice is driven by technical reasons. Finally, it is worth underlining that the number of lags used as input for the neural networks is a hyperparameter set by the user that can strongly impact the overall result. Below, the model by \cite{barrera2022learning} is used as a comparative benchmark.\\
\linebreak
Another approach to ES estimation with a different choice for the loss function is proposed by Patton et al. \cite{patton2019dynamic}. In this case, both ES and VaR are jointly estimated by minimizing the following loss function:
\begin{equation}
\label{eq:0208_1938}
    \mathcal{L}^\theta_P(\bm{\hat{e}}, \bm{\hat{q}}, \bm{y}) := \frac{1}{T} \sum_{t=1}^T \left( \frac{\hat{q}_t}{\hat{e}_t} - \frac{\hat{q}_t-y_t}{\theta\hat{e}_t}\mathbbm{1}_{\{y_t\le\hat{q}_t\}} + \log\left(-\hat{e}_t\right) \right)
\end{equation}
The $\mathcal{L}^\theta_P$ loss can be viewed as a special case of the jointly elicitable losses discussed in \cite{fissler2016higher}. The hypothesis class used for $\bm{\hat{q}}$ and $\bm{\hat{e}}$ is the Generalized Autoregressive Score (GAS) model \cite{creal2013generalized}. Unlike the GAS standard approach, here the forcing variable is derived from the loss in Eq. \ref{eq:0208_1938} rather than from the log-likelihood. That is, the innovation term is obtained from the product between the inverse of the Hessian and the gradient of the loss function with respect to the model factors, net of algebraic simplifications. Specifically, the models proposed in \cite{patton2019dynamic} include the one-factor and the two-factor models, which we indicate with \textbf{GAS1} and \textbf{GAS2}, respectively. The former uses a latent factor $k_t$ and computes the target variables $\hat{q}_t$ and $\hat{e}_t$ as $k_t$ functions:
\begin{equation}
\label{eq:0208_1820}
    \hat{q}_t = ae^{k_t}; \quad \hat{e}_t = be^{k_t} \quad {\text with} \quad \ k_t = \beta k_{t-1} + \frac{\gamma}{\hat{e}_{t-1}}\left( \frac{1}{\theta}\mathbbm{1}_{\{y_{t-1}\le\hat{q}_{t-1}\}}y_{t-1}-\hat{e}_{t-1} \right)
\end{equation}
where the innovation term in the last equation is the so-called score, namely a term proportional to the derivative of the loss function in Eq. \eqref{eq:0208_1938} as a function of $k_t$, see \cite{creal2013generalized} for more details. In particular, notice that the ratio $\frac{\hat{q}_t}{\hat{e}_t}=\frac{a}{b}$ is constant for the one-factor model.\\
The two-factor model describes the factors as the target variables VaR and ES directly, that is:
\begin{equation}
\label{eq:0208_1821}
    \left[\begin{array}{c}\hat{q}_{t} \\ \hat{e}_t \end{array}\right] = \left[\begin{array}{c}w_1 \\ w_2 \end{array}\right] + \left[\begin{array}{c}b_1 \hat{q}_{t-1} \\ b_2 \hat{e}_{t-1} \end{array}\right] + \left[\begin{array}{cc} a_{1,1} & a_{1,2} \\ a_{2,1} & a_{2,2} \end{array}\right] \left[\begin{array}{c} \hat{q}_{t-1}\left(\theta - \mathbbm{1}_{\{y_{t-1}\le\hat{q}_{t-1}\}}\right) \\ \frac{y_{t-1}}{\theta}\mathbbm{1}_{\{y_{t-1}\le \hat{q}_{t-1}\}} - \hat{e}_{t-1} \end{array}\right]
\end{equation}
where the innovation term is the score computed in the bivariate framework.\\
The parameters in Eqs. \eqref{eq:0208_1820} and \eqref{eq:0208_1821} are estimated by minimizing the loss function in Eq. \eqref{eq:0208_1938}\textcolor{black}{. It is important to stress that proper constraints on the parameters space are added} to ensure the monotonicity condition $\hat{e}_t<\hat{q}_t$. Below, we use the model\textcolor{black}{s in} \cite{patton2019dynamic} as a second comparative benchmark.\\
\linebreak
\textcolor{black}{Another noticeable approach for the joint VaR and ES forecasting is proposed in \cite{taylor2019forecasting}. It is based on the link between quantile regression and the Asymmetric Laplace (AL) distribution highlighted in \cite{koenker1999goodness}. Specifically, the estimation problem turns out to be equivalent to maximizing the following probability density function:
\begin{equation}
    \label{eq:250203_1600}
    \frac{\theta - 1}{\hat{e}_t} \exp\left\{ \frac{(\tilde{y}_t - \hat{q}_t)(\theta - \mathbbm{1}_{\{\tilde{y}_t \le \hat{q}_t\}})}{\theta\hat{e}_t} \right\}
\end{equation}
where $\{\tilde{y}_t\}_{t=1}^T$ is obtained by removing the (historical) mean from $\bm{y}$. The VaR estimator $\hat{q}_t$ has the same functional form as the CAViaR in Eq. \eqref{eq:0127_1913}. Two proposals are considered for the ES estimator. The first is $\hat{e}_t = \gamma \hat{q}_t$ with $\gamma > 1$. We name this approach \textbf{AL-Multi} as the ES estimator is obtained by the VaR with a constant multiplicative factor. The second way is to define $\hat{e}_t = \hat{q}_t - x_t$, with
\begin{equation}
    \label{eq:250203_1601}
    x_t = \left[ \gamma_0 + \gamma_1 (\hat{q}_{t-1} - \tilde{y}_{t-1}) + \gamma_2 x_{t-1} \right] \mathbbm{1}_{\{\tilde{y}_{t-1}\le\hat{q}_{t-1}\}} + x_{t-1} \mathbbm{1}_{\{\tilde{y}_{t-1}>\hat{q}_{t-1}\}}
\end{equation}
The coefficients are forced to be non-negative to satisfy the monotonicity constraint. We name this approach \textbf{AL-AR}, as the difference between VaR and ES is modeled as an AR process.}
\\
\linebreak
One last approach for estimating the ES relates to the definition itself of the ES as the tail mean: $ES_t(\theta):=\mathbb{E}_{t-1}[y_t | y_t\le VaR_t(\theta)]$. One might consider approximating the mean with the sample average, which involves averaging the quantile estimates for the $n$ quantiles $VaR(\theta_j)$, $j=1,\cdots,n$\footnote{\textcolor{black}{In the empirical application, we work with $n=10$. Specifically, $n$ is far smaller than the time series length $T$.}}. The confidence levels $\theta_j$ form an equi-spaced partition of the tail $(0,\theta]$, that is: $0<\theta_1<\cdots<\theta_n=\theta$. This method is suggested in \cite{emmer2013best} and \cite{kratz2018multinomial}. In the following, we have implemented it by estimating the quantiles using both the CAViaR (we name this approach \textbf{K-CAViaR}) or the Quantile Regression Neural Network (giving rise to the \textbf{K-QRNN} approach) -that is, a feedforward neural network trained to minimize the Pinball loss in \eqref{eq:0127_1911}. Observe that no monotonicity constraints are explicitly imposed, so, theoretically, this approach is subject to the above-mentioned crossing quantiles problem. However, the key intuition is that, if the quantile estimations for different $\theta_j$ are only affected by numerical errors, these should be uncorrelated and averaged out. On the other side, if the estimations are subject to finite-sample biases, these are positively correlated, thus monotonicity is even more likely to hold.
\linebreak
\textcolor{black}{Other works follow the approach in \cite{taylor2019forecasting} based on maximizing the AL likelihood. A noteworthy example is} \cite{merlo2021forecasting}\textcolor{black}{, which} generalizes \textcolor{black}{the Taylor approach} to the multivariate case. The authors in \cite{fortin2023forecasting} compare univariate and multivariate ES predictions. Specifically, they attempt to predict the weekly ES of a stock portfolio. The univariate approach directly works with the portfolio time series. Instead, the multivariate approach proposes a regression analysis for the portfolio against the Fama–French and momentum factors. Then, a copula model is used to learn the portfolio ES based on these factors. A noteworthy conclusion suggests that the univariate approach performs the best. This is mainly due to the high variance of estimation errors in the multivariate approach, which leads to wrong forecasts. The study in \cite{fuentes2023forecasting} attempts to model the arrival intensity and magnitude of tail events by using a score-driven approach. The proposed model is then tested on a real-world dataset, showing promising results as the event dynamics seem to capture the major financial crisis. That is, there are spikes in the arrivals and magnitude forecasts corresponding to well-recognized turbulent financial periods, such as the 2008 crisis. \textcolor{black}{More recently, \cite{garcia2025measuring} proposes a joint scoring function methodology applied to the market risk of European banks, conditioning the daily price returns to environmental and transition risk.}
\linebreak
\textcolor{black}{Lastly, it is worth mentioning a branch of literature that, even if not directly linked to the context of this paper, is gaining increasing relevance in ES estimation: the Extreme Value Theory (EVT). Works in EVT assume asymptotic behavior as the tails become even deeper, so they are focused on very low values of $\theta$, usually smaller than $0.01$, which are beyond the scope of this paper}. For example, the authors of \cite{daouia2021expecthill} propose an estimator for the tail index, namely a statistical index characterizing heavy-tailed distributions. The same index is then used for the estimation of the expectiles of a distribution, connected to the ES inference. The estimator is then exploited to study the SPX index under different market regimes, showing that the fitted parameters strongly change when looking at the periods before, during, and after the 2008 crisis. A similar perspective is followed in \cite{nicolau2023tail}. The peculiarity of this work is considering covariates in the estimation of the tail index. \textcolor{black}{Finally, other works focus on dealing with the time-varying tail index and its impact on the market risk and its dynamics; see \cite{hautsch2020multivariate, massacci2017tail, shen2022modeling}}.

\section{The CAESar Model}
\label{sec:model}
\begin{figure}[h]
	\centering
	\includegraphics[width=0.9\linewidth]{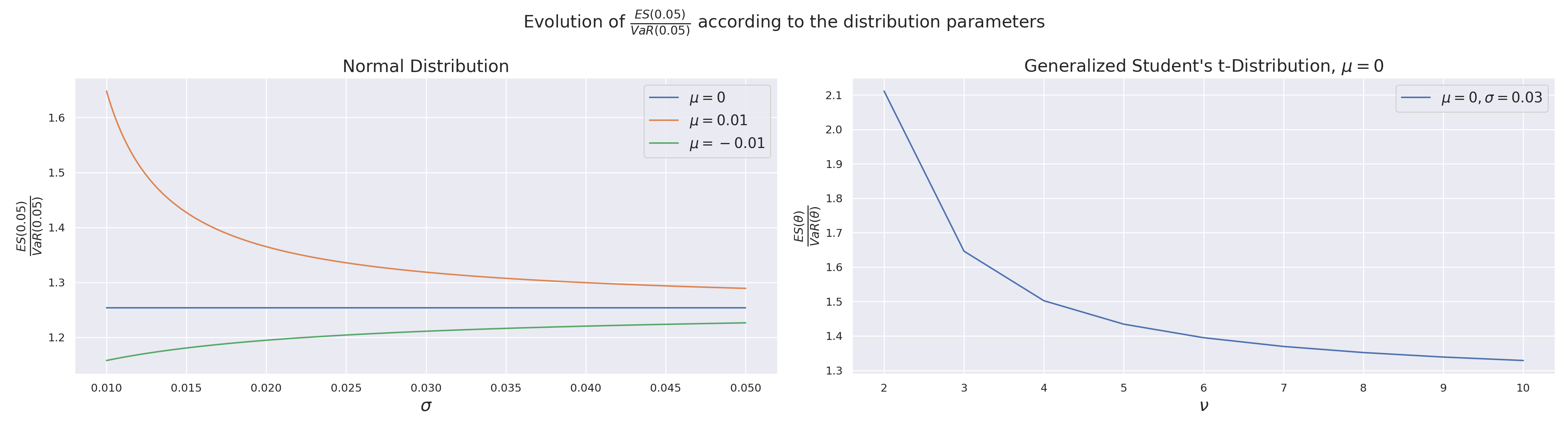}
	\caption{Evolution of the ratio $\frac{Var_t}{ES_t}$ for $\theta=0.05$ as a function of the distribution parameters. In the left plot, the ratio for a standard normal distribution is plotted as a function of the volatility $\sigma$. The right plot shows the ratio for a centred generalized Student's t distribution against the degrees of freedom $\nu$.}
	\label{img:ratio_evolution}
\end{figure}

In this Section, we present the CAESar model. First, we describe the general setting, then we highlight the differences with existing competitors, and finally, we discuss the general idea behind the optimization scheme. More details can be found in the Supplementary Material, sections \ref{app:ms} and \ref{app:os}, which provide a deeper insight into the modeling choices.\\
\linebreak
The main idea behind CAESar is to generalize the CAViaR regression to convert it into a joint (VaR, ES) estimator. Specifically, both the targets $\hat{q}_t$ and $\hat{e}_t$ are regressed against multi-dimensional non-linear transformations of the asset value at previous time steps $t-i, \ i=1,\cdots,p$. We indicate these transformations with $\bm{f}$ for the quantile and $\bm{g}$ for the ES. Furthermore, $\hat{q}_t$ and $\hat{e}_t$ are also regressed on previous estimates $\hat{q}_{t-j}, \ j=1,\cdots,u$ and $\hat{e}_{t-l}, \ l=1,\cdots,v$ to account for heteroskedastic effects. The general form of the proposed estimator is
\begin{align}
\label{eq:0501_2029}
\hat{q}_t = \beta_0 + \sum_{i=1}^p \bm{\beta}_{(i-1)d+1:id} \bm{\cdot} \bm{f}(y_{t-i}) + \sum_{j=1}^u \beta_{j+pd} \hat{q}_{t-j} + \sum_{l=1}^v\beta_{l+pd+u} \hat{e}_{t-l} \\
\hat{e}_t = \gamma_0 + \sum_{i=1}^p \bm{\gamma}_{(i-1)d+1:id} \bm{\cdot} \bm{g}(y_{t-i}) + \sum_{j=1}^u \gamma_{j+pd} \hat{q}_{t-j} + \sum_{l=1}^v\gamma_{l+pd+u} \hat{e}_{t-l}
\end{align}
where two distinct sets of parameters $\bm{\beta}=(\beta_0,\cdots,\beta_{pd+u+v})$ and $\bm{\gamma}=(\gamma_0,\cdots,\gamma_{pd+u+v})$ describe the dynamics of $\bm{\hat{q}}$ and $\bm{\hat{e}}$, respectively. As before, $\bm{\beta}_{(i-1)d+1:id}$ represents the d-dimensional vector $(\beta_{(i-1)d+1}, \cdots, \beta_{id})$ and $\bm{\gamma}_{(i-1)d+1:id}$ is for $(\gamma_{(i-1)d+1}, \cdots, \gamma_{id})$. For simplicity, it is convenient to choose $\bm{f}\equiv\bm{g}$ and to set $u=v$. Thus, the joint VaR-ES estimator can be read as:
\begin{equation}
\label{eq:0502_1200}
\left[\begin{array}{c} \hat{q}_t \\ \hat{e}_t \end{array}\right] = \left[\begin{array}{c} \beta_0 \\ \gamma_0 \end{array}\right] + \sum_{i=1}^p \left[\begin{array}{c} \bm{\beta}_{(i-1)d+1:id} \\  \bm{\gamma}_{(i-1)d+1:id} \end{array}\right] \bm{f}(y_{t-i}) + \sum_{j=1}^u \left[\begin{array}{cc} \beta_{pd+j} & \beta_{pd+u+j} \\ \gamma_{pd+j} & \gamma_{pd+u+j} \end{array}\right] \left[\begin{array}{c} \hat{q}_{t-j} \\ \hat{e}_{t-j} \end{array}\right]
\end{equation}
It is worth noting that the specification we propose for the ES is similar to the GAS models in \cite{patton2019dynamic} (Eqs. \eqref{eq:0208_1820} and \eqref{eq:0208_1821}) \textcolor{black}{and Taylor's models in \cite{taylor2019forecasting}} as they are both autoregressive models. However, our formulation is more general and less limiting. As such, it should be able to better capture the tail behavior. In fact, the one-factor model in \eqref{eq:0208_1820} \textcolor{black}{and the \textbf{AL-Multi} model} assume that the ratio $\frac{Var_t}{ES_t}$ is constant. However, in principle, this ratio can be time-varying, depending on the evolution of the shape of the returns' conditional distribution, while the previous assumption might lead to model misspecification errors. For instance, the ratio for a non-centered normal distribution changes as a function of the volatility. Moreover, the ratio for a centered generalized Student's t distribution varies with the degrees of freedom. These considerations are illustrated in Figure \ref{img:ratio_evolution}.
\\
The GAS two-factor model in \eqref{eq:0208_1821} \textcolor{black}{and the \textbf{AL-AR} model} address this issue but restrict the information set to the magnitude of $y_t$ in the case of a tail event, i.e., when the realization is below the quantile. \textcolor{black}{For example, this can be a limitation in a regime-switching framework. Indeed, \textbf{GAS2} and \textbf{AL-AR} models slowly adapt to changes in the market as their ES estimator does not take into account information about the return when it does not break the quantile. Instead, CAESar can update its forecast faster, thus outperforming its competitors in such a scenario. This finding is corroborated by a simulation study in Section \ref{app:srs} of the Supplementary Material.}\\
\linebreak
On the other side, the model proposed by Barrera et al. in \cite{barrera2022learning} exploits neural networks to catch non-linear dependencies in the data. However, it uses dense layers, so it exploits information only from the more recent time steps -that is, the lags used as input of the network. In contrast, CAESar exploits an exponentially weighted moving average that keeps memory of the entire time series' past history. Moreover, the model estimation (see below) is far simpler and less prone to hyperparameter misspecification. A possible drawback of our approach refers to the high sensibility to the choice of the initial parameters -similar to the case of GAS models in \cite{patton2019dynamic}. To mitigate this issue, we propose a three-step approach for the minimization of the loss in \eqref{eq:0208_1938}. First the $\bm{VaR}$ prediction, $\bm{\hat{q}}$, is obtained using the CAViaR regression. Then, in the second step, quantile information is used to construct the $\bm{ES}$ estimator $\bm{\hat{e}}$. This is done following the idea in \cite{barrera2022learning}, that is by modelling the residual time series $\bm{\hat{r}}$:
\begin{equation}
\label{eq:0501_2030}
\hat{r}_t = \hat{e}_t - \hat{q}_t \quad {\text with} \quad \ \hat{r}_t = \tilde{\gamma_0} + \sum_{i=1}^p \bm{\tilde{\gamma}}_{(i-1)d+1:id} \bm{\cdot} \bm{f}(y_{t-i}) + \sum_{j=1}^u \tilde{\gamma}_{j+pd} \hat{q}_{t-j} + \sum_{l=1}^u\tilde{\gamma}_{l+pd+u} \hat{r}_{t-l}
\end{equation}
\textcolor{black}{where $\bm{\tilde{\gamma}}=(\tilde{\gamma}_0,\cdots,\tilde{\gamma}_{pd + 2u})$ is a set of parameters related to $\bm{\beta}$ and $\bm{\gamma}$ in Eq. \eqref{eq:0501_2029}.} As for the loss function, it is derived from the joint elicitability of $VaR$ and $ES$. Specifically, Eq. \eqref{eq:0127_1912} is used with an added soft constraint term to enforce the monotonicity condition $\hat{e}_t<\hat{q}_t$:
\begin{equation}
    \label{eq:0227_1200}
    \mathcal{L}^\theta_r(\bm{\hat{r}}, \bm{y}; \bm{VaR}) := \mathcal{L}^\theta_B(\bm{\hat{r}}, \bm{y}; \bm{VaR}) + \lambda_r \sum_{t=1}^T (\hat{r}_t)^+
\end{equation}
The regularization parameter $\lambda_r$ is manually set to 10. The uniqueness of the minimizer is a consequence of the uniqueness of the $\mathcal{L}^\theta_B(\bm{\hat{r}}, \bm{y}; \bm{VaR})$ minimizer -see Lemma A.2 in \cite{barrera2022learning}. In the last step, the previously estimated coefficients are used as a suitable starting point for minimizing the loss in \eqref{eq:0208_1938}. Indeed, by rearranging Eq. \eqref{eq:0501_0922} and \eqref{eq:0501_2030}, the formulation for $\hat{e}_t$ can be recovered:
\begin{equation*}
    \begin{split}
    \hat{e}_t = (\tilde{\gamma_0}+\beta_0) & + \sum_{i=1}^p \left( \left(\bm{\tilde{\gamma}} + \bm{\beta}\right)_{(i-1)d+1:id} \bm{\cdot} \bm{f}(y_{t-i}) \right) + \\
    & + \sum_{j=1}^u \left( \tilde{\gamma}_{pd+j} + \beta_{pd+j} - \tilde{\gamma}_{pd+u+j} \right)\hat{q}_{t-j} + \sum_{j=1}^u \gamma_{pd+u+j}\hat{e}_{t-j}
    \end{split}
\end{equation*}
By comparing the previous expression with \eqref{eq:0502_1200}, we can recover the $\gamma_j$ coefficients as:\\
\begin{equation*}
    \gamma_j = \begin{cases} \tilde{\gamma}_j + \beta_j & j=0,\cdots,pd \\ \tilde{\gamma}_j + \beta_j - \tilde{\gamma}_{j+u} & j=pd+1,\cdots,pd+u \\ \tilde{\gamma}_j & j=pd+u+1,\cdots,pd+2u \end{cases}
\end{equation*}\\
The optimal coefficients for $\bm{\hat{q}}$ and $\bm{\hat{e}}$ obtained in the previous steps serve as the starting point for the final step. We jointly optimize the two estimators coefficients in Eq. \eqref{eq:0502_1200}. The additional ones $\beta_{pd+u+1}, \cdots, \beta_{pd+2u}$ (i.e. those absent in the original CAViaR formulation in Eq. \eqref{eq:0501_0922}) are initialized to zero. The loss function used is the one in Eq. \eqref{eq:0208_1938}, with the addition of soft constraints ensuring (i) the monotonicity condition and (ii) the non-positivity of the quantile, which is a straightforward assumption for financial returns:
\begin{equation}
    \label{eq:0227_1201}
    \mathcal{L}^\theta_{q,e}(\bm{\hat{e}}, \bm{\hat{q}}, \bm{y}) := \mathcal{L}^\theta_P(\bm{\hat{e}}, \bm{\hat{q}}, \bm{y}) + \lambda_e \sum_{t=1}^T (\hat{e}_t - \hat{q}_t)^+ + \lambda_q \sum_{t=1}^T (\hat{q}_t)^+
\end{equation}
In the experimental stage, both $\lambda_q$ and $\lambda_e$ are set to 10. At this point, one may wonder whether the updating in the quantile predictor leads to a degradation in the VaR estimate. We show that this is not the case, as discussed in Section \ref{sec:main_res}, but there is a slight improvement in the forecasting performance.\\
\linebreak
Thus far, the CAESar regression has been discussed in the most general setting. However, for practical implementations, we suggest focusing on a particular specification: the AS-(1,1) model -i.e. AS specification for $\bm{f}$ and $p=u=1$. Indeed, the same considerations previously made for the CAViaR hold here, as the AS specification is able to capture asymmetries in the return distribution. Finally, a Markovian specification of the model, namely using the information on the previous time lag only, reduces both the training time and the overfitting risk; at the same time, empirical applications corroborate this modeling choice, displaying very good forecasting performances. The estimator adopted in the applications below is:
\begin{equation}
\label{eq:0226_1701}
    \left[\begin{array}{c} \hat{q}_t \\ \hat{e}_t \end{array}\right] = \left[\begin{array}{c} \beta_0 \\ \gamma_0 \end{array}\right] + \left[\begin{array}{cc} \beta_1 & \beta_2 \\ \gamma_1 & \gamma_2 \end{array}\right] \left[\begin{array}{c} (y_{t-1})^+ \\ (y_{t-1})^- \end{array}\right] + \left[\begin{array}{cc} \beta_3 & \beta_4 \\ \gamma_3 & \gamma_4 \end{array}\right] \left[\begin{array}{c} \hat{q}_{t-1} \\ \hat{e}_{t-1} \end{array}\right]
\end{equation}
A discussion on the adopted specifications for both the CAESar and CAViaR models can be found in the Supplementary Material \ref{app:ms}. And empirical justification for using the cross terms in Eq. \eqref{eq:0226_1701} is also in Supplementary Material \ref{app:ms}. The detailed report of the optimization scheme is presented in Supplementary Material \ref{app:os}. The Python implementation of the proposed methodology is available online; see the ``Data and Code Availability'' Section below.
\section{Statistical Tests}
\label{sec:tests}
As discussed in the Introduction, assessing the effectiveness of an ES estimation model is a challenging task due to the lack of a universally accepted test. Furthermore, comparing different models to determine whether one statistically outperforms the other is not straightforward. An attempt to clarify this is made in \cite{deng2021backtesting}, where existing tests are surveyed and classified. Following their taxonomy, we evaluate our proposal using two separate classes of tests. Some tests directly evaluate the ES approximation, aiming to assess whether the model's estimation is coherent with the definition of the ES itself. Within this context, there is no need for a loss function as an evaluation metric. However, it is still not clear how to study the limiting behavior of the considered statistics. Hence, a bootstrap approach is used to approximate the test p-value  (see \cite{tibshirani1993introduction} for further details). On the other side, some testing procedures have been introduced for model comparison. Such a comparison is typically carried out using the previously discussed loss functions. In other words, these tests are based on the concept of joint elicitability of VaR and ES.
\subsection{Statistical Tests - Direct Approximation}
\label{subs:stat_dir_app}
The first category of tests is those that directly evaluate the ES approximation. Specifically, four statistical tests are conducted. We consider first Christoffersen's conditional coverage test \cite{christoffersen1998evaluating}, which only tests for the VaR forecast. It jointly takes into account the independence of the events $\{y_t < \hat{q}_t\}_{t=1}^T$ and the tail property. Under the null of a correctly specified quantile dynamic, the following statistic is asymptotically distributed as a chi-square with 2 degrees of freedom:
\begin{equation}
    -2\left\{ \log\left( \frac{\theta^{T-v} \; (1-\theta)^v}{\left(1-\hat{\theta}\right)^{T-v} \hat{\theta}^v} \right) + \log\left( \frac{(1-\hat{\theta})^{T-v} \; \hat{\theta}^v}{(1-\theta_0)^{T-v-v_{0}} \; \theta_0^{v_0} \; (1-\theta_1)^{v-v_1} \; \theta_1^{v_1}} \right) \right\}
\end{equation}
where $v$ is the number of violations; $\hat{\theta}:=\frac{v}{T}$ is the empirical violations frequency corresponding to the estimates $\bm{\hat{q}}$; $v_0$ (and $v_1$) are the number of violations conditional on a non-violation (violation) at the previous timestep; $\theta_0:=\frac{v_0}{T-v}$ and $\theta_1:=\frac{v_1}{v}$ are the conditional violation probabilities. Next, we look at the McNeil and Frey test \cite{mcneil2000estimation}, which is a one-sided test for the null hypothesis of no risk underestimation. The intuition is testing the definition of ES as a tail mean. In particular, from the index set $\mathcal{T}=\{1,\cdots,T\}$, we extract the subset of quantile violations $\tilde{\mathcal{T}}=\{t\in\mathcal{T} \ s.t. \ y_t<\hat{q}_t\}$. Then, on this subset, we use a bootstrapping approach to test the null hypothesis that $y_t-\hat{e}_t$ has a non-negative mean, i.e. $H_0: \mathbb{E}\left[ y_t-\hat{e}_t | t\in\tilde{\mathcal{T}} \right]\ge0$. Specifically, we generate 10,000 samples (as detailed in \cite{deng2021backtesting}) to estimate the ``empirical'' p-value for the bootstrapped distribution of the statistic.\\
\linebreak
Acerbi and Szekely \cite{acerbi2014back} proposed two similar tests using the $Z_1$ and $Z_2$ statistics. In both cases, we consider a two-sided test for the null hypothesis of correct  ES predictions. Specifically, both statistics are derived by rearranging the ES definition. For $Z_1$:
\begin{equation}
    ES_t = \mathbb{E}\left[ y_t | y_t \le VaR_t \right] \implies \mathbb{E}\left[ \frac{y_t}{ES_t} \bigg| y_t \le VaR_t \right] = 1
\end{equation}
For $Z_2$, we use the conditional probability formula and the definition of VaR to obtain:
\begin{equation}
    ES_t = \mathbb{E}\left[ y_t | y_t \le VaR_t \right] = \mathbb{E}\left[ \frac{y_t}{\theta} \mathbbm{1}_{\{y_t \le VaR_t\}} \right] \implies  \mathbb{E}\left[ \frac{y_t}{\theta \; ES_t} \mathbbm{1}_{\{y_t \le VaR_t\}} \right] = 1
\end{equation}
Based on the previous equations, the following statistics are defined:
\begin{equation}
    Z_1 := \frac{1}{|\tilde{\mathcal{T}}|}\sum_{t\in\tilde{\mathcal{T}}}\frac{y_t}{\hat{e}_t}; \quad\quad Z_2 := \frac{1}{T} \sum_{t\in\tilde{\mathcal{T}}} \frac{y_t}{\theta \; \hat{e}_t}.
\end{equation}
In both cases, the null hypothesis assumes the statistic to be equal to 1. We employ a standard bootstrap approach to translate this into a testing procedure. Again, 10,000 bootstrap samples are generated to evaluate the ``empirical'' p-value for the bootstrapped distribution of the statistic. Finally, it is worth noting that, according to the empirical analysis carried out in \cite{deng2021backtesting}, in our setting (six years of training/validation and one year of testing for each fold), the $Z_1$ and McNeil and Frey tests are expected to exhibit the best size and power.
\subsection{Statistical Tests - Model Comparison}
We discuss here a number of testing procedures for comparing ES estimators. The underlying null hypothesis is about equal predictive accuracy in comparing two ES predictors. A rejection of the null signals one model outperforming the other (alternative hypothesis). It is important to underline that this kind of test tends to be 'conservative', resulting in a low number of rejections \cite{deng2021backtesting}, a phenomenon also observed in the applications below, see Section \ref{sec:experiments}.\\
Here, we consider four different testing procedures. First, we apply the Diebold-Mariano test \cite{diebold2002comparing}, which is a well-known two-sided statistical test used to compare two time series forecasting estimators, A and B. The null hypothesis is that A and B are equivalent. The focus lies on the time series defined by the difference between the losses associated with the two predictors, under the assumption that such a difference is a stationary process. The statistic is then defined as the mean value of the difference, normalized by the sum of the autocorrelations up to a certain lag. Furthermore, we adjust the statistic with the small-sample size correction discussed in \cite{harvey1997testing}, a method also known as the Harvey adjustment.\\
The second test we consider is the corrected resampled Student's t-test, introduced in \cite{nadeau2003inference}. This statistical test is specifically designed for frameworks where cross-validation is used, as in our case (further details on the experimental setup are in the next section). It is the only test where we do not perform an iteration for each fold. Rather, we conduct the test only once. The population used is made up of the losses among the folds. As for the test statistic, it is obtained from the Student's t-test by properly rescaling the variance. This scaling, which is smaller than 1, embodies the fact that, when performing cross-validation, the variance in the predictions (and thus, in the losses) is also affected by the variance in the train set. As the test is designed to take into account only the volatility in the out-of-sample data, we have to filter the total variance to remove its train component. This task is achieved by the scaling factor.
\linebreak
The third considered test is the Loss Difference test. It aims to determine if the difference between the losses obtained by two predictors (A and B) is statistically smaller than 0. In other words, the null hypothesis is that the loss difference $\mathcal{L}\left( \bm{\hat{e}}^A, \bm{\hat{q}}^A, \bm{y} \right)-\mathcal{L}\left( \bm{\hat{e}}^B, \bm{\hat{q}}^B, \bm{y} \right)$ is greater or equal to 0. If the null is rejected, we can say that the predictor $A$ outperforms $B$. This test is performed by considering the population of the loss at time $t \ \forall t\in\mathcal{T}$, and studying its mean with a one-sided bootstrap test, following the same rationale as in the McNeil and Frey test.\\
The last approach is based on the Encompassing test, which generally indicates a broad class of testing procedures aimed to determine if predictor A outperforms predictor B by studying a linear combination of the two predictors \cite{kicsinbay2010use}. Specifically, in our case, the idea is to introduce a new predictor, C, defined as a linear combination of A and B: $C=\alpha A + \beta B$. The non-negative weights $\alpha$ and $\beta$ are determined through an Ordinary Least Squares regression applied to the first half of the data in the test set. The forecast provided by C is then evaluated in the second half of the data. As long as the information in B is already incorporated in A, adding B should only add noise. Thus, the ultimate prediction provided by C should be inferior to that of A. Thus, we perform a bootstrapping exercise similar to the Loss Difference test, but now we compare predictors A and C. If the null is rejected, then the addition of predictor B only adds noise, indicating that A outperforms B.
\section{Results}
\label{sec:experiments}
In this section, we assess the validity of the proposed framework by comparing the CAESar model with different benchmarks both on synthetic and real-world datasets. We consider \textcolor{black}{seven} competitor models of the proposed CAESar model:
\begin{itemize}[label=$\triangleright$]
    \item \textbf{K-CAViaR} Based on the definition of the $ES(\theta)$ as the mean value of the quantiles below the target confidence $\theta$, we construct an $ES(\theta)$ estimator by averaging the CAViaR predictions for $n=10$ equispaced quantiles $VaR(\theta_j) \ j=1,\cdots,n$ with $0<\theta_1<\cdots<\theta_n=\theta$. This is the same approach proposed in \cite{kratz2018multinomial}. The implementation is a simple extension of the CAViaR methodology; the CAViaR code is available online at \url{http://www.simonemanganelli.org/Simone/Research.html}.
    \item \textbf{BCGNS} We use the BCGNS as presented in \cite{barrera2022learning} by using the code available online at \url{https://github.com/BouazzaSE/Learning-VaR-and-ES}. The model is fed by the previous 25 lags of each timestep, as done in the original paper.
    \item \textbf{K-QRNN} This method mirrors the K-CAViaR approach. One single network is trained, with $n=10$ outputs, one for each quantile $VaR(\theta_j) \ j=1,\cdots,n$. To preserve the coherence with the other neural network model, i.e. the BCGNS, we use as input 25 lagged values.
     \item \textbf{GAS1} and \textbf{GAS2} We employ both the GAS one-factor and two-factor models proposed in \cite{patton2019dynamic}. The methodology has been implemented by ourselves and is available online; see Data and Code Availability Section below.
     \item \textcolor{black}{\textbf{AL-Multi} and \textbf{AL-AR} These are the Asymmetric Laplace based methods proposed in \cite{taylor2019forecasting}. Both use the AS specification for the VaR estimator. The former assumes the ES is a multiple of VaR. The latter exploits an AR formulation to describe the difference between VaR and ES.}
\end{itemize}
The details of the comparison are in the following subsections.
\subsection{Simulation Study}
\label{subs:sim_st}
First, we carry out a simulation experiment to assess whether the predictions provided by the ES models are coherent with the ground truth. Specifically, we use the GARCH as the data-generating process, with both Gaussian and Student's t innovations. Specifically:
\begin{equation}
    GARCH(1,1): \quad y_t = \sigma_t \varepsilon_t \quad with \quad \sigma_t = \omega + \beta \sigma_{t-1}^2 + \gamma \varepsilon_{t-1}^2
\end{equation}
The coefficients of the process with Gaussian innovation are $(\omega, \beta, \gamma)$. When considering the t innovations, also the degrees of freedom $\nu$ have to be considered. These parameters are fitted from the real data (see next section for details). Specifically, we fit the parameters on the SPX (I set of coefficients), FTSE (II set of coefficients), and DAX (III set of coefficients). Thus, there are two different innovations and, for each of them, three sets of coefficients, i.e. we consider six data-generating processes. For each of them, we generate 20 series made up of 1,750 points, which roughly correspond to seven years, divided into 1,500 observations in the train set and 250 in the test set.\\
\linebreak
The algorithms are tested on three different confidence levels $\theta$, namely $\theta=0.05, 0.025, 0.01$. Specifically, the ground truth for the Gaussian innovations reads:
\begin{equation}
   e_t = - \frac{\sigma_t}{\theta} f_N\left( F^{-1}_N(1-\theta) \right)
\end{equation}
where $F_N$ and $f_N$ are, respectively, the cumulative and probability density functions of a standard normal distribution. Instead, as for the Student's t innovations:
\begin{equation}
   e_t = - \sigma_t \frac{\nu + \left( F^{-1}_T(\theta) \right)^2}{(\nu-1)\theta} \frac{\Gamma\left(\frac{\nu+1}{2}\right)}{\Gamma\left(\frac{\nu}{2}\right)\sqrt{\pi\nu}}\left(1+\frac{\left(F^{-1}_T(\theta)\right)^2}{\nu}\right)^{-\frac{\nu+1}{2}}
\end{equation}
where $F_T$ is the cumulative distribution function of a standard t-distribution. The ES forecasts are then compared with the ground truth risk measures using both the Mean Absolute Error (MAE) and the Root Mean Squared Error (RMSE). Moreover, the loss functions $\mathcal{L}^\theta_B$ (Eq. \eqref{eq:0127_1912}) and $\mathcal{L}^\theta_P$ (Eq. \eqref{eq:0208_1938}) are evaluated.
\begin{table}
  \centering
  \vspace{-2cm}
  \begin{adjustwidth}{-1.8cm}{}
  \tiny{
\begin{tabular}{cc|c|c|c|c|c|c||c|c|c|c|c|c|c}
\toprule
\multirow{2}{*}{$\bm{\theta = 0.05}$} & & \multicolumn{3}{c|}{\textbf{N}} & \multicolumn{3}{c||}{\textbf{t}} & & \multicolumn{3}{c|}{\textbf{N}} & \multicolumn{3}{c}{\textbf{t}} \\
     & & \textbf{I} & \textbf{II} & \textbf{III} & \textbf{I} & \textbf{II} & \textbf{III} & & \textbf{I} & \textbf{II} & \textbf{III} & \textbf{I} & \textbf{II} & \textbf{III} \\
\midrule
\multirow{2}{*}{\textbf{CAESar}} & MAE & \underline{0.002} & \underline{0.0022} & \underline{0.0017} & 0.3423 & \underline{0.0282} & \underline{0.0081} & $\mathcal{L}_{B}^{\theta}$ & \textbf{0.0007} & 0.001 & \textbf{0.0005} & \underline{25.9028} & 0.9505 & \underline{0.0267} \\
 & RMSE & \underline{0.002} & \textbf{0.002} & \underline{0.002} & 0.342 & \underline{0.028} & \underline{0.008} & $\mathcal{L}_{P}^{\theta}$ & \underline{0.781} & \underline{0.955} & \textbf{0.657} & 3.369 & \underline{2.362} & \underline{1.633} \\
\midrule
\multirow{2}{*}{\textbf{K-CAViaR}} & MAE & \textbf{0.0015} & \textbf{0.0017} & \textbf{0.0012} & 0.1317 & 0.0345 & 0.0096 & $\mathcal{L}_{B}^{\theta}$ & \underline{0.0007} & \textbf{0.0009} & \underline{0.0005} & 28.2971 & 1.2366 & 0.0285 \\
 & RMSE & \textbf{0.001} & \underline{0.002} & \textbf{0.001} & 0.132 & 0.034 & 0.01 & $\mathcal{L}_{P}^{\theta}$ & \textbf{0.779} & \textbf{0.948} & \underline{0.658} & \underline{3.247} & 2.372 & 1.634 \\
\midrule
\multirow{2}{*}{\textbf{BCGNS}} & MAE & 0.0297 & 0.0332 & 0.0356 & 0.829 & 0.1453 & 0.0542 & $\mathcal{L}_{B}^{\theta}$ & 0.0047 & 0.0066 & 0.0062 & 97.4735 & 3.3069 & 0.071 \\
 & RMSE & 0.03 & 0.033 & 0.036 & 0.829 & 0.145 & 0.054 & $\mathcal{L}_{P}^{\theta}$ & 2.553 & 2.987 & 2.365 & 5.578 & 5.779 & 3.643 \\
\midrule
\multirow{2}{*}{\textbf{K-QRNN}} & MAE & 0.0083 & 0.0085 & 0.0065 & 0.9368 & 0.1767 & 0.0371 & $\mathcal{L}_{B}^{\theta}$ & 0.0008 & \underline{0.0009} & 0.0006 & 198.3513 & 10.0393 & 0.065 \\
 & RMSE & 0.008 & 0.009 & 0.007 & 0.937 & 0.177 & 0.037 & $\mathcal{L}_{P}^{\theta}$ & 0.94 & 1.076 & 0.795 & 9.001 & 5.187 & 2.693 \\
\midrule
\multirow{2}{*}{\textbf{GAS1}} & MAE & 0.0892 & 0.0291 & 0.081 & 0.3503 & 0.0684 & 0.0222 & $\mathcal{L}_{B}^{\theta}$ & 0.0013 & 0.0376 & 0.4944 & 46.5417 & \underline{0.7035} & 0.0375 \\
 & RMSE & 0.089 & 0.029 & 0.081 & 0.35 & 0.068 & 0.022 & $\mathcal{L}_{P}^{\theta}$ & 1.36 & 1.54 & 1.514 & 4.028 & 2.46 & 1.846 \\
\midrule
\multirow{2}{*}{\textbf{GAS2}} & MAE & 0.0199 & 0.2465 & 0.0044 & \underline{0.1288} & 0.0724 & 0.0385 & $\mathcal{L}_{B}^{\theta}$ & 0.0037 & 0.1283 & 0.0007 & 92.1 & 0.9577 & 0.6581 \\
 & RMSE & 0.02 & 0.247 & 0.004 & \underline{0.129} & 0.072 & 0.039 & $\mathcal{L}_{P}^{\theta}$ & 1.543 & 2.529 & 1.463 & 4.182 & 3.795 & 2.225 \\
\midrule
\multirow{2}{*}{\textbf{AL-Multi}} & MAE & 0.0634 & 0.2128 & 0.0716 & \textbf{0.1078} & \textbf{0.0186} & \textbf{0.0062} & $\mathcal{L}_{B}^{\theta}$ & 0.0139 & 0.0069 & 0.0056 & \textbf{21.7296} & \textbf{0.06} & \textbf{0.0262} \\
 & RMSE & 0.063 & 0.213 & 0.072 & \textbf{0.108} & \textbf{0.019} & \textbf{0.006} & $\mathcal{L}_{P}^{\theta}$ & 1.104 & 1.581 & 0.941 & \textbf{3.054} & \textbf{2.36} & \textbf{1.621} \\
\midrule
\multirow{2}{*}{\textbf{AL-AR}} & MAE & 0.0252 & 0.0208 & 0.0582 & 0.2278 & 0.0377 & 0.0187 & $\mathcal{L}_{B}^{\theta}$ & 0.2103 & 0.1489 & 1.0672 & 30.5149 & 1.2649 & 0.0271 \\
 & RMSE & 0.025 & 0.021 & 0.058 & 0.228 & 0.038 & 0.019 & $\mathcal{L}_{P}^{\theta}$ & 3.168 & 2.124 & 1.251 & 3.837 & 3.636 & 1.636 \\
\bottomrule
\end{tabular}

\vspace{1em}

\begin{tabular}{cc|c|c|c|c|c|c||c|c|c|c|c|c|c}
\toprule
\multirow{2}{*}{$\bm{\theta = 0.025}$} & & \multicolumn{3}{c|}{\textbf{N}} & \multicolumn{3}{c||}{\textbf{t}} & & \multicolumn{3}{c|}{\textbf{N}} & \multicolumn{3}{c}{\textbf{t}} \\
 & & \textbf{I} & \textbf{II} & \textbf{III} & \textbf{I} & \textbf{II} & \textbf{III} & & \textbf{I} & \textbf{II} & \textbf{III} & \textbf{I} & \textbf{II} & \textbf{III} \\
\midrule
\multirow{2}{*}{\textbf{CAESar}} & MAE & \underline{0.0022} & \underline{0.0025} & \underline{0.0018} & 0.2286 & \underline{0.0385} & \underline{0.011} & $\mathcal{L}_{B}^{\theta}$ & \textbf{0.0011} & \underline{0.0016} & \underline{0.0009} & 57.4856 & 3.1664 & \underline{0.083} \\
 & RMSE & \textbf{0.002} & \textbf{0.002} & \textbf{0.002} & 0.229 & \underline{0.038} & \underline{0.011} & $\mathcal{L}_{P}^{\theta}$ & \textbf{0.916} & \textbf{1.081} & \textbf{0.795} & 3.458 & \underline{2.559} & 1.833 \\
\midrule
\multirow{2}{*}{\textbf{K-CAViaR}} & MAE & \textbf{0.0021} & \textbf{0.0023} & \textbf{0.0017} & 0.1932 & 0.0443 & 0.0135 & $\mathcal{L}_{B}^{\theta}$ & \underline{0.0011} & 0.0018 & 0.0009 & 66.0381 & 4.279 & 0.0881 \\
 & RMSE & \underline{0.002} & \underline{0.002} & \underline{0.002} & 0.193 & 0.044 & 0.013 & $\mathcal{L}_{P}^{\theta}$ & \underline{0.919} & \underline{1.102} & \underline{0.796} & \underline{3.448} & 2.568 & \underline{1.829} \\
\midrule
\multirow{2}{*}{\textbf{BCGNS}} & MAE & 0.0183 & 0.0175 & 0.0153 & 0.8765 & 0.1474 & 0.0442 & $\mathcal{L}_{B}^{\theta}$ & 0.0045 & 0.0065 & 0.0033 & 267.4624 & 10.6298 & 0.1653 \\
 & RMSE & 0.018 & 0.018 & 0.015 & 0.876 & 0.147 & 0.044 & $\mathcal{L}_{P}^{\theta}$ & 2.724 & 2.786 & 2.809 & 8.081 & 4.927 & 5.39 \\
\midrule
\multirow{2}{*}{\textbf{K-QRNN}} & MAE & 0.0105 & 0.0104 & 0.0081 & 1.065 & 0.2252 & 0.0485 & $\mathcal{L}_{B}^{\theta}$ & 0.0011 & \textbf{0.0012} & \textbf{0.0008} & 426.7518 & 36.681 & 0.1815 \\
 & RMSE & 0.01 & 0.01 & 0.008 & 1.065 & 0.225 & 0.048 & $\mathcal{L}_{P}^{\theta}$ & 1.102 & 1.223 & 0.948 & 21.988 & 6.652 & 2.705 \\
\midrule
\multirow{2}{*}{\textbf{GAS1}} & MAE & 0.0129 & 0.0491 & 0.0115 & \textbf{0.1083} & 0.0884 & 0.0358 & $\mathcal{L}_{B}^{\theta}$ & 0.0025 & 0.6331 & 0.005 & 21.8599 & 1.2833 & 0.0845 \\
 & RMSE & 0.013 & 0.049 & 0.012 & \textbf{0.108} & 0.088 & 0.036 & $\mathcal{L}_{P}^{\theta}$ & 1.358 & 2.921 & 1.131 & 3.793 & 2.994 & 2.131 \\
\midrule
\multirow{2}{*}{\textbf{GAS2}} & MAE & 0.0766 & 0.1956 & 0.2288 & 0.3838 & 0.2137 & 0.0957 & $\mathcal{L}_{B}^{\theta}$ & 0.0063 & 0.1549 & 0.0683 & \textbf{0.4267} & 0.2934 & 0.1117 \\
 & RMSE & 0.077 & 0.196 & 0.229 & 0.384 & 0.214 & 0.096 & $\mathcal{L}_{P}^{\theta}$ & 1.61 & 2.216 & 1.691 & 4.86 & 3.284 & 2.782 \\
\midrule
\multirow{2}{*}{\textbf{AL-Multi}} & MAE & 0.088 & 0.1424 & 0.2001 & 0.2028 & \textbf{0.0255} & \textbf{0.0086} & $\mathcal{L}_{B}^{\theta}$ & 0.2847 & 0.1784 & 0.1865 & \underline{7.0212} & \textbf{0.1321} & \textbf{0.068} \\
 & RMSE & 0.088 & 0.142 & 0.2 & 0.203 & \textbf{0.026} & \textbf{0.009} & $\mathcal{L}_{P}^{\theta}$ & 1.634 & 1.798 & 1.511 & \textbf{3.423} & \textbf{2.542} & \textbf{1.808} \\
\midrule
\multirow{2}{*}{\textbf{AL-AR}} & MAE & 0.0168 & 0.0487 & 0.0145 & \underline{0.1624} & 0.0534 & 0.0227 & $\mathcal{L}_{B}^{\theta}$ & 0.4676 & 1.0839 & 0.2601 & 7.2126 & \underline{0.1399} & 0.2723 \\
 & RMSE & 0.017 & 0.049 & 0.014 & \underline{0.162} & 0.053 & 0.023 & $\mathcal{L}_{P}^{\theta}$ & 1.656 & 1.374 & 0.932 & 4.52 & 2.625 & 3.581 \\
\bottomrule
\end{tabular}

\vspace{1em}

\begin{tabular}{cc|c|c|c|c|c|c||c|c|c|c|c|c|c}
    \toprule
    \multirow{2}{*}{$\bm{\theta = 0.01}$} & & \multicolumn{3}{c|}{\textbf{N}} & \multicolumn{3}{c||}{\textbf{t}} & & \multicolumn{3}{c|}{\textbf{N}} & \multicolumn{3}{c}{\textbf{t}} \\
     & & \textbf{I} & \textbf{II} & \textbf{III} & \textbf{I} & \textbf{II} & \textbf{III} & & \textbf{I} & \textbf{II} & \textbf{III} & \textbf{I} & \textbf{II} & \textbf{III} \\
    \hline
    \multirow{2}{*}{\textbf{CAESar}} & MAE & \textbf{0.0028} & \textbf{0.0036} & \underline{0.003} & 0.2857 & 0.0625 & \underline{0.0172} & $\mathcal{L}_{B}^{\theta}$ & 0.0022 & 0.0031 & 0.0015 & 174.7012 & 15.5411 & 0.2973 \\
 & RMSE & \textbf{0.003} & \textbf{0.004} & \textbf{0.003} & 0.286 & 0.062 & \underline{0.017} & $\mathcal{L}_{P}^{\theta}$ & \textbf{1.056} & \textbf{1.223} & \textbf{0.937} & 3.729 & 2.829 & \textbf{2.171} \\
\midrule
\multirow{2}{*}{\textbf{K-CAViaR}} & MAE & \underline{0.0032} & \underline{0.0042} & \textbf{0.0028} & 0.3341 & 0.0672 & 0.0242 & $\mathcal{L}_{B}^{\theta}$ & 0.0022 & \underline{0.003} & \underline{0.0014} & 288.1228 & 22.8529 & 0.426 \\
 & RMSE & \underline{0.003} & \underline{0.004} & \underline{0.003} & 0.334 & 0.067 & 0.024 & $\mathcal{L}_{P}^{\theta}$ & \underline{1.083} & \underline{1.239} & \underline{0.937} & \underline{3.681} & 2.839 & \underline{2.181} \\
\midrule
\multirow{2}{*}{\textbf{BCGNS}} & MAE & 0.0119 & 0.0133 & 0.0096 & 1.0183 & 0.1483 & 0.04 & $\mathcal{L}_{B}^{\theta}$ & 0.0147 & 0.0145 & 0.0102 & 672.7932 & 20.6229 & 0.5798 \\
 & RMSE & 0.012 & 0.013 & 0.01 & 1.018 & 0.148 & 0.04 & $\mathcal{L}_{P}^{\theta}$ & 2.001 & 2.778 & 1.8 & 7.743 & 4.573 & 3.316 \\
\midrule
\multirow{2}{*}{\textbf{K-QRNN}} & MAE & 0.0129 & 0.0129 & 0.01 & 1.3303 & 0.257 & 0.0552 & $\mathcal{L}_{B}^{\theta}$ & \textbf{0.0013} & \textbf{0.0013} & \textbf{0.001} & 2320.0497 & 219.8131 & 0.6772 \\
 & RMSE & 0.013 & 0.013 & 0.01 & 1.33 & 0.257 & 0.055 & $\mathcal{L}_{P}^{\theta}$ & 1.27 & 1.391 & 1.111 & 18.16 & 11.147 & 3.494 \\
\midrule
\multirow{2}{*}{\textbf{GAS1}} & MAE & 0.0263 & 0.0584 & 0.0201 & \textbf{0.1313} & 0.1911 & 0.0581 & $\mathcal{L}_{B}^{\theta}$ & 0.0684 & 0.6865 & 0.1794 & 67.8674 & 0.5957 & \underline{0.2118} \\
 & RMSE & 0.026 & 0.058 & 0.02 & \textbf{0.131} & 0.191 & 0.058 & $\mathcal{L}_{P}^{\theta}$ & 1.435 & 6.035 & 1.767 & 4.474 & 3.964 & 2.416 \\
\midrule
\multirow{2}{*}{\textbf{GAS2}} & MAE & 0.4575 & 0.3327 & 0.1983 & 0.4976 & 0.3689 & 0.2655 & $\mathcal{L}_{B}^{\theta}$ & 0.1853 & 0.3492 & 0.0177 & \textbf{0.9937} & 1.3381 & 0.3063 \\
 & RMSE & 0.458 & 0.333 & 0.198 & 0.498 & 0.369 & 0.266 & $\mathcal{L}_{P}^{\theta}$ & 2.23 & 2.204 & 1.661 & 4.902 & 4.783 & 3.573 \\
\midrule
\multirow{2}{*}{\textbf{AL-Multi}} & MAE & 0.0167 & 0.1428 & 0.0127 & \underline{0.2232} & \textbf{0.0407} & \textbf{0.0138} & $\mathcal{L}_{B}^{\theta}$ & \underline{0.0021} & 0.0294 & 0.011 & \underline{5.5847} & \textbf{0.4227} & 0.3856 \\
 & RMSE & 0.017 & 0.143 & 0.013 & \underline{0.223} & \textbf{0.041} & \textbf{0.014} & $\mathcal{L}_{P}^{\theta}$ & 1.088 & 1.743 & 1.061 & \textbf{3.651} & \underline{2.796} & 2.184 \\
\midrule
\multirow{2}{*}{\textbf{AL-AR}} & MAE & 0.0758 & 0.0527 & 0.1414 & 0.3033 & \underline{0.057} & 0.0494 & $\mathcal{L}_{B}^{\theta}$ & 0.0026 & 1.6301 & 2.7171 & 47.4437 & \underline{0.4266} & \textbf{0.1457} \\
 & RMSE & 0.076 & 0.053 & 0.141 & 0.303 & \underline{0.057} & 0.049 & $\mathcal{L}_{P}^{\theta}$ & 1.79 & 1.37 & 2.679 & 4.049 & \textbf{2.592} & 3.063 \\
    \bottomrule
  \end{tabular}
}
\end{adjustwidth}
  \caption{Simulation results. The algorithms are compared over six different generating processes. All the generating processes are based on GARCH, both with Gaussian innovations (\textbf{N}) and t innovations (\textbf{t}). The coefficients of the process are fitted on the SPX (\textbf{I} set of coefficients), FTSE (\textbf{II} set), or DAX (\textbf{III} set). The value in each cell represents the mean loss of the row predictor across the twenty time series generated according to the column process. The best result is in bold, the second to best is underlined.}
    \label{tab:sim_study}
\end{table}\\
The results are displayed in Table \ref{tab:sim_study}. The MAE and RMSE values indicate that the best accuracy results are achieved by CAESar, K-CAViaR\textcolor{black}{, and AL-Multi}. Specifically, \textcolor{black}{AL-Multi better works with t-generated time series. Furthermore,} K-CAViaR outperforms \textcolor{black}{CAESar} for $\theta=0.05$, \textcolor{black}{the opposite is true} for $\theta=0.01$, and there is a balanced situation for middle $\theta$. It is worth noting that, as stated in \cite{patton2019dynamic}, the GAS models exhibit strong instability in the parameters optimization stage. Specifically, we have found that for some series, the dynamic given by these models is strongly unrealistic. Thus, in the comparison, we have not kept into account these series when computing the GAS losses.\\
If we look at the $\mathcal{L}^\theta_B$ and $\mathcal{L}^\theta_P$ values, we recover a predominance of CAESar, K-CAViaR\textcolor{black}{, and AL-Multi}. We also find that often, one of them outperforms the others according to a loss, but the situation is reverted according to the other loss. Nonetheless, comparing the best algorithms according to the two classes of metrics (MAE and RMSE on one side, $\mathcal{L}^\theta_B$ and $\mathcal{L}^\theta_P$ on the other side), we find that the $\mathcal{L}^\theta_P$ used by Patton et al. in \cite{patton2019dynamic} provides results which are more similar to those of MAE and RMSE. Such an incoherence between the $\mathcal{L}^\theta_B$ and $\mathcal{L}^\theta_P$ losses can be explained by taking in mind that they are designed for different purposes. Indeed, $\mathcal{L}^\theta_P$ is constructed to jointly estimate the VaR and ES. Instead, $\mathcal{L}^\theta_B$ is defined to estimate the ES given the knowledge of the VaR. However, we do not have the true VaR, but just an estimate for it. This estimate could sometimes not be correct or even show a large error. Thus, in such a context, the value expressed by the $\mathcal{L}^\theta_B$ loss may not be indicative of the true predictor's capability of approximating the ES. To verify this, we evaluate the $\mathcal{L}^\theta_B$ loss by using the actual $VaR_t$ instead of its estimate $\hat{q}_t$. Then, we count how many times the Barrera and Patton losses agree on which algorithm is the best and the second to best. This happens 19 times over a sample of 20 time series. On the contrary, when using the quantile estimate $\hat{q}_t$, the agreement number is reduced to 9, thus confirming the above intuition.\\
\linebreak
\textcolor{black}{Finally, a further simulation study is presented in Section \ref{app:srs} of the Supplementary Material. Here, we analyze the autoregressive models \textbf{CAESar}, \textbf{GAS2}, and \textbf{AL-AR} in a regime-switching framework. The results clearly show the predominance of CAESar due to its ability to rapidly catch changes in the market.}
\subsection{Empirical application}
\label{sec:main_res}

\subsubsection{Investigated datasets and estimation procedure}
As for the real-world dataset, we use daily data from 10 market indices. Specifically, we consider: Standard \& Poor 500 (\textbf{SPX}); IPC Mexico (\textbf{MXX}); BOlsa de Valores do Estado de São PAulo (\textbf{BOVESPA}); Cotation Assistée en Continu 40 (\textbf{CAC}); Korea Composite Stock Price Index (\textbf{KOSPI}); Iberia Index 35 (\textbf{IBEX}); Nihon Keizai Shinbun 225 (\textbf{NIKKEI}); Hang Seng Index (\textbf{HSI}); Deutscher Aktienindex 40 (\textbf{DAX}); Financial Times Stock Exchange 100 (\textbf{FTSE}). The dataset spans from 01-07-1993 to 30-06-2023, and it is divided into 24 block folds, each one seven-year long, obtained by rolling a seven-year-long window over the whole period covered by the dataset and shifting it by one year time by time. Then, each fold is split into training and test sets. We define the training set as the first-part six-year-long time series, which is further divided into five and one-year segments when a validation set is needed. The test set covers the last year. It is worth highlighting that each fold is considered an individual experiment. However, we aggregate the out-of-sample results when discussing the performance in the next subsections.

\subsubsection{Main results}
\begin{figure}[h]
{\color{black}
	\centering
	\includegraphics[width=0.9\linewidth]{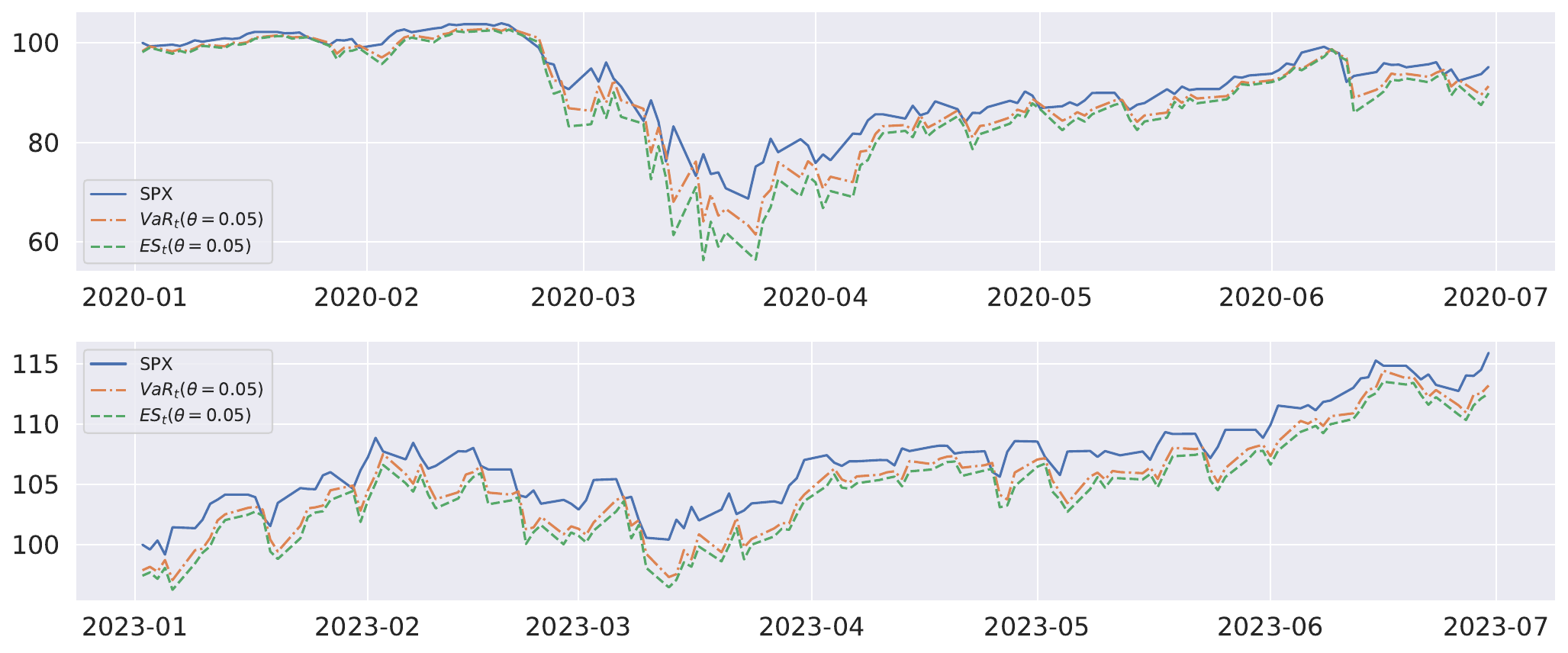}
	\caption{Example of CAESar forecast on the SPX index. The blue line represents the price (normalized to start at $100$); the orange line is for $VaR_t(\theta)$; and the green line is for $ES_t(\theta)$. The probability level $\theta = 0.05$ is considered. The upper subplot shows the COVID pandemic; the lower subplot is for normal market activity (the last period in our dataset).}
	\label{img:example_spx}
}
\end{figure}

\begin{figure}[h]
{\color{black}
	\centering
	\includegraphics[width=0.9\linewidth]{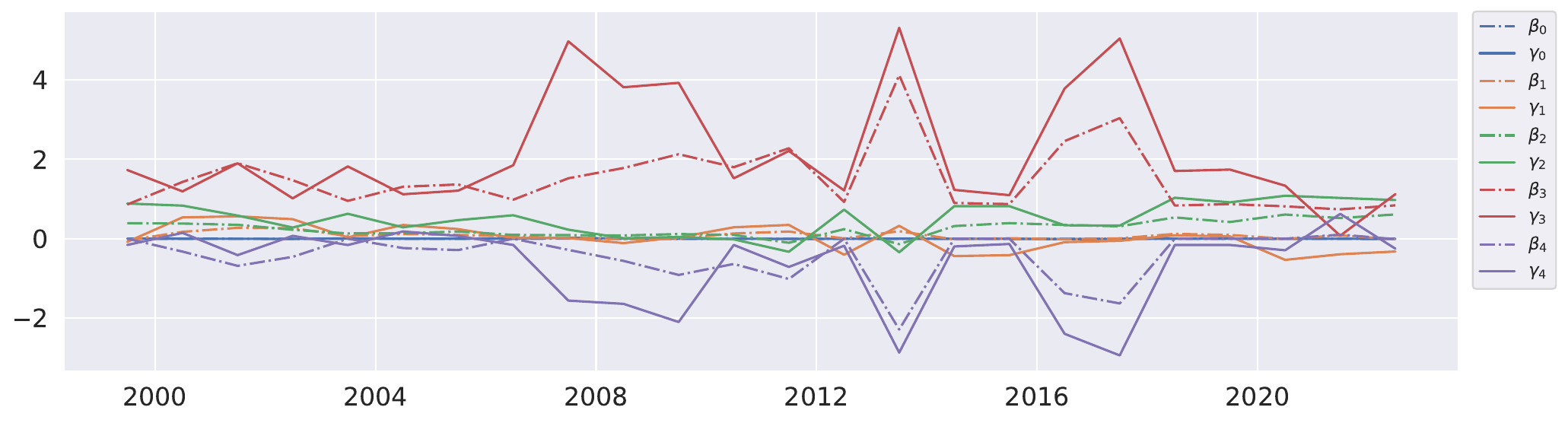}
	\caption{Example of CAESar forecast on the SPX index. The blue line represents the price (normalized to start at $100$); the orange line is for $VaR_t(\theta)$; and the green line is for $ES_t(\theta)$. The probability level $\theta = 0.05$ is considered. The upper subplot shows the COVID pandemic; the lower subplot is for normal market activity (the last period in our dataset).}
	\label{img:coeffs_spx}
}
\end{figure}

\begin{figure}[h]
{\color{black}
	\centering
	\includegraphics[width=0.9\linewidth]{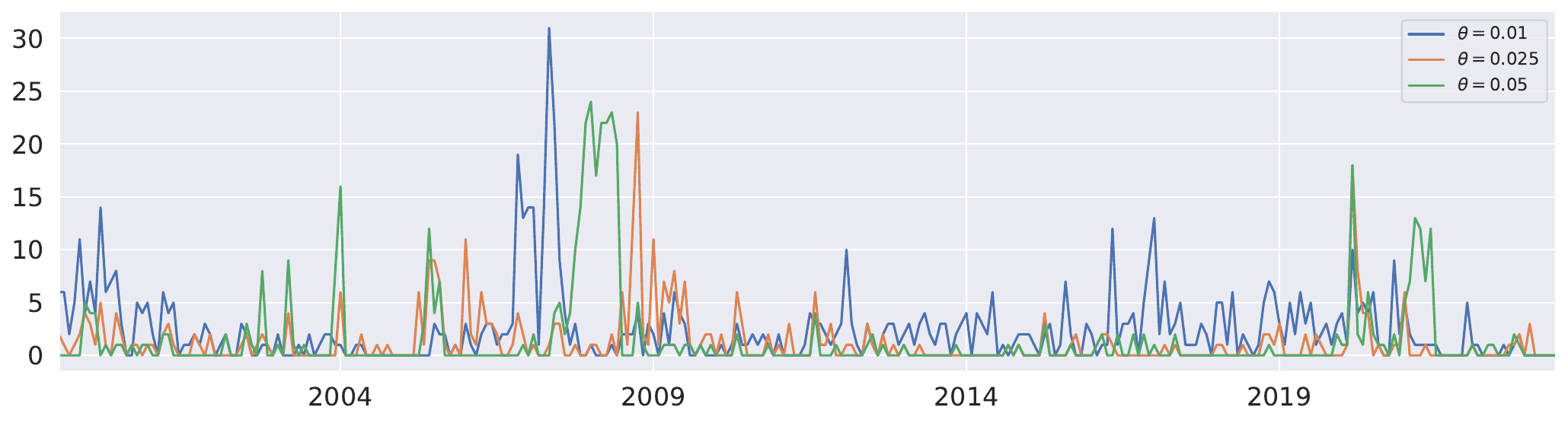}
	\caption{CAESar violations of the monotonicity condition $\hat{e}_t < \hat{q}_t$; monthly time series. Each point represents the total number of violations in the month, summing all the assets in the dataset (thus, each month has approximately $210 \sim 220$ observations). Each line is for a different $\theta$ value.}
	\label{img:mono_violations}
}
\end{figure}

\textcolor{black}{First, we discuss some results about the CAESar forecasts. Later, we show the comparison between the different models. Figure \ref{img:example_spx} presents an example of the risk measures forecast provided by CAESar. Two six-month periods are examined: one involves the financial crisis during the COVID pandemic; the other is a stable market. During the market turmoil, several quantile breaks are observable.}
\\
\textcolor{black}{The parameters fitted to the SPX time series are in Figure \ref{img:coeffs_spx}. According to our rolling block validation, the parameters are updated once per year. Specifically, $\beta_0$ and $\gamma_0$ are approximately flattened to $0$; their scale is far smaller than the other coefficients. The previous-lag price coefficients are relatively flat, too. Ultimately, the autoregressive is the most relevant component, both for the magnitude and the variance. Both the VaR and ES estimators are positively correlated with the previous lag estimation of the VaR; negatively correlated with the previous ES. The ES is typically more exposed than the VaR: both $\gamma_3 > \beta_3$ and $\gamma_4 > \beta_4$.}
\\
\textcolor{black}{Moreover, we analyze the worthiness of the effectiveness of the soft constraints used to prevent violations of the monotonicity conditions $\hat{q}_t < 0$ and $\hat{e}_t < \hat{q}_t$. In the whole testing period and over the $10$ indices considered in our dataset and the different $\theta$ values of our analysis, the condition on the quantile is violated only once, on 26 March 2020, when estimating the quantile at the $0.025$ probability level. Instead, the percentage of violations of the joint VaR-ES condition is $1.1\%$ with $\theta=0.01$, $0.6\%$ with $\theta=0.025$, and $0.7\%$ with $\theta=0.05$. The monthly time series of violations (aggregated over all the considered assets) is in Figure \ref{img:mono_violations}. As shown, the violations cluster during the Great Financial Crisis, mostly caused by the extreme turmoil of that period. Another notable rise, especially in $\theta=0.05$, is noticeable during the COVID pandemic.}
\\
As a first inspection of the models performance, we use the statistical tests previously described to evaluate the goodness of the ES fit. Specifically, we work with the Christoffersen's conditional coverage test, the McNeil and Frey test, the $Z_1$ statistic from \cite{acerbi2014back}, and the $Z_2$ statistic. It is worth to underline that these tests are performed asset by asset, fold by fold. So, for every asset, the test is performed 24 times, for a total of 240 iterations. Table \ref{tab:mnf_as_test} shows the ratio of rejection of the null hypothesis. The lower the ratio, the better the model. The confidence level of each test is set to 0.05.
\begin{table}
    \centering
    \begin{adjustwidth}{-1.4cm}{}
    \small{
    \begin{tabular}{c|cccc|cccc|cccc}
        \toprule
        & \multicolumn{4}{c|}{$\bm{\theta = 0.05}$} &  \multicolumn{4}{c|}{$\bm{\theta = 0.025}$} &  \multicolumn{4}{c}{$\bm{\theta = 0.01}$} \\
        & $\mathbf{CC}$ & $\mathbf{MNF}$ & $\mathbf{AS1}$ & $\mathbf{AS2}$ & $\mathbf{CC}$ & $\mathbf{MNF}$ & $\mathbf{AS1}$ & $\mathbf{AS2}$ & $\mathbf{CC}$ & $\mathbf{MNF}$ & $\mathbf{AS1}$ & $\mathbf{AS2}$ \\
        \midrule
        \textbf{CAESar} & 0.09 & \textbf{0.09} & \underline{0.14} & 0.18 & \textbf{0.04} & \textbf{0.13} & \textbf{0.16} & \textbf{0.11} & \textbf{0.05} & \textbf{0.31} & \textbf{0.32} & \textbf{0.19} \\
        \textbf{K-CAViaR} & 0.11 & 0.2 & 0.16 & 0.21 & 0.08 & \underline{0.19} & \underline{0.19} & 0.14 & \textbf{0.05} & \underline{0.4} & \underline{0.4} & \underline{0.22} \\
        \textbf{BCGNS} & \textbf{0.0} & 0.25 & \textbf{0.1} & \textbf{0.06} & \textbf{0.04} & 0.3 & 0.19 & \underline{0.13} & 0.09 & 0.55 & 0.4 & 0.32 \\
        \textbf{K-QRNN} & 0.19 & 0.43 & 0.37 & 0.48 & 0.17 & 0.53 & 0.52 & 0.44 & 0.09 & 0.71 & 0.66 & 0.59 \\
        \textbf{GAS1} & \underline{0.07} & \underline{0.1} & 0.34 & \underline{0.16} & 0.06 & 0.2 & 0.36 & 0.15 & 0.16 & 0.49 & 0.5 & 0.27 \\
        \textbf{GAS2} & 0.13 & 0.26 & 0.49 & 0.57 & 0.12 & 0.42 & 0.52 & 0.41 & 0.07 & 0.62 & 0.62 & 0.47 \\
        \textbf{AL-Multi} & 0.1 & 0.27 & 0.26 & 0.31 & 0.07 & 0.26 & 0.25 & 0.21 & 0.05 & 0.45 & 0.45 & 0.26 \\
        \textbf{AL-AR} & 0.1 & 0.37 & 0.41 & 0.36 & 0.1 & 0.4 & 0.45 & 0.38 & 0.09 & 0.58 & 0.62 & 0.4 \\
        \bottomrule
    \end{tabular}
    }
  \caption{Direct Approximation Tests - Results from Christoffersen's conditional coverage test \cite{mcneil2000estimation} (\textbf{CC)}, McNeil and Frey test \cite{mcneil2000estimation} (\textbf{MNF)}, and Acerbi and Szekely tests \cite{acerbi2014back}, both with the Z1 (\textbf{AS1}) and Z2 statistics (\textbf{AS2}). A separate test is conducted for each asset and each fold, focusing only on the out-of-sample predictions. Each cell value represents the ratio of null hypothesis rejections (with a threshold of 0.05). The best result (i.e., the lowest number of rejections) is in bold, and the second-best is underlined.}
    \label{tab:mnf_as_test}
    \end{adjustwidth}
\end{table}\\
For large values of $\theta$, several models perform well. Specifically, for $\theta=0.05$, the three top-performing algorithms are CAESar, BCGNS, and GAS1. The former shows superior performances in the McNeil and Frey (\textbf{MNF}) and the $Z_1$ (\textbf{AS1}) tests. On the other side, BCGNS obtains good performances in both the conditional coverage test and the Acerbi and Szekely statistics. Instead, the GAS1 model performs well in the McNeil and Frey and $Z_2$ tests (\textbf{AS2}). For $\theta=0.025$, the situation for BCGNS is almost the same. Instead, CAESar achieves good results in all the tests, while GAS1 shows a degradation of its performance and performs well only in the \textbf{MNF} test. Instead, for $\theta=0.01$, the best results on both tests are obtained almost always by CAESar and the second-best by the K-CAViaR. As pointed out in Section \ref{subs:stat_dir_app}, the discrepancy between the tests' results can be explained by taking into account the empirical study carried out in \cite{deng2021backtesting}. They compare several direct approximation tests applied to different time series lengths in order to have a picture of their finite-sample properties. The ultimate result, when using the length the most similar to that of our folds, is that the \textbf{AS1} test is the most stable both regarding size and power, followed by the \textbf{MNF} test. Finally, observe that the models performance tend to degrade (i.e., the number of rejections increases) when $\theta$ is small. This trend is coherent with the ES definition. Indeed, it averages values in the tail, and we may assume that the deep tail is caused by very extreme and rare (hence unpredictable) events. However, they are softened by the less severe events near the central body of the distribution. As the confidence value $\theta$ decreases, the contribution of these mitigating values vanishes, making the prediction problem even tougher. This leads to worsened performance. Similar results are also observed when working with stock data, as shown in the Supplementary Material \ref{app:sd}. However, as the relative performances of CAESar improve (i.e., its results look better than the ones of the competitors), we can conclude that it experiences less degradation in performance. Similar results are also observed when working with stock data, as shown in the Supplementary Material \ref{app:sd}.
\subsubsection{Models Comparison}
\begin{table}
  \centering
  \vspace{-2cm}
  \begin{adjustwidth}{-0.7cm}{}
\tiny{\begin{tabular}{cc|c|c|c|c|c|c|c|c|c|c}
\toprule
$\bm{\theta = 0.05}$ & & \textbf{SPX} & \textbf{MXX} & \textbf{BOVESPA} & \textbf{CAC} & \textbf{KOSPI} & \textbf{IBEX} & \textbf{NIKKEI} & \textbf{HSI} & \textbf{DAX} & \textbf{FTSE} \\
\midrule
\multirow{2}{*}{\textbf{CAESar}} & $\mathcal{L}_{B}^{\theta}$ & \underline{0.0023} & \textbf{0.0022} & \underline{0.0045} & 0.0025 & \underline{0.0033} & \textbf{0.0032} & \textbf{0.0031} & \textbf{0.0025} & \textbf{0.0024} & \underline{0.0019} \\
 & $\mathcal{L}_{P}^{\theta}$ & \underline{0.847} & \textbf{0.901} & \textbf{1.25} & 0.985 & \textbf{1.01} & \underline{1.074} & \textbf{1.061} & \textbf{1.05} & \textbf{0.998} & \underline{0.797} \\
\midrule
\multirow{2}{*}{\textbf{K-CAViaR}} & $\mathcal{L}_{B}^{\theta}$ & \textbf{0.0022} & \underline{0.0023} & 0.006 & \underline{0.0024} & 0.0035 & \underline{0.0032} & \underline{0.0031} & \underline{0.0028} & 0.0025 & \textbf{0.0017} \\
 & $\mathcal{L}_{P}^{\theta}$ & \textbf{0.844} & \underline{0.908} & \underline{1.286} & \textbf{0.978} & \underline{1.054} & \textbf{1.039} & \underline{1.092} & \underline{1.061} & \underline{1.005} & \textbf{0.792} \\
\midrule
\multirow{2}{*}{\textbf{BCGNS}} & $\mathcal{L}_{B}^{\theta}$ & 0.0111$^*$ & 0.0091$^*$ & 0.0117 & 0.0224$^*$ & 0.0071 & 0.0101$^*$ & 0.0159$^*$ & 0.0136$^*$ & 0.0158$^*$ & 0.018$^*$ \\
 & $\mathcal{L}_{P}^{\theta}$ & 2.44$^*$ & 2.571$^*$ & 8.451$^*$ & 5.984$^*$ & 2.204$^*$ & 2.168$^*$ & 4.249$^*$ & 3.126$^*$ & 15.583$^*$ & 2.516$^*$ \\
\midrule
\multirow{2}{*}{\textbf{K-QRNN}} & $\mathcal{L}_{B}^{\theta}$ & 0.0049 & 0.0026 & 0.0064 & 0.0048 & 0.0041 & 0.0048 & 0.0048 & 0.0038 & 0.0045 & 0.0037 \\
 & $\mathcal{L}_{P}^{\theta}$ & 1.152 & 1.071 & 1.427 & 1.303 & 1.218 & 1.26 & 1.23 & 1.282 & 1.339$^*$ & 1.104 \\
\midrule
\multirow{2}{*}{\textbf{GAS1}} & $\mathcal{L}_{B}^{\theta}$ & 0.0057 & 0.0024 & \textbf{0.0036} & 0.0043 & \textbf{0.0028} & 0.0033 & 0.0038 & 0.0035 & 0.0031 & 0.0022 \\
 & $\mathcal{L}_{P}^{\theta}$ & 1.216 & 0.985 & 1.31 & 1.72$^*$ & 1.154 & 1.296 & 1.166 & 1.108 & 1.071 & 0.944 \\
\midrule
\multirow{2}{*}{\textbf{GAS2}} & $\mathcal{L}_{B}^{\theta}$ & 0.0057 & 0.0183$^*$ & 0.005 & 0.0268$^*$ & 0.05$^*$ & 0.0075 & 0.0033 & 0.0544$^*$ & 0.0099$^*$ & 0.0086$^*$ \\
 & $\mathcal{L}_{P}^{\theta}$ & 1.313$^*$ & 1.718$^*$ & 1.656$^*$ & 1.672$^*$ & 1.775$^*$ & 2.065$^*$ & 1.568$^*$ & 2.484$^*$ & 1.957$^*$ & 1.484$^*$ \\
\midrule
\multirow{2}{*}{\textbf{AL-Multi}} & $\mathcal{L}_{B}^{\theta}$ & 0.0069$^*$ & 0.0194$^*$ & 0.0327$^*$ & 0.0162$^*$ & 0.0037 & 0.0032 & 0.0245$^*$ & 0.0029 & \underline{0.0024} & 0.0091$^*$ \\
 & $\mathcal{L}_{P}^{\theta}$ & 1.171 & 1.573$^*$ & 1.472 & 1.522$^*$ & 1.353 & 1.556$^*$ & 1.601$^*$ & 1.232 & 1.215 & 1.789$^*$ \\
\midrule
\multirow{2}{*}{\textbf{AL-AR}} & $\mathcal{L}_{B}^{\theta}$ & 0.0154$^*$ & 0.0214$^*$ & 0.0245$^*$ & \textbf{0.0019} & 0.0201$^*$ & 0.0141$^*$ & 0.0423$^*$ & 0.021$^*$ & 0.0857$^*$ & 0.0903$^*$ \\
 & $\mathcal{L}_{P}^{\theta}$ & 1.308$^*$ & 1.084 & 1.581$^*$ & \underline{0.979} & 1.14 & 2.153$^*$ & 1.4$^*$ & 1.7$^*$ & 2.229$^*$ & 2.404$^*$ \\
\bottomrule
\end{tabular}}

\vspace{1em}

\begin{tabular}{cc|c|c|c|c|c|c|c|c|c|c}
\toprule
$\bm{\theta = 0.025}$ & & \textbf{SPX} & \textbf{MXX} & \textbf{BOVESPA} & \textbf{CAC} & \textbf{KOSPI} & \textbf{IBEX} & \textbf{NIKKEI} & \textbf{HSI} & \textbf{DAX} & \textbf{FTSE} \\
\midrule
\multirow{2}{*}{\textbf{CAESar}} & $\mathcal{L}_{B}^{\theta}$ & \textbf{0.0051} & 0.0047 & 0.0191 & 0.0058 & \textbf{0.0091} & 0.0075 & 0.0072 & \textbf{0.0048} & \underline{0.0056} & \textbf{0.0033} \\
 & $\mathcal{L}_{P}^{\theta}$ & \underline{1.064} & \textbf{1.096} & \underline{1.475} & \textbf{1.148} & \underline{1.275} & \textbf{1.191} & \textbf{1.243} & \textbf{1.255} & \textbf{1.169} & \underline{0.965} \\
\midrule
\multirow{2}{*}{\textbf{K-CAViaR}} & $\mathcal{L}_{B}^{\theta}$ & \underline{0.0054} & 0.0057 & 0.0188 & \underline{0.0057} & \underline{0.0093} & \textbf{0.0072} & \underline{0.0068} & 0.0103 & 0.0057 & \underline{0.0033} \\
 & $\mathcal{L}_{P}^{\theta}$ & \textbf{1.06} & \underline{1.119} & 1.487 & \underline{1.151} & \textbf{1.27} & \underline{1.191} & \underline{1.268} & \underline{1.261} & \underline{1.172} & \textbf{0.96} \\
\midrule
\multirow{2}{*}{\textbf{BCGNS}} & $\mathcal{L}_{B}^{\theta}$ & 0.0149$^*$ & 0.0106 & 0.0228 & 0.0164$^*$ & 0.017 & 0.0171 & 0.0187$^*$ & 0.0158$^*$ & 0.0159$^*$ & 0.0128$^*$ \\
 & $\mathcal{L}_{P}^{\theta}$ & 5.263$^*$ & 2.218$^*$ & 2.536$^*$ & 2.023$^*$ & 2.523$^*$ & 1.97$^*$ & 4.342$^*$ & 1.84$^*$ & 3.0$^*$ & 2.807$^*$ \\
\midrule
\multirow{2}{*}{\textbf{K-QRNN}} & $\mathcal{L}_{B}^{\theta}$ & 0.016$^*$ & 0.0061 & 0.0212 & 0.0133 & 0.0116 & 0.014 & 0.0133 & 0.0103 & 0.0112 & 0.0114$^*$ \\
 & $\mathcal{L}_{P}^{\theta}$ & 1.466$^*$ & 1.294 & 1.693 & 1.518$^*$ & 1.492 & 1.492 & 1.467 & 1.517 & 1.507$^*$ & 1.379$^*$ \\
\midrule
\multirow{2}{*}{\textbf{GAS1}} & $\mathcal{L}_{B}^{\theta}$ & 0.0068 & \underline{0.0043} & \underline{0.018} & \textbf{0.0054} & 0.0117 & 0.0204 & 0.0129 & \underline{0.0085} & 0.0467$^*$ & 0.0055 \\
 & $\mathcal{L}_{P}^{\theta}$ & 1.25 & 1.328 & 1.886$^*$ & 1.415 & 1.401 & 1.376 & 1.505 & 1.443 & 1.636$^*$ & 1.386$^*$ \\
\midrule
\multirow{2}{*}{\textbf{GAS2}} & $\mathcal{L}_{B}^{\theta}$ & 0.0232$^*$ & 0.0242$^*$ & 0.0382 & 0.0417$^*$ & 0.0637$^*$ & 0.0114 & 0.0744$^*$ & 0.0238$^*$ & 0.0225$^*$ & 0.0113$^*$ \\
 & $\mathcal{L}_{P}^{\theta}$ & 2.65$^*$ & 2.288$^*$ & 1.696 & 1.975$^*$ & 2.451$^*$ & 1.551$^*$ & 1.887$^*$ & 2.303$^*$ & 1.834$^*$ & 2.187$^*$ \\
\midrule
\multirow{2}{*}{\textbf{AL-Multi}} & $\mathcal{L}_{B}^{\theta}$ & 0.0213$^*$ & \textbf{0.0042} & \textbf{0.0171} & 0.0058 & 0.0235 & \underline{0.0073} & \textbf{0.0052} & 0.0116$^*$ & \textbf{0.0053} & 0.0039 \\
 & $\mathcal{L}_{P}^{\theta}$ & 1.591$^*$ & 1.277 & \textbf{1.429} & 1.424 & 1.596 & 1.366 & 1.486 & 1.598 & 1.546$^*$ & 2.192$^*$ \\
\midrule
\multirow{2}{*}{\textbf{AL-AR}} & $\mathcal{L}_{B}^{\theta}$ & 0.0419$^*$ & 0.0058 & 0.0317 & 0.0066 & 0.0127 & 0.011 & 0.0478$^*$ & 0.0682$^*$ & 0.0512$^*$ & 0.1061$^*$ \\
 & $\mathcal{L}_{P}^{\theta}$ & 2.702$^*$ & 1.381 & 1.561 & 1.467 & 1.791 & 1.468 & 1.878$^*$ & 2.176$^*$ & 1.683$^*$ & 1.886$^*$ \\
\bottomrule
\end{tabular}

\vspace{1em}

\begin{tabular}{cc|c|c|c|c|c|c|c|c|c|c}
\toprule
$\bm{\theta = 0.01}$ & & \textbf{SPX} & \textbf{MXX} & \textbf{BOVESPA} & \textbf{CAC} & \textbf{KOSPI} & \textbf{IBEX} & \textbf{NIKKEI} & \textbf{HSI} & \textbf{DAX} & \textbf{FTSE} \\
\midrule
\multirow{2}{*}{\textbf{CAESar}} & $\mathcal{L}_{B}^{\theta}$ & \underline{0.0224} & 0.0188 & 0.0451 & 0.0273 & \textbf{0.0353} & 0.026 & 0.0236 & 0.0212 & 0.0333 & 0.011 \\
 & $\mathcal{L}_{P}^{\theta}$ & \textbf{1.427} & \textbf{1.376} & \textbf{1.652} & \underline{1.394} & \textbf{1.493} & \textbf{1.389} & \textbf{1.482} & \underline{1.476} & \underline{1.461} & \underline{1.191} \\
\midrule
\multirow{2}{*}{\textbf{K-CAViaR}} & $\mathcal{L}_{B}^{\theta}$ & 0.0286 & 0.0191 & 0.0441 & \underline{0.0225} & \underline{0.0364} & \underline{0.0249} & \textbf{0.0214} & \textbf{0.0153} & 0.0331 & \textbf{0.0095} \\
 & $\mathcal{L}_{P}^{\theta}$ & 1.554 & \underline{1.412} & \underline{1.756} & \textbf{1.387} & \underline{1.541} & \underline{1.394} & \underline{1.499} & \textbf{1.433} & \textbf{1.441} & \textbf{1.172} \\
\midrule
\multirow{2}{*}{\textbf{BCGNS}} & $\mathcal{L}_{B}^{\theta}$ & 0.0743$^*$ & 0.0341 & 0.1062 & 0.044 & 0.0777 & 0.0688 & 0.067$^*$ & 0.0628$^*$ & 0.0487 & 0.0382$^*$ \\
 & $\mathcal{L}_{P}^{\theta}$ & 3.48$^*$ & 2.262$^*$ & 2.335$^*$ & 1.993$^*$ & 2.539$^*$ & 2.161$^*$ & 2.603$^*$ & 2.438$^*$ & 2.222$^*$ & 2.131$^*$ \\
\midrule
\multirow{2}{*}{\textbf{K-QRNN}} & $\mathcal{L}_{B}^{\theta}$ & 0.0631$^*$ & 0.0131 & 0.0941 & 0.0453 & 0.0396 & 0.0504 & 0.0426 & 0.037 & 0.0355 & 0.0378$^*$ \\
 & $\mathcal{L}_{P}^{\theta}$ & 1.815 & 1.508 & 2.049$^*$ & 1.77 & 1.782 & 1.735 & 1.752 & 1.811 & 1.717 & 1.659$^*$ \\
\midrule
\multirow{2}{*}{\textbf{GAS1}} & $\mathcal{L}_{B}^{\theta}$ & 0.0248 & 0.0131 & 0.0496 & 0.0704 & 0.0921 & 0.0304 & 0.0325 & 0.049 & 0.1029 & 0.0675$^*$ \\
 & $\mathcal{L}_{P}^{\theta}$ & \underline{1.443} & 1.78 & 2.046$^*$ & 2.103$^*$ & 1.774 & 1.82$^*$ & 2.203$^*$ & 2.246$^*$ & 1.612 & 1.555 \\
\midrule
\multirow{2}{*}{\textbf{GAS2}} & $\mathcal{L}_{B}^{\theta}$ & 0.0479 & 0.034 & \underline{0.0131} & \textbf{0.0211} & 0.0602 & 0.0533 & 0.0383 & 0.0319 & 0.0387 & 0.0385$^*$ \\
 & $\mathcal{L}_{P}^{\theta}$ & 2.312$^*$ & 1.84 & 2.026$^*$ & 1.903$^*$ & 2.297$^*$ & 1.766$^*$ & 2.235$^*$ & 1.962$^*$ & 1.9 & 2.153$^*$ \\
\midrule
\multirow{2}{*}{\textbf{AL-Multi}} & $\mathcal{L}_{B}^{\theta}$ & 0.07$^*$ & \underline{0.0122} & 0.0184 & 0.0396 & 0.0371 & 0.0272 & 0.0279 & \underline{0.0169} & \textbf{0.012} & \underline{0.0097} \\
 & $\mathcal{L}_{P}^{\theta}$ & 2.32$^*$ & 1.682 & 1.888 & 1.646 & 1.604 & 1.549 & 1.887$^*$ & 1.487 & 1.791 & 1.291 \\
\midrule
\multirow{2}{*}{\textbf{AL-AR}} & $\mathcal{L}_{B}^{\theta}$ & \textbf{0.0117} & \textbf{0.0086} & \textbf{0.0081} & 0.0521 & 0.0399 & \textbf{0.0218} & \underline{0.0229} & 0.0211 & \underline{0.0169} & 0.0161 \\
 & $\mathcal{L}_{P}^{\theta}$ & 4.079$^*$ & 1.545 & 1.983 & 2.195$^*$ & 2.133 & 2.831$^*$ & 1.563 & 1.489 & 2.608$^*$ & 2.287$^*$ \\
\bottomrule
\end{tabular}
  \caption{Out-of-sample loss. It is shown the mean value across the 24 folds. $\mathcal{L}^\theta_B$ represents the loss used in \cite{barrera2022learning} (Eq. \ref{eq:0127_1912}), while $\mathcal{L}^\theta_P$ denotes the loss in \cite{patton2019dynamic} (Eq. \eqref{eq:0208_1938}). Different confidence levels are studied: $\theta=0.05, 0.025, 0.01$. The best result is highlighted in bold, and the second best is underlined. The $^*$ is used for those predictors whose mean loss is outside the CAESar confidence interval with one standard deviation.}
    \label{tab:ae_sum}
    \end{adjustwidth}
\end{table}
We corroborate the previous findings by directly comparing CAESar with the benchmark models. The results for the indices dataset at different levels of $\theta$ are shown in Table \ref{tab:ae_sum}. The table reports the mean value of the out-of-sample loss across the 24 folds, considering both losses in Eqs. \eqref{eq:0127_1912} and \eqref{eq:0208_1938}) used in \cite{barrera2022learning} and \cite{patton2019dynamic}. For the latter, we have used the percentage returns to obtain values coherent with \cite{patton2019dynamic}. As the table indicates, better results are typically achieved using the CAESar and K-CAViaR approaches. Specifically, CAESar overcomes the competitor more frequently, it is outside the top two algorithms fewer times, and it displays better performances when using the Patton loss $\mathcal{L}_P^\theta$ for evaluation. However, for some time series, GAS1 \textcolor{black}{and AL based} approach\textcolor{black}{es} also perform well. Widely, GAS1 seems to very often outperform GAS2. This is in line with the findings in \cite{patton2019dynamic}.

Next, we evaluate the proposed ES estimators using the Diebold-Mariano test, which is based on the loss function. As we use two losses in this work, the test is repeated twice. Furthermore, it is applied once for each asset and fold. The null hypothesis is the statistical equivalence of the two ES estimators. Table \ref{tab:dbtest_mig} shows the number of times this hypothesis is rejected in the Diebold-Mariano test (with a confidence level equal to 0.05). The left value in each cell indicates the number of times CAESar has statistically outperformed the competitor, while the right value is the count of the competitor wins. For a quick overview, we have only reported the sum of the number of rejections for each asset. The results clearly show the supremacy of CAESar, which outperforms all the competitors, by considering both losses and for all $\theta$ levels. Finally, the Loss Difference, the Encompassing, and the corrected resampled t-test previously described are shown in Supplementary Material \ref{app:eld}. The results are in line with the Diebold-Mariano test.\\
\begin{table}
  \centering
    \begin{adjustwidth}{-1.0cm}{}
\small{
    \begin{tabular}{cc|c|c|c|c|c|c|c}
    \toprule
    & & \textbf{K-CAViaR} & \textbf{BCGNS} & \textbf{K-QRNN} & \textbf{GAS1} & \textbf{GAS2} & \textbf{AL-Multi} & \textbf{AL-AR} \\
    \midrule
    \multirow{2}{*}{$\bm{\theta = 0.05}$} & $\mathcal{L}_{B}^{\theta}$ & 20 / 0 & 176 / 0 & 27 / 0 & 52 / 0 & 78 / 0 & 32 / 0 & 104 / 0 \\
    & $\mathcal{L}_{P}^{\theta}$ & 28 / 15 & 163 / 0 & 134 / 3 & 74 / 2 & 128 / 5 & 60 / 9 & 115 / 7 \\
    \midrule
    \multirow{2}{*}{$\bm{\theta = 0.025}$} & $\mathcal{L}_{B}^{\theta}$ & 26 / 0 & 99 / 0 & 22 / 0 & 41 / 0 & 82 / 0 & 23 / 0 & 107 / 0 \\
    & $\mathcal{L}_{P}^{\theta}$ & 29 / 24 & 133 / 2 & 134 / 3 & 102 / 5 & 126 / 0 & 44 / 23 & 120 / 12 \\
    \midrule
    \multirow{2}{*}{$\bm{\theta = 0.01}$} & $\mathcal{L}_{B}^{\theta}$ & 38 / 0 & 40 / 0 & 19 / 0 & 57 / 0 & 65 / 0 & 18 / 0 & 117 / 0 \\
    & $\mathcal{L}_{P}^{\theta}$ & 54 / 15 & 70 / 8 & 122 / 2 & 114 / 6 & 119 / 4 & 58 / 21 & 124 / 14 \\
    \bottomrule
    \end{tabular}
    }
  \caption{Model Comparison Tests - Diebold-Mariano test with confidence 0.05. The results obtained across various assets, for a given probability level $\theta$, have been aggregated by summing them. Each cell value is in the form $g/b$, where $g$ represents the number of folds where CAESar statistically outperforms the competitor, and $b$ is the contrary.}
    \label{tab:dbtest_mig}
    \end{adjustwidth}
\end{table}

\begin{figure}[h]
	\centering
	\includegraphics[width=0.9\linewidth]{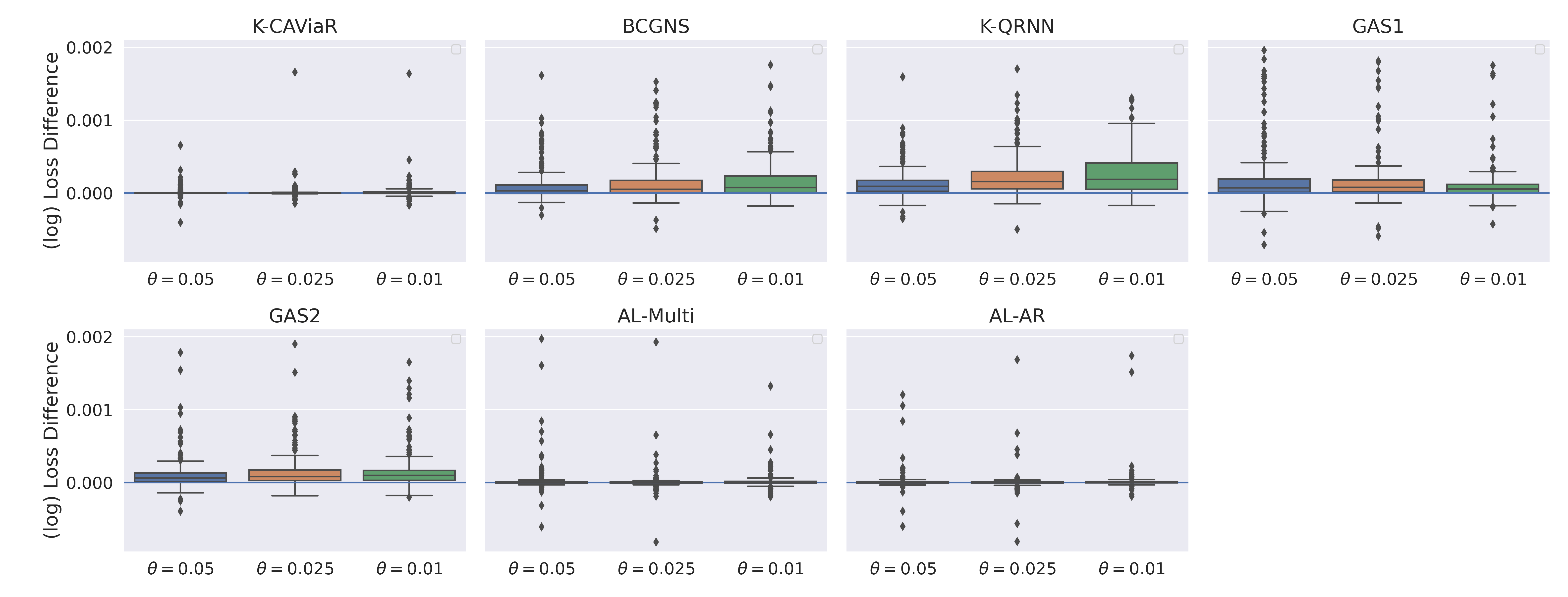}
	\caption{Boxplot of the Pinball loss difference. CAESar is compared with all the competitors in terms of quantile estimates. That is, the difference between the competitor and CAESar's loss is plotted. Positive values mean that CAESar outperforms the competitor. The y-axis is scaled, to improve the readability, according to the map: $x\rightarrow sign(x)\log(1+|x|)$.}
	\label{img:pinball_loss_diff_box}
\end{figure}

\begin{figure}[h]
	\centering
	\includegraphics[width=0.9\linewidth]{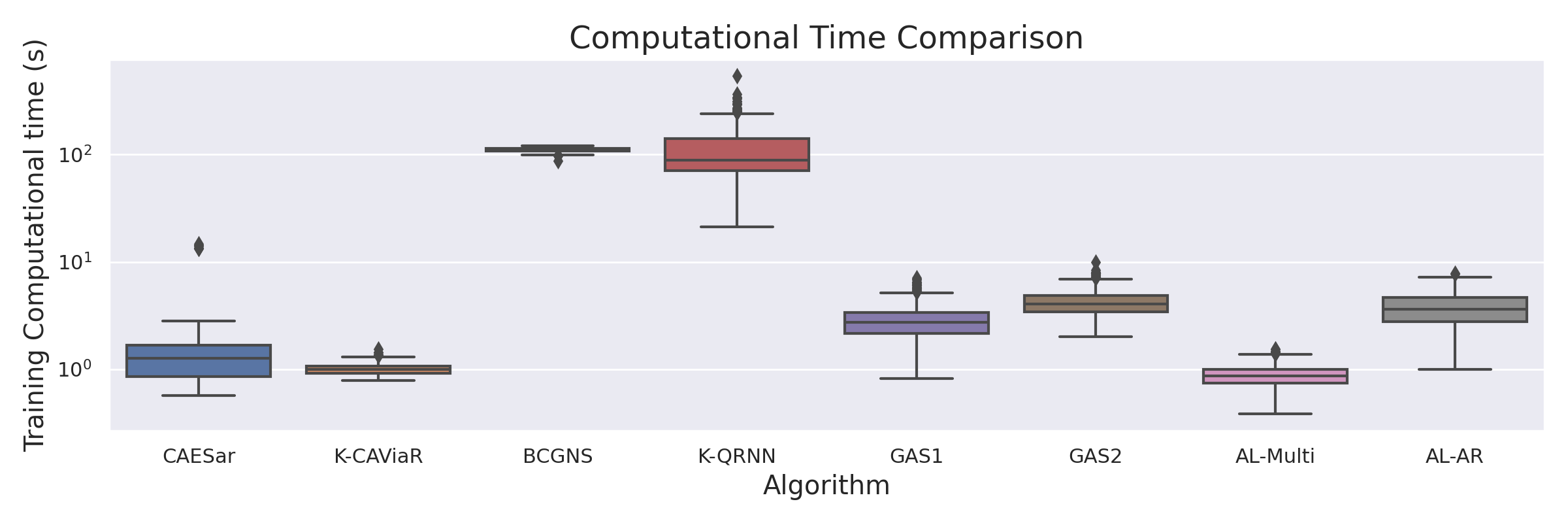}
	\caption{Comparison of the computational time required for each algorithm. The plot is in log scale. The training time is measured in seconds. Every box summarizes a 720-dimensional vector containing 10 assets, 24 folds, and 3 $\theta$ values.}
	\label{img:comp_time}
\end{figure}

As previously discussed in Section \ref{sec:model}, it is interesting to understand what happens to the quantile estimate when we update the $\bm{\hat{q}}$ parameters in the third step of the CAESar optimization procedure. Specifically, a degradation in the VaR forecast capability could undermine CAESar's reliability. Actually, we find out this is not the case. Specifically, we study the Pinball loss (Eq. \eqref{eq:0127_1911}) of all the competing methods. For each confidence level $\theta$, asset, and fold, we compute the mean loss difference between the competitor and CAESar. Figure \ref{img:pinball_loss_diff_box} shows the boxplots of these differences -one for each probability level $\theta$. Specifically, for every competitor $Comp$, the boxplots show the distribution of the vector $\left[\mathcal{L}_q^{\theta}(\bm{\hat{q}}^{Comp}_{\theta, a,f}, \bm{y}_{a,f}) - \mathcal{L}_q^{\theta}(\bm{\hat{q}}^{CAESar}_{\theta, a,f}, \bm{y}_{a,f})\right]_{a\in A, \ f\in F}$, $\forall \theta\in\{0.05, 0.025, 0.01\}$ where $a\in A=\{SPX, MXX, \cdots\}$ represents the asset, and $f\in F=\{1,\cdots,24\}$ is for the fold. $\bm{\hat{q}}^{CAESar}_{\theta, a,f}$ is for the CAESar prediction, while $\bm{\hat{q}}^{Comp}_{\theta, a,f}$ is for the competitor forecast.
\\
Three qualitative conclusions can be drawn. The first one is that conditional autoregressive models overcome the competitors even when looking at the quantile estimate. Then, neural network models (BCGNS and K-QRNN) outperform GAS \textcolor{black}{and AL based} models. Finally, keep in mind that, by construction, the CAESar quantile estimate before the joint updating step is exactly the K-CAViaR quantile estimate (as both of them are simply the CAViaR estimate at the confidence level $\theta$). Thus, from the upper-left plot, we can observe how the VaR forecast by CAESar is not degraded by the update in the $\bm{\hat{q}}$ parameters. On the contrary, it is even slightly improved.\\
\linebreak
Finally, Figure \ref{img:comp_time} shows the average computational times of the different methods. The times are computed on a Ubuntu notebook equipped with a \textit{13th Gen Intel Core™ i7-13620H x 16} processor and a \textit{NVIDIA GeForce RTX 4060} video card. The CAESar \textcolor{black}{and AL based} model\textcolor{black}{s are} parallelized over 3 processes, one for each random vector of coefficients used as the starting point in the optimization step (more details in Supplementary Material \ref{app:os}). The K-CAViaR approach exploits 10 processes, one for each CAViaR model corresponding to a quantile $\theta_j$. As for the GAS1 and GAS2 models, we initialize their parameters to those provided by the authors, as their performance is highly dependent on the choice of initial coefficients (as pointed out in \cite{patton2019dynamic}). Therefore, multiprocessing is not necessary. Finally, the K-QRNN and BCGNS models are trained on the GPU.
\section{Conclusions}
\label{sec:conclusion}
Forecasting the ES in a dynamic context is known to be a hard task for several reasons. One of them is the intrinsic complexity of financial time series, which are characterized by heteroskedasticity and fat tails, which complicates the modelling and forecasting process. Moreover, the fact that ES is not elicitable further adds to the challenge. This lack has historically slowed down the research in this direction, at least until the finding that ES and VaR are jointly elicitable. This breakthrough has opened new doors in the field. Despite this progress, a noticeable gap still exists between the theoretical works and the practical needs of the industry for reliable models. Adding to these challenges are the issues associated with ES backtesting, mainly driven by the lack of elicitability. Furthermore, the fact that ES is a latent variable -meaning it is not directly observed from the data- adds another layer of complexity to its backtesting. Thus, backtesting is still an open point, and a universally recognized statistical procedure to assess models' performance in forecasting the ES of a conditional distribution is missing.\\
\linebreak
We contribute to this discussion by proposing CAESar, a three-step approach for predicting the ES at a given time, conditional on the information available at the previous time. CAESar draws significant inspiration from the CAViaR approach to quantile regression, a statistical technique used to estimate the conditional quantile. Indeed, the specification for the $\bm{\hat{e}}$ estimator is designed as the asymmetric slope with two autoregressive addends: one for the previous step quantile and the other for the previous step ES. This allows for a robust and dynamic estimation of risk. The CAESar coefficients optimization is in three steps. First, the VaR coefficients are optimized via CAViaR. Then, the $\bm{\hat{e}}$ coefficients are trained while the VaR model remains fixed. Finally, the two estimators are jointly trained. In both steps, a soft constraint is added to the loss function to ensure the monotonicity condition - that is, the ES estimate has to be less than the quantile forecast.\\
\linebreak
To assess the effectiveness of our proposal, we have conducted extensive testing comparing CAESar with several state-of-the-art approaches. Most of the effort has been put into the daily index dataset. We collected a dataset comprising daily observations from 10 stock indices over the last 30 years. This dataset was divided into block folds, and several models were compared. Specifically, three types of evaluation were performed. Firstly, we have compared the model loss functions. Next, we have conducted statistical tests to assess whether one model statistically outperforms another based on out-of-sample losses. Then, we performed loss-agnostic tests to directly evaluate the ES estimator (i.e., the coherence between the approximated values and the ES definition). The main finding is that CAESar outperforms the other models throughout the comparative analysis we present. In particular, looking at the direct approximation tests' rejection frequency, CAESar almost always displays the lowest value. Moreover, the gap is far more evident in the very tail of the returns' distribution (low values of $\theta$), when all other ES models start to fail. It is worth noting that both BCGNS (that is, a neural network-based approach) and K-CAViaR (which approximates the expectation in the ES definition with the sample mean of quantile forecasting at different confidence levels) models sometimes display good forecasting performance, but the results for these models are not stable and strongly depend on the value of $\theta$. Additionally, we conducted experiments on a dataset composed of stocks in the U.S. banking sector. Once again, CAESar has on average superior performance.\\
\linebreak
There are two primary directions for further research. On the algorithmic side, \textcolor{black}{a first improvement would be to extend the proposed formulation to handle jump processes. Specifically, this can be achieved by using realized estimators as additional regressors. Then,} the next step would be to predict ES for multiple correlated assets. This would involve updating CAESar to handle multivariate time series, providing valuable insights into tail risk networks. Furthermore, moving from the ES of single assets to that of a portfolio is key. In this context, CAESar could serve as a tool to assist investors and risk managers in adjusting portfolio weights to minimize overall tail risk. On the theoretical side, it would be useful to gain a deeper understanding of the proposed estimator, particularly its asymptotic properties. This could enhance our understanding of the dynamics captured by CAESar.
%
%

\section*{Data and Code Availability}
The data utilized in this paper are publicly accessible. Specifically, index data have been downloaded from three sources: Investing \url{https://www.investing.com/}, Yahoo Finance \url{https://finance.yahoo.com}, and MarketWatch \url{https://www.marketwatch.com/}. The three sources have been used together to complete any gaps in the time series. As for the banking sector stocks, data have been obtained from Investing and Yahoo Finance. The CAESar code, along with the codes for competing strategies and model comparison, is available online on GitHub (and not disclosed here as the journal's policy on author identification). The code can be provided upon request at any moment.

\section*{Acknowledgments}
We would like to thank the organizers and participants of the Quantitative Finance Workshop 2025 and of the International Fintech Research Conference. PM acknowledges financial support under the National Recovery and Resilience Plan (PNNR), Mission 4, Component 2, Investment 1.1, Call for tender No. 104 published on 2.2.2022 by the Italian Ministry of University and Research (MUR), funded by the European Union – NextGenerationEU– Project Title ‘‘Realized Random Graphs: A New Econometric Methodology
for the Inference of Dynamic Networks’’ – CUP B53D23010100001 - Grant Assignment Decree No. 2022MRSYB7 by the Italian Ministry of University and Research (MUR).

\bibliographystyle{abbrv}
\bibliography{caesar_bib}

\newpage

\clearpage

\setcounter{affn}{0}
\begin{frontmatter}



\title{\\Supplementary Material}


\end{frontmatter}

\clearpage
\setcounter{page}{2}

\bigskip

\newpage

\appendix
\renewcommand{\appendixname}{}

\section{Supplementary Material - Model Specifications}
\label{app:ms}

In Section \ref{sec:model}, we have discussed a general formulation for the CAESar regressor, as we have done for the CAViaR in Section \ref{sec:problem}. Nonetheless, we specifically propose to use the (1,1) model with Asymmetric Slope (AS) for $y_t$ transformation. Indeed, as it is a popular choice for CAViaR in the literature, we expect it to work very well also for CAESar. The reasons behind such a success are twofold. As for the $\bm{f}$ and $\bm{g}$ functions specification, Asymmetric Slope is the only one (among those proposed in \cite{engle2004caviar}) that can efficiently capture the asymmetry in market movements. Indeed, the other popular alternatives, namely the Symmetric Absolute Value (SAV) and the indirect GARCH specifications, are not able to do this, typically resulting in poorer performance, or at least not overcoming AS. Specifically, the SAV (1,1) specification assumes symmetry in the information contained in positive and negative market returns, thus resulting in:
\begin{equation}
\label{eq:0213_1755}
\hat{q}_t = \beta_0 + \beta_1 |y_{t-1}| + \beta_2 \hat{q}_{t-1} + \beta_3 \hat{e}_{t-1} \quad and \quad \hat{e}_t = \gamma_0 + \gamma_1 |y_{t-1}| + \gamma_2 \hat{q}_{t-1} + \gamma_3 \hat{e}_{t-1}
\end{equation}
On the other hand, there is the GARCH (1,1):
\begin{equation}
\label{eq:0213_1756}
\hat{q}_t = -\sqrt{\beta_0 + \beta_1 y_{t-1}^2 + \beta_2 \hat{q}_{t-1}^2 + \beta_3 \hat{e}_{t-1}^2} \quad and \quad \hat{e}_t = -\sqrt{\gamma_0 + \gamma_1 y_{t-1}^2 + \gamma_2 \hat{q}_{t-1}^2 + \gamma_3 e_{t-1}^2}
\end{equation}
Observe that according to the formulation in Eq. \eqref{eq:0501_0922}, with a slight abuse of notation, we are applying this specification to compute the squared target. To corroborate our choice, we repeat the experiments using both the AS, SAV, and GARCH (1,1) specifications. The results are in Table \ref{tab:spec_comp} and show that the other specifications do not significantly outperform AS.
\begin{table}
    \centering
    \begin{adjustwidth}{-.8cm}{}
\tiny{
\begin{tabular}{cc|c|c|c|c|c|c|c|c|c|c}
\toprule
$\bm{\theta = 0.05}$ & & \textbf{SPX} & \textbf{MXX} & \textbf{BOVESPA} & \textbf{CAC} & \textbf{KOSPI} & \textbf{IBEX} & \textbf{NIKKEI} & \textbf{HSI} & \textbf{DAX} & \textbf{FTSE} \\
\midrule
\multirow{2}{*}{\textbf{CAESar (AS)}} & $\mathcal{L}_{B}^{\theta}$ & \textbf{0.0023} & \textbf{0.0022} & \textbf{0.0045} & \underline{0.0025} & \textbf{0.0033} & \underline{0.0032} & \underline{0.0031} & \textbf{0.0025} & \textbf{0.0024} & \textbf{0.0019} \\
 & $\mathcal{L}_{P}^{\theta}$ & \textbf{0.847} & \textbf{0.901} & \textbf{1.25} & \textbf{0.985} & \textbf{1.01} & 1.074 & \textbf{1.061} & \textbf{1.05} & \textbf{0.998} & \textbf{0.797} \\
\midrule
\multirow{2}{*}{\textbf{CAESar (SAV)}} & $\mathcal{L}_{B}^{\theta}$ & \underline{0.0028} & 0.0025 & \underline{0.0049} & 0.0032 & \underline{0.0037} & 0.0035 & 0.0032 & \underline{0.003} & \underline{0.0028} & \underline{0.0021} \\
 & $\mathcal{L}_{P}^{\theta}$ & \underline{0.994} & \underline{0.923} & \underline{1.266} & \underline{1.026} & \underline{1.071} & \textbf{1.057} & \underline{1.097} & \underline{1.055} & \underline{1.036} & \underline{0.831} \\
\midrule
\multirow{2}{*}{\textbf{CAESar (G)}} & $\mathcal{L}_{B}^{\theta}$ & 0.0055 & \underline{0.0022} & 0.0058 & \textbf{0.0023} & 0.005 & \textbf{0.0025} & \textbf{0.0026} & 0.0037 & 0.0035 & 0.0024 \\
 & $\mathcal{L}_{P}^{\theta}$ & 1.025 & 1.003 & 1.356 & 1.261 & 1.396 & \underline{1.073} & 1.113 & 1.393 & 1.152 & 1.15 \\
\bottomrule
\end{tabular}
\vspace{1em}
\begin{tabular}{cc|c|c|c|c|c|c|c|c|c|c}
\toprule
$\bm{\theta = 0.025}$ & & \textbf{SPX} & \textbf{MXX} & \textbf{BOVESPA} & \textbf{CAC} & \textbf{KOSPI} & \textbf{IBEX} & \textbf{NIKKEI} & \textbf{HSI} & \textbf{DAX} & \textbf{FTSE} \\
\midrule
\multirow{2}{*}{\textbf{CAESar (AS)}} & $\mathcal{L}_{B}^{\theta}$ & \textbf{0.0051} & \textbf{0.0047} & \textbf{0.0191} & \textbf{0.0058} & \textbf{0.0091} & \underline{0.0075} & 0.0072 & \textbf{0.0048} & \underline{0.0056} & \underline{0.0033} \\
 & $\mathcal{L}_{P}^{\theta}$ & \underline{1.064} & \textbf{1.096} & \textbf{1.475} & \textbf{1.148} & \textbf{1.275} & \textbf{1.191} & \textbf{1.243} & 1.255 & \textbf{1.169} & \textbf{0.965} \\
\midrule
\multirow{2}{*}{\textbf{CAESar (SAV)}} & $\mathcal{L}_{B}^{\theta}$ & \underline{0.0051} & \underline{0.0049} & \underline{0.0194} & \underline{0.0071} & \underline{0.011} & 0.0095 & \underline{0.0068} & 0.0056 & 0.0065 & 0.004 \\
 & $\mathcal{L}_{P}^{\theta}$ & \textbf{1.061} & \underline{1.12} & \underline{1.549} & \underline{1.19} & \underline{1.278} & 1.248 & \underline{1.277} & \underline{1.222} & \underline{1.201} & \underline{1.0} \\
\midrule
\multirow{2}{*}{\textbf{CAESar (G)}} & $\mathcal{L}_{B}^{\theta}$ & 0.014 & 0.005 & 0.0464 & 0.0148 & 0.0111 & \textbf{0.0065} & \textbf{0.0059} & \underline{0.0052} & \textbf{0.0054} & \textbf{0.0031} \\
 & $\mathcal{L}_{P}^{\theta}$ & 1.376 & 1.165 & 2.113 & 1.468 & 1.592 & \underline{1.196} & 1.289 & \textbf{1.2} & 1.258 & 1.137 \\
\bottomrule
\end{tabular}
\vspace{1em}
\begin{tabular}{cc|c|c|c|c|c|c|c|c|c|c}
\toprule
$\bm{\theta = 0.01}$ & & \textbf{SPX} & \textbf{MXX} & \textbf{BOVESPA} & \textbf{CAC} & \textbf{KOSPI} & \textbf{IBEX} & \textbf{NIKKEI} & \textbf{HSI} & \textbf{DAX} & \textbf{FTSE} \\
\midrule
\multirow{2}{*}{\textbf{CAESar (AS)}} & $\mathcal{L}_{B}^{\theta}$ & \underline{0.0224} & 0.0188 & \textbf{0.0451} & \textbf{0.0273} & \underline{0.0353} & \textbf{0.026} & 0.0236 & \underline{0.0212} & 0.0333 & \textbf{0.011} \\
 & $\mathcal{L}_{P}^{\theta}$ & \underline{1.427} & \textbf{1.376} & \textbf{1.652} & \textbf{1.394} & \textbf{1.493} & \textbf{1.389} & \underline{1.482} & \underline{1.476} & \textbf{1.461} & \textbf{1.191} \\
\midrule
\multirow{2}{*}{\textbf{CAESar (SAV)}} & $\mathcal{L}_{B}^{\theta}$ & \textbf{0.018} & \textbf{0.0135} & 0.0882 & \underline{0.0306} & 0.0378 & \underline{0.0328} & \textbf{0.0202} & \textbf{0.0178} & \underline{0.0251} & \underline{0.0123} \\
 & $\mathcal{L}_{P}^{\theta}$ & \textbf{1.334} & \underline{1.391} & 1.862 & \underline{1.432} & 2.078 & \underline{1.429} & \textbf{1.474} & \textbf{1.44} & 1.509 & \underline{1.195} \\
\midrule
\multirow{2}{*}{\textbf{CAESar (G)}} & $\mathcal{L}_{B}^{\theta}$ & 0.0563 & \underline{0.0172} & \underline{0.061} & 0.1043 & \textbf{0.0251} & 0.0494 & \underline{0.0207} & 0.0518 & \textbf{0.0206} & 0.0135 \\
 & $\mathcal{L}_{P}^{\theta}$ & 1.613 & 1.501 & \underline{1.706} & 1.984 & \underline{1.62} & 1.609 & 1.5 & 2.018 & \underline{1.478} & 1.335 \\
\bottomrule
\end{tabular}
}
    \caption{Comparison of the three specifications (SAV, AS, and GARCH) with one lag for both the asset value and previous quantile and expected shortfall on the index dataset. The same specification is used both for the CAViaR regressor in the first step and for the $\bm{\hat{r}}$ and $\bm{\hat{e}}$ models in the last steps. The mean value across the 24 folds is displayed. The best result is in bold, the second to best is underlined.}
    \label{tab:spec_comp}
    \end{adjustwidth}
\end{table}\\
\linebreak
As for the SAV and GARCH specifications, the former seems to perform slightly better for $\theta=0.01$, while the latter works better with higher $\theta$. However, none of them can consistently overcome the AS specification, which shows superior performances for all $\theta$ and almost all the considered assets, being always able to give performances close to the best one.\\
\linebreak
Another point when choosing the CAESar model is related to the lags to consider. The choice is related to the observations lag $p$ and the previous estimates lag $u$. We suggest to use $p=u=1$. Indeed, it is the cheapest. Moreover, it avoids the overfitting risk, and its estimates are often the most accurate. We empirically show this in Table \ref{tab:n_lags_comp}, where the AS-(1,1) is compared with the AS-(1,2), AS-(2,1), and AS(2,2) specifications.
\begin{table}
    \centering
    \begin{adjustwidth}{-.8cm}{}
\tiny{    
\begin{tabular}{cc|c|c|c|c|c|c|c|c|c|c}
\toprule
$\bm{\theta = 0.05}$ & & \textbf{SPX} & \textbf{MXX} & \textbf{BOVESPA} & \textbf{CAC} & \textbf{KOSPI} & \textbf{IBEX} & \textbf{NIKKEI} & \textbf{HSI} & \textbf{DAX} & \textbf{FTSE} \\
\midrule
\multirow{3}{*}{\textbf{CAESar (1,1)}} & $\mathcal{L}_{B}^{\theta}$ & \underline{0.0023} & \underline{0.0022} & \textbf{0.0045} & \textbf{0.0025} & \underline{0.0033} & \textbf{0.0032} & \textbf{0.0031} & \textbf{0.0025} & \textbf{0.0024} & \textbf{0.0019} \\
 & $\mathcal{L}_{P}^{\theta}$ & \underline{0.847} & \textbf{0.901} & \underline{1.25} & \underline{0.985} & \textbf{1.01} & \underline{1.074} & \textbf{1.061} & \underline{1.05} & \textbf{0.998} & \textbf{0.797} \\
 & $C.T.$ & 1.3 & 2.5 & 2.0 & 1.5 & 1.3 & 2.0 & 1.9 & 1.4 & 1.3 & 1.4 \\
\midrule
\multirow{3}{*}{\textbf{CAESar (1,2)}} & $\mathcal{L}_{B}^{\theta}$ & 0.0023 & 0.0024 & 0.0062 & \underline{0.0026} & 0.0035 & \underline{0.0034} & \underline{0.0032} & 0.0027 & 0.0027 & \underline{0.0019} \\
 & $\mathcal{L}_{P}^{\theta}$ & 0.867 & 1.009 & 1.29 & 1.073 & 1.115 & 1.114 & 1.099 & 1.405 & \underline{1.013} & 0.825 \\
 & $C.T.$ & 5.0 & 5.1 & 4.4 & 4.4 & 2.8 & 5.6 & 3.1 & 4.9 & 4.2 & 4.3 \\
\midrule
\multirow{3}{*}{\textbf{CAESar (2,1)}} & $\mathcal{L}_{B}^{\theta}$ & \textbf{0.0022} & \textbf{0.002} & \underline{0.0047} & 0.0026 & \textbf{0.0032} & 0.0035 & 0.0032 & \underline{0.0025} & 0.0048 & 0.0019 \\
 & $\mathcal{L}_{P}^{\theta}$ & \textbf{0.846} & \underline{0.91} & \textbf{1.247} & \textbf{0.984} & \underline{1.046} & \textbf{1.037} & \underline{1.086} & \textbf{1.035} & 1.488 & \underline{0.805} \\
 & $C.T.$ & 4.8 & 7.1 & 7.3 & 5.3 & 6.3 & 7.8 & 4.5 & 6.2 & 5.4 & 3.3 \\
\midrule
\multirow{3}{*}{\textbf{CAESar (2,2)}} & $\mathcal{L}_{B}^{\theta}$ & 0.0023 & 0.0023 & 0.0062 & 0.0027 & 0.0037 & 0.0039 & 0.0033 & 0.0028 & \underline{0.0025} & 0.002 \\
 & $\mathcal{L}_{P}^{\theta}$ & 0.882 & 0.925 & 1.324 & 1.011 & 1.056 & 1.32 & 1.105 & 1.059 & 1.097 & 0.898 \\
 & $C.T.$ & 14.2 & 16.8 & 16.0 & 11.3 & 10.7 & 13.9 & 10.5 & 14.6 & 11.9 & 8.7 \\
\bottomrule
\end{tabular}

\vspace{1em}

\begin{tabular}{cc|c|c|c|c|c|c|c|c|c|c}
\toprule
$\bm{\theta = 0.025}$ & & \textbf{SPX} & \textbf{MXX} & \textbf{BOVESPA} & \textbf{CAC} & \textbf{KOSPI} & \textbf{IBEX} & \textbf{NIKKEI} & \textbf{HSI} & \textbf{DAX} & \textbf{FTSE} \\
\hline
\multirow{3}{*}{\textbf{CAESar (1,1)}} & $\mathcal{L}_{B}^{\theta}$ & \textbf{0.0051} & \underline{0.0047} & 0.0191 & \underline{0.0058} & 0.0091 & \underline{0.0075} & 0.0072 & \underline{0.0048} & \textbf{0.0056} & \textbf{0.0033} \\
 & $\mathcal{L}_{P}^{\theta}$ & \textbf{1.064} & 1.096 & 1.475 & \textbf{1.148} & 1.275 & \textbf{1.191} & \textbf{1.243} & 1.255 & \underline{1.169} & \textbf{0.965} \\
 & $C.T.$ & 1.2 & 1.6 & 1.3 & 1.2 & 1.3 & 1.6 & 1.4 & 1.5 & 1.0 & 1.2 \\
\midrule
\multirow{3}{*}{\textbf{CAESar (1,2)}} & $\mathcal{L}_{B}^{\theta}$ & 0.007 & 0.0051 & \textbf{0.0127} & 0.0067 & \underline{0.0082} & 0.0331 & 0.0079 & 0.0054 & 0.0081 & 0.008 \\
 & $\mathcal{L}_{P}^{\theta}$ & 1.193 & 1.133 & \textbf{1.442} & \underline{1.17} & \textbf{1.211} & 1.611 & 1.562 & \underline{1.227} & 6.456 & 2.296 \\
 & $C.T.$ & 3.7 & 3.5 & 3.5 & 3.2 & 4.8 & 3.9 & 3.5 & 3.0 & 3.6 & 5.1 \\
\midrule
\multirow{3}{*}{\textbf{CAESar (2,1)}} & $\mathcal{L}_{B}^{\theta}$ & \underline{0.0053} & \textbf{0.0042} & 0.0156 & 0.0063 & \textbf{0.0076} & \textbf{0.0073} & \textbf{0.0068} & \textbf{0.0044} & \underline{0.0061} & \underline{0.0035} \\
 & $\mathcal{L}_{P}^{\theta}$ & \underline{1.079} & \textbf{1.061} & \underline{1.451} & 1.419 & \underline{1.221} & \underline{1.194} & \underline{1.272} & \textbf{1.195} & \textbf{1.168} & \underline{0.988} \\
 & $C.T.$ & 6.8 & 3.3 & 7.4 & 5.4 & 4.7 & 8.9 & 4.8 & 5.7 & 4.6 & 3.8 \\
\midrule
\multirow{3}{*}{\textbf{CAESar (2,2)}} & $\mathcal{L}_{B}^{\theta}$ & 0.0086 & 0.0049 & \underline{0.0131} & \textbf{0.0055} & 0.012 & 0.0083 & \underline{0.0069} & 0.0057 & 0.0063 & 0.004 \\
 & $\mathcal{L}_{P}^{\theta}$ & 3.891 & \underline{1.083} & 1.459 & 1.206 & 1.413 & 1.256 & 1.283 & 1.321 & 1.224 & 1.027 \\
 & $C.T.$ & 8.5 & 11.9 & 11.7 & 11.0 & 9.1 & 15.2 & 8.8 & 11.3 & 8.1 & 8.6 \\
\bottomrule
\end{tabular}

\vspace{1em}

\begin{tabular}{cc|c|c|c|c|c|c|c|c|c|c}
\toprule
$\bm{\theta = 0.01}$ & & \textbf{SPX} & \textbf{MXX} & \textbf{BOVESPA} & \textbf{CAC} & \textbf{KOSPI} & \textbf{IBEX} & \textbf{NIKKEI} & \textbf{HSI} & \textbf{DAX} & \textbf{FTSE} \\
\midrule
\multirow{3}{*}{\textbf{CAESar (1,1)}} & $\mathcal{L}_{B}^{\theta}$ & \textbf{0.0224} & 0.0188 & \underline{0.0451} & 0.0273 & 0.0353 & \underline{0.026} & \textbf{0.0236} & 0.0212 & 0.0333 & \textbf{0.011} \\
 & $\mathcal{L}_{P}^{\theta}$ & \textbf{1.427} & 1.376 & \textbf{1.652} & \underline{1.394} & \textbf{1.493} & \textbf{1.389} & \textbf{1.482} & \underline{1.476} & 1.461 & \textbf{1.191} \\
 & $C.T.$ & 1.8 & 1.3 & 1.3 & 1.0 & 1.1 & 1.2 & 1.3 & 1.4 & 0.9 & 1.2 \\
\midrule
\multirow{3}{*}{\textbf{CAESar (1,2)}} & $\mathcal{L}_{B}^{\theta}$ & \underline{0.0239} & \textbf{0.0116} & 0.1044 & \textbf{0.0212} & \textbf{0.0278} & 0.0541 & 0.0267 & \underline{0.0179} & \textbf{00256} & 0.0166 \\
 & $\mathcal{L}_{P}^{\theta}$ & \underline{1.473} & \underline{1.352} & 1.893 & \textbf{1.373} & 1.619 & 1.564 & 1.521 & 1.551 & \textbf{1.394} & 1.374 \\
 & $C.T.$ & 3.4 & 3.2 & 3.0 & 5.0 & 3.8 & 4.1 & 3.1 & 3.1 & 2.6 & 3.5 \\
\midrule
\multirow{3}{*}{\textbf{CAESar (2,1)}} & $\mathcal{L}_{B}^{\theta}$ & 0.0657 & \underline{0.0129} & \textbf{0.043} & 0.0403 & 0.0449 & \textbf{0.0231} & \underline{0.0251} & \textbf{0.0154} & \underline{0.0306} & \underline{0.0128} \\
 & $\mathcal{L}_{P}^{\theta}$ & 1.74 & \textbf{1.338} & \underline{1.674} & 1.56 & 1.658 & \underline{1.411} & \underline{1.495} & \textbf{1.434} & \underline{1.414} & \underline{1.233} \\
 & $C.T.$ & 4.7 & 4.5 & 6.1 & 3.1 & 6.0 & 6.8 & 4.2 & 4.3 & 3.3 & 3.5 \\
\midrule
\multirow{3}{*}{\textbf{CAESar (2,2)}} & $\mathcal{L}_{B}^{\theta}$ & 0.0336 & 0.0158 & 0.0944 & \underline{0.0238} & \underline{0.0293} & 0.029 & 0.0314 & 0.0251 & 0.031 & 0.016 \\
 & $\mathcal{L}_{P}^{\theta}$ & 1.555 & 1.415 & 1.881 & 1.48 & \underline{1.57} & 1.46 & 1.63 & 20.276 & 2.127 & 1.379 \\
 & $C.T.$ & 7.0 & 8.0 & 9.2 & 10.8 & 10.0 & 10.5 & 8.3 & 5.8 & 6.4 & 5.9 \\
\bottomrule
\end{tabular}
}
    \caption{Comparison between different numbers of lags for the AS specification. The specifications are compared by means of the $\mathcal{L}_B^\theta$ loss, the $\mathcal{L}_B^\theta$ loss, and the computational time ($C.T.$). The best result is in bold, and the second-best is underlined.}
    \label{tab:n_lags_comp}
    \end{adjustwidth}
\end{table}\\
As shown in the table, the (1,1) model is by far the fastest. Furthermore, its performances are often at the top. Indeed, for $\theta=0.05$, even when it is not the best specification, its performance is close to the best one. Instead, for lower $\theta$, sometimes it is clearly overcome by the (1,2) or, more frequently, the (2,1) specifications. However, if we look at the mean behaviour, it looks pointless to consider over-complicated models. Interestingly, the (2,2) model exhibits the worst performance. This can be due to both misspecification or issues in the parameters optimization process due to the increased complexity of the model.\\
\linebreak
Finally, it is worth observe we have taken for granted the cross terms in the $\hat{q}_t$ and $\hat{e}_t$ estimators in Eq. \eqref{eq:0502_1200}. That is, we have assumed that $\hat{q}_t$ depends on $\hat{e}_{t-1}$ and $\hat{e}_t$ on $\hat{q}_{t-1}$. To verify this assumption, we compare CAESar with its \textbf{No Cross} version. The parameters of the latter are optimized by only looking at the $\mathcal{L}_P^\theta$ loss, as the optimization of $\mathcal{L}_B^\theta$ would necessarily require $\hat{e}_t$ dependent on $\hat{q}_{t-1}$. The result is shown in Table \ref{tab:cross_terms_comp}.

\begin{table}[h]
\centering
\tiny{
\begin{tabular}{cc|c|c|c|c|c|c|c|c|c|c}
\toprule
$\bm{\theta = 0.05}$ & & \textbf{SPX} & \textbf{MXX} & \textbf{BOVESPA} & \textbf{CAC} & \textbf{KOSPI} & \textbf{IBEX} & \textbf{NIKKEI} & \textbf{HSI} & \textbf{DAX} & \textbf{FTSE} \\
\midrule
\multirow{2}{*}{\textbf{CAESar}} & $\mathcal{L}_{B}^{\theta}$ & 0.0023 & \textbf{0.0022} & \textbf{0.0045} & 0.0025 & \textbf{0.0033} & \textbf{0.0032} & \textbf{0.0031} & \textbf{0.0025} & 0.0024 & 0.0019 \\
 & $\mathcal{L}_{P}^{\theta}$ & \textbf{0.847} & 0.901 & \textbf{1.25} & \textbf{0.985} & \textbf{1.01} & 1.074 & \textbf{1.061} & 1.05 & \textbf{0.998} & 0.797 \\
\midrule
\multirow{2}{*}{\textbf{No Cross}} & $\mathcal{L}_{B}^{\theta}$ & \textbf{0.0021} & 0.0023 & 0.006 & \textbf{0.0024} & 0.0034 & 0.0033 & 0.0032 & 0.0026 & \textbf{0.0023} & \textbf{0.0018} \\
 & $\mathcal{L}_{P}^{\theta}$ & 0.862 & \textbf{0.892} & 1.266 & 0.988 & 1.049 & \textbf{1.026} & 1.085 & \textbf{1.023} & 1.003 & \textbf{0.793} \\
\bottomrule
\end{tabular}

\vspace{1em}

\begin{tabular}{cc|c|c|c|c|c|c|c|c|c|c}
\toprule
$\bm{\theta = 0.025}$ & & \textbf{SPX} & \textbf{MXX} & \textbf{BOVESPA} & \textbf{CAC} & \textbf{KOSPI} & \textbf{IBEX} & \textbf{NIKKEI} & \textbf{HSI} & \textbf{DAX} & \textbf{FTSE} \\
\midrule
\multirow{2}{*}{\textbf{CAESar}} & $\mathcal{L}_{B}^{\theta}$ & \textbf{0.0051} & \textbf{0.0047} & 0.0191 & 0.0058 & \textbf{0.0091} & 0.0075 & 0.0072 & \textbf{0.0048} & \textbf{0.0056} & \textbf{0.0033} \\
 & $\mathcal{L}_{P}^{\theta}$ & \textbf{1.064} & \textbf{1.096} & \textbf{1.475} & \textbf{1.148} & \textbf{1.275} & \textbf{1.191} & \textbf{1.243} & \textbf{1.255} & \textbf{1.169} & \textbf{0.965} \\
\midrule
\multirow{2}{*}{\textbf{No Cross}} & $\mathcal{L}_{B}^{\theta}$ & 0.0062 & 0.0051 & \textbf{0.0187} & \textbf{0.0057} & 0.0093 & \textbf{0.0071} & \textbf{0.0069} & 0.0126 & 0.0057 & 0.0033 \\
 & $\mathcal{L}_{P}^{\theta}$ & 1.174 & 1.133 & 1.489 & 1.148 & 3.054 & 1.196 & 1.258 & 1.256 & 1.173 & 0.985 \\
\bottomrule
\end{tabular}

\vspace{1em}

\begin{tabular}{cc|c|c|c|c|c|c|c|c|c|c}
\toprule
$\bm{\theta = 0.01}$ & & \textbf{SPX} & \textbf{MXX} & \textbf{BOVESPA} & \textbf{CAC} & \textbf{KOSPI} & \textbf{IBEX} & \textbf{NIKKEI} & \textbf{HSI} & \textbf{DAX} & \textbf{FTSE} \\
\midrule
\multirow{2}{*}{\textbf{CAESar}} & $\mathcal{L}_{B}^{\theta}$ & \textbf{0.0224} & 0.0188 & \textbf{0.0451} & 0.0273 & \textbf{0.0353} & 0.026 & \textbf{0.0236} & 0.0212 & 0.0333 & 0.011 \\
 & $\mathcal{L}_{P}^{\theta}$ & \textbf{1.427} & 1.376 & \textbf{1.652} & \textbf{1.394} & \textbf{1.493} & \textbf{1.389} & \textbf{1.482} & 1.476 & 1.461 & \textbf{1.191} \\
\midrule
\multirow{2}{*}{\textbf{No Cross}} & $\mathcal{L}_{B}^{\theta}$ & 0.0289 & \textbf{0.0128} & 0.0592 & \textbf{0.0225} & 0.036 & \textbf{0.0246} & 0.0245 & \textbf{0.0175} & \textbf{0.0246} & \textbf{0.0096} \\
 & $\mathcal{L}_{P}^{\theta}$ & 1.682 & \textbf{1.328} & 1.847 & 1.404 & 1.672 & 1.402 & 1.504 & \textbf{1.471} & \textbf{1.416} & 1.228 \\
\bottomrule
\end{tabular}
}
    \caption{Comparison of CAESar with AS-(1,1) specification and the \textbf{No Cross} specification, i.e., that obtained from Eq. \eqref{eq:0226_1701} by forcing $\beta_4=\gamma_3=0$, that means no cross dependency between the VaR and ES estimators. The best result is in bold.}
    \label{tab:cross_terms_comp}
\end{table}
Even if the results are not completely unidirectional, the table still shows that the specification that takes into account cross terms roughly outperforms the other. This suggests that the intrinsic relationship between VaR and ES is not merely related to elicitability, as information on the previous estimate of one of them could improve the prediction for the other, suggesting a kind of Granger causality between them.
\newpage
\section{Supplementary Material - Optimization Scheme}
\label{app:os}

In this Supplementary Material section, we discuss the optimization scheme developed to learn the CAESar parameters. First, we outline the optimization routine. Then, the choice of the loss functions is discussed. CAESar's first step involves obtaining the CAViaR quantile estimator. This is accomplished using the code available on the personal website of its author\footnote{\url{simonemanganelli.org}}. As for the second step, we follow this guideline:
\begin{enumerate}
    \item The $\hat{e}_0$ estimate, which will serve as the starting point in the CAESar loop, is obtained using the empirical ES computed over the first 10\% of points in the training set.
    \item As for the initial guess of the coefficients vector $\bm{\gamma}=(\gamma_0,\cdots,\gamma_4)$, 100 random initializations are drawn from uniform (between -1 and 1) and standard normal distributions. Then, they are sorted according to the $\mathcal{L}^\theta_r(\bm{\hat{r}}, \bm{y}; \bm{VaR}(\theta))$ loss, and only the top three are kept.
    \item Each of the three $\bm{\gamma}$ initial guesses is used as the starting point for optimization. Specifically, a series of six sequential optimizations are performed. Sequential here means the output of one optimization is used as the starting point for the next. The optimizer used is the Sequential Least SQuares Programming \cite{kraft1988software}. Note that three processes are simultaneously run, one for each starting $\bm{\gamma}$ choice.
\end{enumerate}
After the last point, we have three optimal $\bm{\gamma}$, each one obtained from a different starting point. They are sorted by means of the $\mathcal{L}^\theta_r(\bm{\hat{r}}, \bm{y}; \bm{VaR}(\theta))$ loss, and the most promising one is selected as the output for the entire step. Lastly, in the third step, we repeat the previous routine with using the output from the previous step as the only starting and the loss $\mathcal{L}^\theta_{q,e}(\bm{\hat{e}}, \bm{\hat{q}}, \bm{y})$ instead of $\mathcal{L}^\theta_r(\bm{\hat{r}}, \bm{y}; \bm{VaR}(\theta))$.\\
\linebreak
Three alternatives have been considered for the loss function used in the training process. These include training only with the loss in Eq. \eqref{eq:0227_1200}, training with the loss in Eq. \eqref{eq:0227_1201}, and training with both losses. The loss function in Eq. \eqref{eq:0227_1200} involves only the optimization of the ES coefficients, so it is much faster. On the other side, we expect the loss function in Eq. \eqref{eq:0227_1201} to be more precise as it refines the VaR prediction by incorporating information on the expected tail mean from the previous step. So, as standard in the optimization literature, the faster method is initially utilized to achieve a coarse approximation of the solution. This approximation is then used as the starting point for the more precise method.\\
We empirically compare three different versions of CAESar. These include the original version presented in Section \ref{sec:model}, the version optimized based on the loss in Eq. \eqref{eq:0227_1200} (B-CAESar), and the version derived solely from the loss in Eq. \eqref{eq:0227_1201} (P-CAESar). The comparison is shown in Table \ref{tab:comp_caesar_loss}.
\begin{table}
  \centering
  \begin{adjustwidth}{-.6cm}{}
\tiny{
\begin{tabular}{cc|c|c|c|c|c|c|c|c|c|c}
\toprule
$\bm{\theta = 0.05}$ & & \textbf{SPX} & \textbf{MXX} & \textbf{BOVESPA} & \textbf{CAC} & \textbf{KOSPI} & \textbf{IBEX} & \textbf{NIKKEI} & \textbf{HSI} & \textbf{DAX} & \textbf{FTSE} \\
\midrule
\multirow{2}{*}{\textbf{CAESar}} & $\mathcal{L}_{B}^{\theta}$ & 0.0023 & \textbf{0.0022} & \textbf{0.0045} & 0.0025 & \textbf{0.0033} & \textbf{0.0032} & \textbf{0.0031} & \underline{0.0025} & \textbf{0.0024} & 0.0019 \\
 & $\mathcal{L}_{P}^{\theta}$ & 0.847 & \textbf{0.901} & \underline{1.25} & 0.985 & \textbf{1.01} & 1.074 & \textbf{1.061} & \underline{1.05} & \textbf{0.998} & \underline{0.797} \\
\midrule
\multirow{2}{*}{\textbf{B-CAESar}} & $\mathcal{L}_{B}^{\theta}$ & \textbf{0.0022} & \underline{0.0023} & 0.006 & \textbf{0.0024} & 0.0035 & \underline{0.0033} & \underline{0.0032} & 0.0028 & 0.0025 & \textbf{0.0017} \\
 & $\mathcal{L}_{P}^{\theta}$ & \textbf{0.834} & 0.908 & 1.297 & \textbf{0.983} & 1.053 & \underline{1.041} & \underline{1.095} & 1.069 & 1.007 & 0.798 \\
\midrule
\multirow{2}{*}{\textbf{P-CAESar}} & $\mathcal{L}_{B}^{\theta}$ & \underline{0.0022} & 0.0024 & \underline{0.0045} & \underline{0.0024} & \underline{0.0034} & 0.0035 & 0.0032 & \textbf{0.0022} & \underline{0.0024} & \underline{0.0017} \\
 & $\mathcal{L}_{P}^{\theta}$ & \underline{0.844} & \underline{0.902} & \textbf{1.247} & \underline{0.983} & \underline{1.027} & \textbf{1.034} & 1.102 & \textbf{1.018} & \underline{1.003} & \textbf{0.796} \\
\bottomrule
\end{tabular}
\vspace{1em}
\begin{tabular}{cc|c|c|c|c|c|c|c|c|c|c}
\toprule
$\bm{\theta = 0.025}$ & & \textbf{SPX} & \textbf{MXX} & \textbf{BOVESPA} & \textbf{CAC} & \textbf{KOSPI} & \textbf{IBEX} & \textbf{NIKKEI} & \textbf{HSI} & \textbf{DAX} & \textbf{FTSE} \\
\midrule
\multirow{2}{*}{\textbf{CAESar}} & $\mathcal{L}_{B}^{\theta}$ & \textbf{0.0051} & \underline{0.0047} & 0.0191 & \underline{0.0058} & \underline{0.0091} & \underline{0.0075} & \underline{0.0072} & \textbf{0.0048} & \underline{0.0056} & \underline{0.0033} \\
 & $\mathcal{L}_{P}^{\theta}$ & \underline{1.064} & \underline{1.096} & 1.475 & \textbf{1.148} & 1.275 & 1.191 & \textbf{1.243} & \underline{1.255} & \underline{1.169} & 0.965 \\
\midrule
\multirow{2}{*}{\textbf{B-CAESar}} & $\mathcal{L}_{B}^{\theta}$ & \underline{0.0054} & 0.0057 & \underline{0.0188} & \textbf{0.0057} & 0.0093 & \textbf{0.0072} & \textbf{0.0068} & 0.0105 & 0.0057 & 0.0033 \\
 & $\mathcal{L}_{P}^{\theta}$ & \textbf{1.048} & 1.133 & \underline{1.473} & 1.149 & \underline{1.239} & \underline{1.183} & \underline{1.276} & 1.295 & 1.169 & \textbf{0.962} \\
\midrule
\multirow{2}{*}{\textbf{P-CAESar}} & $\mathcal{L}_{B}^{\theta}$ & 0.0055 & \textbf{0.0044} & \textbf{0.0146} & 0.006 & \textbf{0.0081} & 0.0077 & 0.0072 & \underline{0.005} & \textbf{0.0055} & \textbf{0.0032} \\
 & $\mathcal{L}_{P}^{\theta}$ & 1.074 & \textbf{1.071} & \textbf{1.438} & \underline{1.148} & \textbf{1.179} & \textbf{1.17} & 1.281 & \textbf{1.213} & \textbf{1.161} & \underline{0.962} \\
\bottomrule
\end{tabular}
\vspace{1em}
\begin{tabular}{cc|c|c|c|c|c|c|c|c|c|c}
\toprule
$\bm{\theta = 0.01}$ & & \textbf{SPX} & \textbf{MXX} & \textbf{BOVESPA} & \textbf{CAC} & \textbf{KOSPI} & \textbf{IBEX} & \textbf{NIKKEI} & \textbf{HSI} & \textbf{DAX} & \textbf{FTSE} \\
\midrule
\multirow{2}{*}{\textbf{CAESar}} & $\mathcal{L}_{B}^{\theta}$ & \textbf{0.0224} & \underline{0.0188} & \underline{0.0451} & \underline{0.0273} & \textbf{0.0353} & 0.026 & 0.0236 & 0.0212 & 0.0333 & 0.011 \\
 & $\mathcal{L}_{P}^{\theta}$ & \textbf{1.427} & \underline{1.376} & \textbf{1.652} & \underline{1.394} & \textbf{1.493} & \underline{1.389} & \underline{1.482} & 1.476 & \underline{1.461} & 1.191 \\
\midrule
\multirow{2}{*}{\textbf{B-CAESar}} & $\mathcal{L}_{B}^{\theta}$ & 0.0285 & 0.0192 & \textbf{0.0439} & \textbf{0.0223} & \underline{0.0363} & \underline{0.0248} & \textbf{0.0214} & \textbf{0.0153} & \underline{0.0332} & \textbf{0.0094} \\
 & $\mathcal{L}_{P}^{\theta}$ & \underline{1.787} & 1.455 & 1.743 & \textbf{1.374} & \underline{1.498} & \textbf{1.378} & 1.493 & \textbf{1.438} & 1.461 & \textbf{1.166} \\
\midrule
\multirow{2}{*}{\textbf{P-CAESar}} & $\mathcal{L}_{B}^{\theta}$ & \underline{0.0253} & \textbf{0.0105} & 0.0506 & 0.0275 & 0.0398 & \textbf{0.0225} & \underline{0.023} & \underline{0.0157} & \textbf{0.0252} & \underline{0.0103} \\
 & $\mathcal{L}_{P}^{\theta}$ & 1.867 & \textbf{1.261} & \underline{1.698} & 1.397 & 1.591 & 1.407 & \textbf{1.473} & \underline{1.45} & \textbf{1.401} & \underline{1.184} \\
\bottomrule
\end{tabular}
}
  \caption{Comparison between different versions of CAESar. \textbf{CAESar} refers to the version discussed in the main text, Section \ref{sec:model}. \textbf{B-CAESar} is the version trained only with the Barrera loss $\mathcal{L}^\theta_r(\bm{\hat{r}}, \bm{y}; \bm{VaR}(\theta))$. \textbf{P-CAESar} is the version that solely uses the Patton loss $\mathcal{L}^\theta_{q,e}(\bm{\hat{e}}, \bm{\hat{q}}, \bm{y})$. The best result is in bold, while the second to best is underlined.}
    \label{tab:comp_caesar_loss}
    \end{adjustwidth}
\end{table}\\
This comparison shows a balanced situation, with all competing models closely aligned. However, a deeper inspection of the results shows that the current version appears to be the most stable, that is, less susceptible to negative outliers. This is the rationale for our choice.
\newpage
\section{Supplementary Material - Simulated Regimes Switch}
\label{app:srs}
This section analyzes the simulation results in a regime-switching framework. The starting point is the simulation study described in Section \ref{subs:sim_st} of the main text. That is, GARCH processes with Gaussian and Student's t innovations are used. Now, we add a regime-switching component to show the advantage of CAESar in quickly adapting to new regimes. Specifically, for every simulated process, two regimes are considered: bull and bear. The former has a positive mean $\mu_{bull}$ and smaller GARCH coefficients, and the latter has a negative mean and larger GARCH coefficients. Specifically:
\begin{equation}
    \mu_{bull} = -2\mu_{bear} \quad\quad \omega_{bull} = \frac{1}{2}\omega_{bear} \quad\quad \beta_{bull} = \frac{1}{2}\beta_{bear} \quad\quad \gamma_{bull} = \frac{1}{2}\gamma_{bear}
\end{equation}
As for the transition probabilities, we set $bull\rightarrow bear = bear\rightarrow bull = 0.05$. The initial regime is set as $bull$ or $bear$ with equal probabilities. CAESar, GAS2, and Al-AR models are compared in Table \ref{tab:sim_study_rs}, which clearly shows the superiority of CAESar. This result is achieved due to the peculiar functional form of the CAESar ES regressor, which allows it to catch the regime changes rapidly.
  
\begin{table}
  \centering
  \begin{adjustwidth}{-1.6cm}{}
  \tiny{
  \begin{tabular}{cc|c|c|c|c|c|c|c|c|c|c|c|c|c}
\toprule
\multirow{2}{*}{$\bm{\theta = 0.05}$} & & \multicolumn{3}{c|}{\textbf{N}} & \multicolumn{3}{c|}{\textbf{t}} & & \multicolumn{3}{c|}{\textbf{N}} & \multicolumn{3}{c}{\textbf{t}} \\
& & \textbf{I} & \textbf{II} & \textbf{III} & \textbf{I} & \textbf{II} & \textbf{III} & & \textbf{I} & \textbf{II} & \textbf{III} & \textbf{I} & \textbf{II} & \textbf{III} \\
\midrule
\multirow{2}{*}{\textbf{CAESar}} & MAE & \textbf{0.001} & \textbf{0.0012} & \textbf{0.001} & \textbf{0.0007} & \textbf{0.0017} & \textbf{0.0013} & $\mathcal{L}_{B}^{\theta}$ & \textbf{0.0} & \textbf{0.0002} & \textbf{0.0} & \textbf{0.0001} & \textbf{0.0003} & \textbf{0.0002} \\
& RMSE & \textbf{0.001} & \textbf{0.001} & \textbf{0.001} & \textbf{0.001} & \textbf{0.002} & \textbf{0.001} & $\mathcal{L}_{P}^{\theta}$ & \textbf{-9.126} & \textbf{-2.596} & \textbf{-18.607} & \textbf{-4.165} & \underline{-2.596} & \underline{-3.998} \\
\midrule
\multirow{2}{*}{\textbf{GAS2}} & MAE & 0.3088 & 0.2935 & 0.2879 & 0.2632 & 0.4531 & 0.4332 & $\mathcal{L}_{B}^{\theta}$ & 0.0374 & 0.0288 & \underline{0.0286} & \underline{0.0631} & 0.0452 & \underline{0.0444} \\
& RMSE & 0.309 & 0.294 & 0.288 & 0.263 & 0.453 & 0.433 & $\mathcal{L}_{P}^{\theta}$ & \underline{1.611} & \underline{1.676} & 1.697 & 1.149 & 2.548 & 2.551 \\
\midrule
\multirow{2}{*}{\textbf{AL-AR}} & MAE & \underline{0.2242} & \underline{0.0457} & \underline{0.0064} & \underline{0.0978} & \underline{0.1004} & \underline{0.0569} & $\mathcal{L}_{B}^{\theta}$ & \underline{0.0014} & \underline{0.0264} & 0.0495 & 2.242 & \underline{0.0394} & 0.1655 \\
& RMSE & \underline{0.224} & \underline{0.046} & \underline{0.006} & \underline{0.098} & \underline{0.1} & \underline{0.057} & $\mathcal{L}_{P}^{\theta}$ & 4.448 & 6.254 & \underline{1.584} & \underline{-3.201} & \textbf{-3.68} & \textbf{-15.903} \\
\bottomrule
\end{tabular}

\vspace{1em}

\begin{tabular}{cc|c|c|c|c|c|c|c|c|c|c|c|c|c}
\toprule
\multirow{2}{*}{$\bm{\theta = 0.025}$} & & \multicolumn{3}{c|}{\textbf{N}} & \multicolumn{3}{c|}{\textbf{t}} & & \multicolumn{3}{c|}{\textbf{N}} & \multicolumn{3}{c}{\textbf{t}} \\
& & \textbf{I} & \textbf{II} & \textbf{III} & \textbf{I} & \textbf{II} & \textbf{III} & & \textbf{I} & \textbf{II} & \textbf{III} & \textbf{I} & \textbf{II} & \textbf{III} \\
\midrule
\multirow{2}{*}{\textbf{CAESar}} & MAE & \textbf{0.0011} & \textbf{0.0011} & \textbf{0.001} & \textbf{0.0009} & \textbf{0.0018} & \textbf{0.0015} & $\mathcal{L}_{B}^{\theta}$ & \textbf{0.0001} & \textbf{0.0002} & \textbf{0.0001} & \textbf{0.0003} & \textbf{0.0006} & \textbf{0.0003} \\
& RMSE & \textbf{0.001} & \textbf{0.001} & \textbf{0.001} & \textbf{0.001} & \textbf{0.002} & \textbf{0.001} & $\mathcal{L}_{P}^{\theta}$ & \textbf{-8.311} & \textbf{-4.32} & \textbf{-8.987} & \textbf{-1.993} & \textbf{-5.031} & \underline{-4.65} \\
\midrule
\multirow{2}{*}{\textbf{GAS2}} & MAE & 0.533 & 0.4935 & 0.5746 & 0.3651 & 0.5174 & 0.7071 & $\mathcal{L}_{B}^{\theta}$ & \underline{0.1365} & \underline{0.0453} & \underline{0.4753} & \underline{0.0389} & 0.7863 & \underline{0.0654} \\
& RMSE & 0.533 & 0.493 & 0.575 & 0.365 & 0.517 & 0.707 & $\mathcal{L}_{P}^{\theta}$ & \underline{2.273} & \underline{2.107} & \underline{2.463} & \underline{1.429} & 2.127 & 2.869 \\
\midrule
\multirow{2}{*}{\textbf{AL-AR}} & MAE & \underline{0.2235} & \underline{0.1313} & \underline{0.099} & \underline{0.1061} & \underline{0.0432} & \underline{0.0825} & $\mathcal{L}_{B}^{\theta}$ & 1.1328 & 0.6813 & 1.8513 & 12.3794 & \underline{0.476} & 0.8662 \\
& RMSE & \underline{0.223} & \underline{0.131} & \underline{0.099} & \underline{0.106} & \underline{0.043} & \underline{0.082} & $\mathcal{L}_{P}^{\theta}$ & 6.127 & 7.331 & 7.252 & 4.42 & \underline{-4.241} & \textbf{-6.501} \\
\bottomrule
\end{tabular}

\vspace{1em}

\begin{tabular}{cc|c|c|c|c|c|c|c|c|c|c|c|c|c}
\toprule
\multirow{2}{*}{$\bm{\theta = 0.01}$} & & \multicolumn{3}{c|}{\textbf{N}} & \multicolumn{3}{c|}{\textbf{t}} & & \multicolumn{3}{c|}{\textbf{N}} & \multicolumn{3}{c}{\textbf{t}} \\
& & \textbf{I} & \textbf{II} & \textbf{III} & \textbf{I} & \textbf{II} & \textbf{III} & & \textbf{I} & \textbf{II} & \textbf{III} & \textbf{I} & \textbf{II} & \textbf{III} \\
\midrule
\multirow{2}{*}{\textbf{CAESar}} & MAE & \textbf{0.0015} & \textbf{0.0013} & \textbf{0.0014} & \textbf{0.0013} & \textbf{0.0021} & \textbf{0.0017} & $\mathcal{L}_{B}^{\theta}$ & \textbf{0.0006} & \textbf{0.0015} & \textbf{0.0009} & \textbf{0.0011} & \textbf{0.0016} & \textbf{0.001} \\
& RMSE & \textbf{0.001} & \textbf{0.001} & \textbf{0.001} & \textbf{0.001} & \textbf{0.002} & \textbf{0.002} & $\mathcal{L}_{P}^{\theta}$ & \textbf{-2.49} & \textbf{-8.873} & \textbf{-2.655} & \textbf{-0.232} & \textbf{-0.68} & \textbf{-3.924} \\
\midrule
\multirow{2}{*}{\textbf{GAS2}} & MAE & 0.7588 & 1.0565 & 0.7934 & 0.4709 & 0.6539 & 0.6503 & $\mathcal{L}_{B}^{\theta}$ & \underline{0.1195} & \underline{0.0994} & \underline{0.0746} & \underline{0.0933} & \underline{0.0628} & \underline{0.0613} \\
& RMSE & 0.759 & 1.057 & 0.793 & 0.471 & 0.654 & 0.65 & $\mathcal{L}_{P}^{\theta}$ & \underline{2.182} & 3.12 & \underline{2.035} & \underline{1.85} & 3.066 & \underline{2.257} \\
\midrule
\multirow{2}{*}{\textbf{AL-AR}} & MAE & \underline{0.1108} & \underline{0.0052} & \underline{0.0875} & \underline{0.009} & \underline{0.0837} & \underline{0.0721} & $\mathcal{L}_{B}^{\theta}$ & 0.9076 & 4.5673 & 1.4464 & 49.0873 & 2.5932 & 0.8248 \\
& RMSE & \underline{0.111} & \underline{0.005} & \underline{0.087} & \underline{0.009} & \underline{0.084} & \underline{0.072} & $\mathcal{L}_{P}^{\theta}$ & 3.74 & \underline{-7.949} & 4.3 & 2.544 & \underline{2.924} & 2.98 \\
\bottomrule
\end{tabular}
}
\end{adjustwidth}
  \caption{Simulation results with a regime-switching model. All the generating processes are based on GARCH, both with Gaussian innovations (\textbf{N}) and t innovations (\textbf{t}). The coefficients of the process are fitted on the SPX (\textbf{I} set of coefficients), FTSE (\textbf{II} set), or DAX (\textbf{III} set). Two regimes are considered, namely bull and bearish, with equal transition probabilities. The values shown represent the mean loss over twenty time series. The best result is in bold, the second to best is underlined.}
    \label{tab:sim_study_rs}
\end{table}

\section{Supplementary Material - Model Comparison Tests}
\label{app:eld}
This Supplementary Material section shows the results obtained by the Loss Difference, the Encompassing, and the corrected resampled Student's t-test. Both are model comparison tests. The formers are applied asset by asset and fold by fold, while the latter is applied once for each asset. These tests look at the out-of-sample losses to asses whether the compared models are statistically equivalent (null hypothesis). As for the Loss Difference test, the exceptions' number (p-value $<$ 0.05) is reported in Table \ref{tab:ldtest_mig}.
\begin{table}[h]
  \centering
    \tiny{\begin{tabular}{cc|c|c|c|c|c|c|c}
    \toprule
    & & \textbf{K-CAViaR} & \textbf{BCGNS} & \textbf{K-QRNN} & \textbf{GAS1} & \textbf{GAS2} & \textbf{AL-Multi} & \textbf{AL-AR} \\
    \midrule
    \multirow{2}{*}{$\bm{\theta = 0.05}$} & $\mathcal{L}_{B}^{\theta}$ &  37 / 81 & 185 / 12 & 50 / 112 & 97 / 47 & 97 / 50 & 50 / 91 & 123 / 61 \\
     & $\mathcal{L}_{P}^{\theta}$ & 44 / 26 & 228 / 0 & 132 / 5 & 88 / 3 & 127 / 8 & 70 / 29 & 132 / 32 \\
        \midrule
    \multirow{2}{*}{$\bm{\theta = 0.025}$} & $\mathcal{L}_{B}^{\theta}$ & 48 / 70 & 145 / 32 & 54 / 87 & 84 / 53 & 105 / 37 & 42 / 91 & 135 / 73 \\
     & $\mathcal{L}_{P}^{\theta}$ & 40 / 40 & 165 / 2 & 113 / 5 & 115 / 9 & 126 / 5 & 47 / 40 & 130 / 49  \\
        \midrule
    \multirow{2}{*}{$\bm{\theta = 0.01}$} & $\mathcal{L}_{B}^{\theta}$ & 39 / 67 & 107 / 38 & 44 / 83 & 90 / 36 & 71 / 31 & 41 / 80 & 139 / 54 \\
     & $\mathcal{L}_{P}^{\theta}$ & 50 / 31 & 106 / 9 & 99 / 9 & 117 / 13 & 109 / 8  & 62 / 51 & 143 / 55 \\
    \bottomrule
    \end{tabular}}
  \caption{Model Comparison Tests - Loss Difference test with the p-value threshold equal to 0.05. The results obtained across various assets, for a given probability level $\theta$, have been aggregated by summing them. The first value in each cell represents the number of folds where CAESar outperforms the competitor. The second value is the number of folds where the competitor outperforms CAESar.}
    \label{tab:ldtest_mig}
\end{table}\\
From the table, we can recover a certain discrepancy between losses. Indeed, although CAESar generally performs better than K-CAViaR, K-QRNN, and AL-Multi when looking at $\mathcal{L}_P^\theta$, the opposite is true for the $\mathcal{L}_B^\theta$ loss. The intuition behind this is the same as before: as discussed in Section \ref{subs:sim_st} of the main test, the Barrera loss requires the correctness of the quantile estimate, which sometimes may be an incorrect assumption. Finally, the results for the encompassing test are shown in Table \ref{tab:entest_mig}, which follows the same structure and content as the previous tables.
\begin{table}
  \centering
    \tiny{\begin{tabular}{cc|c|c|c|c|c|c|c}
    \toprule
    & & \textbf{K-CAViaR} & \textbf{BCGNS} & \textbf{K-QRNN} & \textbf{GAS1} & \textbf{GAS2} & \textbf{AL-Multi} & \textbf{AL-AR} \\
    \midrule
    \multirow{2}{*}{$\bm{\theta = 0.05}$} & $\mathcal{L}_{B}^{\theta}$ & 3 / 21 & 55 / 0 & 4 / 50 & 13 / 5 & 26 / 17 & 13 / 32 & 100 / 21 \\
    & $\mathcal{L}_{P}^{\theta}$ & 29 / 18 & 127 / 0 & 29 / 18 & 40 / 9 & 26 / 4 & 34 / 17 & 73 / 28 \\
        \midrule
    \multirow{2}{*}{$\bm{\theta = 0.025}$} & $\mathcal{L}_{B}^{\theta}$ & 20 / 21 & 33 / 7 & 18 / 37 & 34 / 19 & 54 / 23 & 25 / 30 & 101 / 31 \\
    & $\mathcal{L}_{P}^{\theta}$ & 23 / 24 & 50 / 5 & 41 / 20 & 51 / 24 & 38 / 20 & 31 / 19 & 67 / 42 \\
        \midrule
    \multirow{2}{*}{$\bm{\theta = 0.01}$} & $\mathcal{L}_{B}^{\theta}$ & 41 / 43 & 34 / 44 & 30 / 73 & 63 / 36 & 57 / 23 & 32 / 59 & 120 / 51 \\
    & $\mathcal{L}_{P}^{\theta}$ & 46 / 41 & 46 / 27 & 57 / 32 & 73 / 35 & 57 / 23 & 50 / 40 & 97 / 54 \\
    \bottomrule
    \end{tabular}}
  \caption{Model Comparison Tests - Encompassing test. The testing procedure and the results presentation are the same as in the Loss Difference test.}
    \label{tab:entest_mig}
\end{table}\\
The test results qualitatively are the same as those obtained by the Loss Difference test. The only difference is in the rejection numbers. Indeed, the Encompassing test is the most prudent test, while the Loss Difference test is the most permissive.\\
\linebreak
Another perspective in comparing models is provided by the corrected resampled Student's t-test. We have applied this test once for each asset. This is because it takes as input a vector where each component represents the mean loss in a fold (as its population is made up of the losses among the folds). Its results are shown in Table \ref{tab:crtest}. The table shows the number of rejections of the null hypothesis about models' equivalence. Each cell represents the number of indexes for which the row predictor overcomes the column one. Therefore, the better models are those with low values in the corresponding column and high values in their row. CAESar and K-CAViaR are confirmed to be the best-performing models, with CAESar slightly outperforming. However, the test often fails to reject the null hypothesis. This aligns with the findings in \cite{deng2021backtesting}, indicating that a testing procedure based on the joint eligibility of VaR and ES requires very strong evidence for rejection.
\begin{table}[h!]
    \centering
    \begin{adjustwidth}{-.1cm}{}
\tiny{
    \begin{tabular}{cc|c|c|c|c|c|c|c|c}
        \toprule
        $\bm{\theta = 0.05}$ & & \textbf{CAESar} & \textbf{K-CAViaR} & \textbf{BCGNS} & \textbf{K-QRNN} & \textbf{GAS1} & \textbf{GAS2} & \textbf{AL-Multi} & \textbf{AL-AR} \\
\midrule
        \textbf{CAESar} & $\mathcal{L}_{B}^{\theta}$ / $\mathcal{L}_{P}^{\theta}$ & - / - & 0 / 0 & 1 / 5 & 0 / 8 & 0 / 0 & 0 / 1 & 0 / 0 & 0 / 0 \\
\textbf{K-CAViaR} & $\mathcal{L}_{B}^{\theta}$ / $\mathcal{L}_{P}^{\theta}$ & 0 / 0 & - / - & 1 / 4 & 0 / 8 & 0 / 0 & 0 / 1 & 0 / 0 & 0 / 0 \\
\textbf{BCGNS} & $\mathcal{L}_{B}^{\theta}$ / $\mathcal{L}_{P}^{\theta}$ & 0 / 0 & 0 / 0 & - / - & 0 / 0 & 0 / 0 & 0 / 0 & 0 / 0 & 0 / 0 \\
\textbf{K-QRNN} & $\mathcal{L}_{B}^{\theta}$ / $\mathcal{L}_{P}^{\theta}$ & 0 / 0 & 0 / 0 & 0 / 4 & - / - & 0 / 0 & 0 / 0 & 0 / 0 & 0 / 0 \\
\textbf{GAS1} & $\mathcal{L}_{B}^{\theta}$ / $\mathcal{L}_{P}^{\theta}$ & 0 / 0 & 0 / 0 & 0 / 0 & 0 / 0 & - / - & 0 / 0 & 0 / 0 & 0 / 0 \\
\textbf{GAS2} & $\mathcal{L}_{B}^{\theta}$ / $\mathcal{L}_{P}^{\theta}$ & 0 / 0 & 0 / 0 & 0 / 0 & 0 / 0 & 0 / 0 & - / - & 0 / 0 & 0 / 0 \\
\textbf{AL-Multi} & $\mathcal{L}_{B}^{\theta}$ / $\mathcal{L}_{P}^{\theta}$ & 0 / 0 & 0 / 0 & 0 / 0 & 0 / 0 & 0 / 0 & 0 / 0 & - / - & 0 / 0 \\
\textbf{AL-AR} & $\mathcal{L}_{B}^{\theta}$ / $\mathcal{L}_{P}^{\theta}$ & 0 / 0 & 0 / 0 & 0 / 0 & 0 / 0 & 0 / 0 & 0 / 0 & 0 / 0 & - / - \\
\bottomrule
\end{tabular}
\vspace{1em}
\begin{tabular}{cc|c|c|c|c|c|c|c|c}
\toprule
$\bm{\theta = 0.025}$ & & \textbf{CAESar} & \textbf{K-CAViaR} & \textbf{BCGNS} & \textbf{K-QRNN} & \textbf{GAS1} & \textbf{GAS2} & \textbf{AL-Multi} & \textbf{AL-AR} \\
\midrule
\textbf{CAESar} & $\mathcal{L}_{B}^{\theta}$ / $\mathcal{L}_{P}^{\theta}$ & - / - & 0 / 0 & 0 / 1 & 0 / 6 & 0 / 0 & 0 / 0 & 0 / 0 & 0 / 0 \\
\textbf{K-CAViaR} & $\mathcal{L}_{B}^{\theta}$ / $\mathcal{L}_{P}^{\theta}$ & 0 / 0 & - / - & 0 / 2 & 0 / 5 & 0 / 0 & 0 / 0 & 0 / 0 & 0 / 0 \\
\textbf{BCGNS} & $\mathcal{L}_{B}^{\theta}$ / $\mathcal{L}_{P}^{\theta}$ & 0 / 0 & 0 / 0 & - / - & 0 / 0 & 0 / 0 & 0 / 0 & 0 / 0 & 0 / 0 \\
\textbf{K-QRNN} & $\mathcal{L}_{B}^{\theta}$ / $\mathcal{L}_{P}^{\theta}$ & 0 / 0 & 0 / 0 & 0 / 0 & - / - & 0 / 0 & 0 / 0 & 0 / 0 & 0 / 0 \\
\textbf{GAS1} & $\mathcal{L}_{B}^{\theta}$ / $\mathcal{L}_{P}^{\theta}$ & 0 / 0 & 0 / 0 & 0 / 0 & 0 / 0 & - / - & 0 / 0 & 0 / 0 & 0 / 0 \\
\textbf{GAS2} & $\mathcal{L}_{B}^{\theta}$ / $\mathcal{L}_{P}^{\theta}$ & 0 / 0 & 0 / 0 & 0 / 0 & 0 / 0 & 0 / 0 & - / - & 0 / 0 & 0 / 0 \\
\textbf{AL-Multi} & $\mathcal{L}_{B}^{\theta}$ / $\mathcal{L}_{P}^{\theta}$ & 0 / 0 & 0 / 0 & 0 / 0 & 0 / 0 & 0 / 0 & 0 / 0 & - / - & 0 / 0 \\
\textbf{AL-AR} & $\mathcal{L}_{B}^{\theta}$ / $\mathcal{L}_{P}^{\theta}$ & 0 / 0 & 0 / 0 & 0 / 0 & 0 / 0 & 0 / 0 & 0 / 0 & 0 / 0 & - / - \\
\bottomrule
\end{tabular}
\vspace{1em}
\begin{tabular}{cc|c|c|c|c|c|c|c|c}
\toprule
$\bm{\theta = 0.01}$ & & \textbf{CAESar} & \textbf{K-CAViaR} & \textbf{BCGNS} & \textbf{K-QRNN} & \textbf{GAS1} & \textbf{GAS2} & \textbf{AL-Multi} & \textbf{AL-AR} \\
\midrule
\textbf{CAESar} & $\mathcal{L}_{B}^{\theta}$ / $\mathcal{L}_{P}^{\theta}$ & - / - & 0 / 0 & 0 / 0 & 0 / 4 & 0 / 0 & 0 / 0 & 0 / 0 & 0 / 0 \\
\textbf{K-CAViaR} & $\mathcal{L}_{B}^{\theta}$ / $\mathcal{L}_{P}^{\theta}$ & 0 / 0 & - / - & 0 / 0 & 0 / 3 & 0 / 0 & 0 / 0 & 0 / 0 & 0 / 0 \\
\textbf{BCGNS} & $\mathcal{L}_{B}^{\theta}$ / $\mathcal{L}_{P}^{\theta}$ & 0 / 0 & 0 / 0 & - / - & 0 / 0 & 0 / 0 & 0 / 0 & 0 / 0 & 0 / 0 \\
\textbf{K-QRNN} & $\mathcal{L}_{B}^{\theta}$ / $\mathcal{L}_{P}^{\theta}$ & 0 / 0 & 0 / 0 & 0 / 0 & - / - & 0 / 0 & 0 / 0 & 0 / 0 & 0 / 0 \\
\textbf{GAS1} & $\mathcal{L}_{B}^{\theta}$ / $\mathcal{L}_{P}^{\theta}$ & 0 / 0 & 0 / 0 & 0 / 0 & 0 / 0 & - / - & 0 / 0 & 0 / 0 & 0 / 0 \\
\textbf{GAS2} & $\mathcal{L}_{B}^{\theta}$ / $\mathcal{L}_{P}^{\theta}$ & 0 / 0 & 0 / 0 & 0 / 0 & 0 / 0 & 0 / 0 & - / - & 0 / 0 & 0 / 0 \\
\textbf{AL-Multi} & $\mathcal{L}_{B}^{\theta}$ / $\mathcal{L}_{P}^{\theta}$ & 0 / 0 & 0 / 0 & 0 / 0 & 0 / 0 & 0 / 0 & 0 / 0 & - / - & 0 / 0 \\
\textbf{AL-AR} & $\mathcal{L}_{B}^{\theta}$ / $\mathcal{L}_{P}^{\theta}$ & 0 / 0 & 0 / 0 & 0 / 0 & 0 / 0 & 0 / 0 & 0 / 0 & 0 / 0 & - / - \\
\bottomrule
    \end{tabular}
    }
  \caption{Model Comparison Tests - Corrected resampled t-test with confidence 0.05. Each cell value represents the number of assets where the row algorithm performs statistically better than column one.}
    \label{tab:crtest}
    \end{adjustwidth}
\end{table}
\newpage
\section{Supplementary Material - Stock Dataset}
\label{app:sd}
To further validate our proposal, we carry out an additional experiment using a stock dataset. Specifically, we select a dataset of stocks from the US banking sector, as done in \cite{keilbar2022modelling}. This new dataset consists of 8 stocks and covers the period from 2000-01-01 to 2023-12-31. Again, we use block cross-validation to evaluate performance, with 6 years for training/validation and 1 year for the test, resulting in 18 folds. The results from ES forecasting are shown in Table \ref{tab:caesar_loss_bank}. As before, both $\mathcal{L}_{B}^{\theta}$ and $\mathcal{L}_{P}^{\theta}$ are shown.
\begin{table}
  \centering
  \vspace{-2cm}
  \tiny{
\begin{tabular}{cc|c|c|c|c|c|c|c|c}
\toprule
$\bm{\theta = 0.05}$ & & \textbf{BAC} & \textbf{BK} & \textbf{C} & \textbf{GS} & \textbf{JPM} & \textbf{MS} & \textbf{STT} & \textbf{WFC} \\
\midrule
\multirow{2}{*}{\textbf{CAESar}} & $\mathcal{L}_{B}^{\theta}$ & \textbf{0.0126} & 0.0138 & \textbf{0.0133} & \underline{0.0087} & \textbf{0.0065} & 0.0262 & \underline{0.0338} & 0.0088 \\
 & $\mathcal{L}_{P}^{\theta}$ & \underline{1.467} & \underline{1.758} & \underline{1.468} & \textbf{1.364} & \textbf{1.253} & 2.59 & 1.563 & \textbf{1.327} \\
\midrule
\multirow{2}{*}{\textbf{K-CAViaR}} & $\mathcal{L}_{B}^{\theta}$ & \underline{0.0128} & 0.0141 & \underline{0.0134} & \textbf{0.0084} & \underline{0.0067} & 0.0258 & 0.0347 & \underline{0.0079} \\
 & $\mathcal{L}_{P}^{\theta}$ & \textbf{1.419} & \textbf{1.627} & 1.483 & \underline{1.408} & \underline{1.26} & 3.345 & \underline{1.547} & \underline{1.337} \\
\midrule
\multirow{2}{*}{\textbf{BCGNS}} & $\mathcal{L}_{B}^{\theta}$ & 0.0532 & 0.0386 & 0.0948 & 0.02 & 0.041 & 0.0431 & 0.0716 & 0.0365 \\
 & $\mathcal{L}_{P}^{\theta}$ & 3.424 & 3.201 & 3.33 & 2.298 & 2.497 & 6.646 & 3.0 & 4.818 \\
\midrule
\multirow{2}{*}{\textbf{K-QRNN}} & $\mathcal{L}_{B}^{\theta}$ & 0.0639 & 0.0257 & 0.075 & 0.0222 & 0.0265 & 0.0433 & 0.0565 & 0.0369 \\
 & $\mathcal{L}_{P}^{\theta}$ & 2.153 & 1.797 & 2.211 & 1.74 & 1.906 & 1.987 & 1.984 & 2.146 \\
\midrule
\multirow{2}{*}{\textbf{GAS1}} & $\mathcal{L}_{B}^{\theta}$ & 0.0268 & \underline{0.0111} & 0.0391 & 0.0143 & 0.0183 & 0.0201 & \textbf{0.0334} & \textbf{0.0074} \\
 & $\mathcal{L}_{P}^{\theta}$ & 1.793 & 2.023 & 1.989 & 1.444 & 1.575 & \textbf{1.532} & 1.853 & 1.499 \\
\midrule
\multirow{2}{*}{\textbf{GAS2}} & $\mathcal{L}_{B}^{\theta}$ & 0.5711 & 0.0255 & 0.1433 & 0.0238 & 0.0415 & 0.0236 & 0.2587 & 0.028 \\
 & $\mathcal{L}_{P}^{\theta}$ & 2.0 & 1.97 & 2.123 & 3.412 & 1.871 & 2.489 & 2.213 & 2.123 \\
\midrule
\multirow{2}{*}{\textbf{AL-Multi}} & $\mathcal{L}_{B}^{\theta}$ & 0.2952 & \textbf{0.006} & 0.1364 & 0.1463 & 0.1301 & \underline{0.0131} & 0.0443 & 0.0088 \\
 & $\mathcal{L}_{P}^{\theta}$ & 1.554 & 1.767 & 1.718 & 2.03 & 1.383 & \underline{1.647} & 1.746 & 1.362 \\
\midrule
\multirow{2}{*}{\textbf{AL-AR}} & $\mathcal{L}_{B}^{\theta}$ & 0.91 & 0.4118 & 0.253 & 0.679 & 0.3346 & \textbf{0.0127} & 0.0447 & 0.9223 \\
 & $\mathcal{L}_{P}^{\theta}$ & 1.537 & 2.504 & \textbf{1.318} & 4.916 & 3.359 & 2.41 & \textbf{1.326} & 1.965 \\
\bottomrule
\end{tabular}

\vspace{1em}

\begin{tabular}{cc|c|c|c|c|c|c|c|c}
\toprule
$\bm{\theta = 0.025}$ & & \textbf{BAC} & \textbf{BK} & \textbf{C} & \textbf{GS} & \textbf{JPM} & \textbf{MS} & \textbf{STT} & \textbf{WFC} \\
\midrule
\multirow{2}{*}{\textbf{CAESar}} & $\mathcal{L}_{B}^{\theta}$ & 0.0425 & 0.0766 & \textbf{0.0292} & \textbf{0.0221} & 0.0159 & \underline{0.0323} & \underline{0.1072} & \underline{0.0182} \\
 & $\mathcal{L}_{P}^{\theta}$ & 2.352 & 2.034 & \textbf{1.679} & \underline{1.597} & \underline{1.476} & \underline{1.714} & 1.962 & \textbf{1.543} \\
\midrule
\multirow{2}{*}{\textbf{K-CAViaR}} & $\mathcal{L}_{B}^{\theta}$ & \underline{0.0315} & 0.0773 & 0.0623 & \underline{0.0236} & \underline{0.0155} & 0.077 & 0.1075 & \textbf{0.0179} \\
 & $\mathcal{L}_{P}^{\theta}$ & \textbf{1.639} & \underline{1.759} & \underline{1.763} & \textbf{1.587} & 1.476 & 2.229 & \textbf{1.769} & \underline{1.544} \\
\midrule
\multirow{2}{*}{\textbf{BCGNS}} & $\mathcal{L}_{B}^{\theta}$ & 0.1456 & 0.0979 & 0.1475 & 0.0459 & 0.0609 & 0.1322 & 0.1908 & 0.0725 \\
 & $\mathcal{L}_{P}^{\theta}$ & 2.587 & 3.35 & 2.849 & 2.757 & 2.051 & 2.832 & 3.311 & 3.084 \\
\midrule
\multirow{2}{*}{\textbf{K-QRNN}} & $\mathcal{L}_{B}^{\theta}$ & 0.2679 & 0.1076 & 0.3197 & 0.0798 & 0.1072 & 0.1852 & 0.2295 & 0.1709 \\
 & $\mathcal{L}_{P}^{\theta}$ & 3.691 & 2.492 & 3.676 & 2.273 & 2.592 & 2.685 & 2.68 & 3.431 \\
\midrule
\multirow{2}{*}{\textbf{GAS1}} & $\mathcal{L}_{B}^{\theta}$ & 0.0413 & \textbf{0.0259} & \underline{0.0402} & 0.0295 & \textbf{0.0116} & 0.117 & 0.1228 & 0.0243 \\
 & $\mathcal{L}_{P}^{\theta}$ & \underline{1.837} & 1.949 & 1.865 & 1.791 & 1.521 & 2.186 & 2.217 & 1.773 \\
\midrule
\multirow{2}{*}{\textbf{GAS2}} & $\mathcal{L}_{B}^{\theta}$ & \textbf{0.0293} & 0.3891 & 0.2874 & 0.0367 & 0.057 & \textbf{0.0257} & 0.2466 & 0.0418 \\
 & $\mathcal{L}_{P}^{\theta}$ & 3.188 & 2.456 & 2.829 & 2.516 & 3.164 & 2.168 & 2.068 & 2.255 \\
\midrule
\multirow{2}{*}{\textbf{AL-Multi}} & $\mathcal{L}_{B}^{\theta}$ & 0.069 & \underline{0.0655} & 0.2684 & 0.0302 & 0.0157 & 0.1163 & \textbf{0.1011} & 0.0439 \\
 & $\mathcal{L}_{P}^{\theta}$ & 2.428 & \textbf{1.676} & 1.877 & 2.807 & \textbf{1.454} & 2.157 & 2.36 & 2.379 \\
\midrule
\multirow{2}{*}{\textbf{AL-AR}} & $\mathcal{L}_{B}^{\theta}$ & 0.7725 & 0.624 & 0.3008 & 0.1197 & 0.0621 & 1.1717 & 0.4551 & 0.2793 \\
 & $\mathcal{L}_{P}^{\theta}$ & 2.961 & 2.778 & 4.845 & 3.908 & 2.111 & \textbf{1.645} & \underline{1.892} & 1.766 \\
\bottomrule
\end{tabular}

\vspace{1em}

\begin{tabular}{cc|c|c|c|c|c|c|c|c}
\toprule
$\bm{\theta = 0.01}$ & & \textbf{BAC} & \textbf{BK} & \textbf{C} & \textbf{GS} & \textbf{JPM} & \textbf{MS} & \textbf{STT} & \textbf{WFC} \\
\midrule
\multirow{2}{*}{\textbf{CAESar}} & $\mathcal{L}_{B}^{\theta}$ & \textbf{0.2374} & 0.4018 & 0.1063 & \underline{0.1102} & \textbf{0.0596} & 0.3967 & 0.4823 & \underline{0.0352} \\
 & $\mathcal{L}_{P}^{\theta}$ & \underline{2.44} & 2.519 & \underline{1.985} & \textbf{1.886} & \underline{1.977} & 2.224 & 2.2 & \textbf{1.748} \\
\midrule
\multirow{2}{*}{\textbf{K-CAViaR}} & $\mathcal{L}_{B}^{\theta}$ & \underline{0.4017} & 0.4149 & 0.1034 & 0.1473 & \underline{0.0722} & 0.3981 & 0.6182 & 0.062 \\
 & $\mathcal{L}_{P}^{\theta}$ & 2.828 & 3.104 & \textbf{1.969} & \underline{1.951} & \textbf{1.82} & 2.211 & 2.23 & 1.817 \\
\midrule
\multirow{2}{*}{\textbf{BCGNS}} & $\mathcal{L}_{B}^{\theta}$ & 0.7912 & 0.3912 & 0.7121 & 0.1945 & 0.2417 & 0.5786 & 0.8724 & 0.2629 \\
 & $\mathcal{L}_{P}^{\theta}$ & 5.339 & 3.434 & 3.387 & 2.471 & 2.318 & 2.877 & 3.309 & 2.565 \\
\midrule
\multirow{2}{*}{\textbf{K-QRNN}} & $\mathcal{L}_{B}^{\theta}$ & 0.564 & 0.4053 & 0.7435 & 0.2382 & 0.3498 & 0.6308 & 0.8121 & 0.3415 \\
 & $\mathcal{L}_{P}^{\theta}$ & 2.54 & 2.828 & 2.695 & 2.267 & 2.898 & 2.504 & 2.77 & 2.537 \\
\midrule
\multirow{2}{*}{\textbf{GAS1}} & $\mathcal{L}_{B}^{\theta}$ & 0.4972 & 0.3312 & 0.098 & \textbf{0.0469} & 0.3558 & 0.4366 & 0.8826 & 0.1189 \\
 & $\mathcal{L}_{P}^{\theta}$ & \textbf{2.426} & \underline{2.14} & 3.583 & 4.111 & 2.006 & 3.073 & 2.77 & 3.538 \\
\midrule
\multirow{2}{*}{\textbf{GAS2}} & $\mathcal{L}_{B}^{\theta}$ & 0.4444 & \textbf{0.0755} & 0.3093 & 0.199 & 0.1359 & \textbf{0.0841} & \textbf{0.1363} & 0.2065 \\
 & $\mathcal{L}_{P}^{\theta}$ & 2.557 & 2.241 & 2.545 & 3.019 & 2.649 & 2.342 & 2.675 & 2.71 \\
\midrule
\multirow{2}{*}{\textbf{AL-Multi}} & $\mathcal{L}_{B}^{\theta}$ & 0.523 & \underline{0.1382} & \textbf{0.0893} & 0.1185 & 0.1268 & \underline{0.2946} & 0.433 & 0.0383 \\
 & $\mathcal{L}_{P}^{\theta}$ & 3.605 & \textbf{1.874} & 2.055 & 2.148 & 2.263 & \underline{2.076} & \textbf{2.178} & 1.823 \\
\midrule
\multirow{2}{*}{\textbf{AL-AR}} & $\mathcal{L}_{B}^{\theta}$ & 1.175 & 0.4842 & \underline{0.0911} & 0.1615 & 0.4197 & 0.3434 & \underline{0.1699} & \textbf{0.0303} \\
 & $\mathcal{L}_{P}^{\theta}$ & 4.249 & 3.206 & 3.169 & 3.424 & 2.853 & \textbf{2.037} & \underline{2.185} & \underline{1.786} \\
\bottomrule
\end{tabular}
}
  \caption{Out-of-sample loss for stock data. The best result is highlighted in bold, the second-best is underlined. The $*$ is for those values that are outside the CAESar mean + one standard deviation confidence interval.}
    \label{tab:caesar_loss_bank}
\end{table}\\
Also in this case, the CAESar, K-CAViaR, and AL-Multi models outperform the competitors (even if the gap between them is narrower). Also, the GAS models perform well, even if they are often overcome. Specifically, this rise in their performance can be explained by taking into account that the tails for indexes are usually thinner than stocks ones \cite{deng2021backtesting}. Furthermore, as seen in Section \ref{sec:experiments}, we recover the inverse dependence of the loss function value by the confidence level $\theta$. Finally, we report the direct approximation test result in Table \ref{tab:d_approx_bank}.
\begin{table}[h]
    \centering
    \begin{adjustwidth}{-1.6cm}{}
    \begin{tabular}{c|cccc|cccc|cccc}
        \toprule
        & \multicolumn{4}{c|}{$\bm{\theta = 0.05}$} &  \multicolumn{4}{c|}{$\bm{\theta = 0.025}$} &  \multicolumn{4}{c}{$\bm{\theta = 0.01}$} \\
        & $\mathbf{CC}$ & $\mathbf{MNF}$ & $\mathbf{AS1}$ & $\mathbf{AS2}$ & $\mathbf{CC}$ & $\mathbf{MNF}$ & $\mathbf{AS1}$ & $\mathbf{AS2}$ & $\mathbf{CC}$ & $\mathbf{MNF}$ & $\mathbf{AS1}$ & $\mathbf{AS2}$ \\
        \midrule
        \textbf{CAESar} & 0.16 & \underline{0.13} & \textbf{0.12} & 0.17 & 0.18 & \textbf{0.1} & \textbf{0.09} & \underline{0.14} & 0.11 & \textbf{0.32} & \textbf{0.26} & 0.20 \\
        \textbf{K-CAViaR} & 0.15 & 0.19 & \textbf{0.12} & 0.16 & 0.17 & \underline{0.14} & \underline{0.14} & \textbf{0.12} & 0.11 & 0.38 & 0.35 & 0.24 \\
        \textbf{BCGNS} & \textbf{0.02} & 0.28 & 0.13 & \textbf{0.12} & \textbf{0.03} & 0.41 & 0.25 & 0.25 & \textbf{0.07} & 0.61 & 0.39 & 0.4 \\
        \textbf{K-QRNN} & 0.34 & 0.43 & 0.43 & 0.53 & 0.22 & 0.48 & 0.48 & 0.37 & 0.17 & 0.62 & 0.58 & 0.41 \\
        \textbf{GAS1} & \underline{0.11} & \textbf{0.12} & 0.31 & \underline{0.13} & \underline{0.09} & 0.2 & 0.31 & 0.14 & \textbf{0.07} & 0.5 & 0.49 & \textbf{0.18} \\
        \textbf{GAS2} & 0.14 & 0.34 & 0.49 & 0.42 & 0.15 & 0.49 & 0.52 & 0.40 & 0.12 & 0.69 & 0.66 & 0.46 \\
        \textbf{AL-Multi} & 0.14 & 0.18 & 0.16 & 0.22 & 0.15 & 0.26 & 0.18 & 0.16 & 0.08 & \underline{0.36} & \underline{0.34} & \underline{0.19} \\
        \textbf{AL-AR} & 0.13 & 0.36 & 0.42 & 0.34 & 0.15 & 0.41 & 0.42 & 0.29 & 0.14 & 0.42 & 0.48 & 0.35 \\
        \bottomrule
    \end{tabular}
  \caption{Direct approximation test for stock data. The table shows the rejection of the null hypothesis in Christoffersen's conditional coverage test (\textbf{CC)}, McNeil and Frey test (\textbf{MNF}) and in the tests based on the Acerbi and Szekely statistics: Z1 (\textbf{AS1}) and Z2 (\textbf{AS2}). As the null hypothesis of all the tests is related to the correct specification of the ES estimator, the lower the rejection rate, the better the model. Finally, the best result is in bold, and the second-best is underlined.}
    \label{tab:d_approx_bank}
    \end{adjustwidth}
\end{table}\\
When using the rejection rates to compare the algorithms, we find an improvement in CAESar performances (with respect to the results in Table \ref{tab:caesar_loss_bank}). Indeed, while the competition with K-CAViaR is still close for $\theta=0.05$, CAESar certainly defeats the competitors for lower confidence levels. Finally, it is worth observing that BCGNS achieves a good result in the \textbf{AS2} test with $\theta=0.05$.

\end{document}